\documentclass[submission, Phys]{SciPost}
\usepackage[utf8]{inputenc}    
\usepackage{float}
\usepackage{amsmath,braket,mathdots,mathtools,amssymb,xcolor,calligra,color}
\usepackage{amsthm} % JJ
\usepackage{bbm}  
\usepackage{cancel}
\usepackage{tikz}
\usetikzlibrary{spy}
\usetikzlibrary{patterns}
\usetikzlibrary{calc,arrows,positioning}
\usetikzlibrary{arrows,shadows}
\usetikzlibrary{decorations.pathmorphing}	% For Feynman 
\usetikzlibrary{decorations.markings}
\usepackage{pgfplots}
\usepackage{pgfplotstable}
\pgfplotsset{compat=1.14}
\usepackage{graphicx,dblfloatfix}
\usepackage{hyperref}
\newcommand{\be}{\begin{equation}}
\newcommand{\ee}{\end{equation}}
\newcommand{\bea}{\begin{eqnarray}}
\newcommand{\eea}{\end{eqnarray}}

\def\Xint#1{\mathchoice
   {\XXint\displaystyle\textstyle{#1}}%
   {\XXint\textstyle\scriptstyle{#1}}%
   {\XXint\scriptstyle\scriptscriptstyle{#1}}%
   {\XXint\scriptscriptstyle\scriptscriptstyle{#1}}%
   \!\int}

\def\XXint#1#2#3{{\setbox0=\hbox{$#1{#2#3}{\int}$}
     \vcenter{\hbox{$#2#3$}}\kern-.5\wd0}}
\def\ddashint{\Xint=}
\def\dashint{\Xint-}

\def\nn{\nonumber\\}

\def\fr#1{(\ref{#1})}
\def\ontop#1#2{\genfrac{}{}{0pt}{}{#1}{#2}}
\def\sfix#1{\texorpdfstring{#1}{Lg}}

\def\lan{{\lambda}_{a,n}}
\def\lbm{{\lambda}_{b,m}}
\def\gam{(1+\frac{2D}{c})}
\def\gamb{\big(1+\frac{2D}{c}\big)}
\def\qp{{q^\prime}}
\def\omegap{{\omega^\prime}}
\def\lp{L'}
\def\suman{\sum_{\substack{n\\\forall k, \, \lambda_{a,n}\neq\lambda_k\\|\lambda_{a,n}|<\Lambda}}}
\def\sumbm{\sum_{\substack{m\\\forall i, \, \lambda_{b,m}
      \neq\lambda_i\\ \lambda_{b,m}\neq\lambda_{a,n}\\|\lambda_{b,m}|<\Lambda}}}

\def\doi{http://dx.doi.org/}
\let\OLDthebibliography\thebibliography
\renewcommand\thebibliography[1]{
  \OLDthebibliography{#1}
  \setlength{\parskip}{0pt}
  \setlength{\itemsep}{0pt plus 0.3ex}
}

\newcommand{\sign}{\,\text{sgn}\,}

\newcommand{\li}{\,\text{Li}\,}

\newcommand{\1}{\,\pmb{1}}
\newcommand{\D}[1]{\text{d}#1}

\begin{document}
%%%%%%%%%%%%%%%%%%%%%%%%%%%%%%%%%%%%%%%
\begin{center}
{\Large\bf{
%Dynamical density correlations in arbitrary macro states for all space
%and time in the repulsive Lieb-Liniger model at order $c^{-2}$
A systematic $1/c$-expansion of form factor sums for dynamical correlations in the Lieb-Liniger model
}}
\end{center}
%%%%%%%%%%%%%%%%%%%%%%%%%%%%%%%%%%%%%%%
\begin{center}
Etienne Granet\textsuperscript{1$\star$} and Fabian H. L. Essler\textsuperscript{1},
\end{center}
\begin{center}
{\bf 1} The Rudolf Peierls Centre for Theoretical Physics, Oxford
University, Oxford OX1 3PU, UK\\
${}^\star$ {\small \sf etienne.granet@physics.ox.ac.uk}
\end{center}
\date{\today}

\section*{Abstract}
{\bf 
We introduce a framework for calculating dynamical correlations in the
Lieb-Liniger model in arbitrary energy eigenstates and for all
space and time, that combines a Lehmann representation with a $1/c$
expansion. The $n^{\rm th}$ term of the expansion is of order $1/c^n$
and takes into account all $\lfloor \tfrac{n}{2}\rfloor+1$
particle-hole excitations over the averaging eigenstate. Importantly,
in contrast to a ``bare" $1/c$ expansion it is uniform in space and
time. The framework is based on a method for taking the
thermodynamic limit of sums of form factors that exhibit non
integrable singularities. We expect our framework to be applicable to
any local operator.

We determine the first three terms of this expansion and obtain
an explicit expression for
the density-density dynamical correlations and the dynamical
structure factor at order $1/c^2$. We apply these to
finite-temperature equilibrium states and non-equilibrium
  steady states after quantum quenches. We recover predictions of
(nonlinear) Luttinger liquid theory and generalized hydrodynamics
in the appropriate limits, and are able to compute sub-leading
corrections to these.

}

\vspace{10pt}
\noindent\rule{\textwidth}{1pt}
\tableofcontents\thispagestyle{fancy}
\noindent\rule{\textwidth}{1pt}
\vspace{10pt}

%\renewcommand\Affilfont{\fontsize{9}{10.8}\itshape}
%\renewcommand{\thesubsection}{\arabic{subsection}}

%\tableofcontents
%%%%%%%%%%%%%%%%%%%%%%%%%%
\section{Introduction}

The Lieb-Liniger model \cite{LiebLiniger63a} is a key paradigm of
integrable many-particle systems \cite{korepin}. Moreover, it is
directly relevant to a range of cold atom experiments both in and out
of equilibrium, see
e.g. \cite{kww-06,HL:Bose07,cetal-12,naegerl15,Bouchoule16,Fabbri15}.  
While the excitation spectrum at zero temperature \cite{Lieb63} and
thermodynamic properties \cite{YangYang69} have been known for a long time, the
exact solution does not provide easy access to correlations functions
as these encode more detailed information about the exact energy
eigenstates. An exception is the case of impenetrable bosons
\cite{Lenard64,Lenard66,VaidyaTracy79a,VaidyaTracy79b,JMMS80,Thacker81,IIK89,KS90,IIK90a,IIK90b,IIKS90,IIK91,IIKV92,Gangardt04},
which can be mapped onto non-interacting fermions. In absence of full
analytic solutions valuable insights on the large space and time
asymptotic behaviours of correlation functions at zero and low
temperatures were gained by combining exact results on spectral
properties obtained from the Bethe Ansatz with with conformal field
theory (CFT) \cite{BIK86,IKR87} and Luttinger liquid theory \cite{haldane,
  cazalilla} and its recent extensions 
\cite{IG08,PAW09,ISG12,P12,shashipanfilcaux,Price17}. The last two
decades then witnessed remarkable progress in the computation of
zero temperature dynamical correlation functions by expressing them in
terms of spectral representations over the energy eigenstates of the
model. On the one hand it became possible to numerically evaluate the
spectral sums to very high precision for large, finite systems
\cite{CC06,CCS07}. On the other hand remarkable analytic progress 
led to a fairly complete understanding of the asymptotic behaviour at
late times and large distances
\cite{kitanineetcformfactor,kozlowskimaillet,kozlowski4}.
In contrast to ground state case and non-interacting theories
\cite{Perk80,Perk84,Perk09,MPS83a,MPS83b,IIKS93,Colomo93,leclair1996,Doyon05,Reyes06,doyongamsa,Steinberg,GKS18,GKSS19,GKS19} 
progress on determining finite temperature correlators in interacting
integrable models has been much more limited. The basic idea in
interacting integrable models has been to 
again use spectral representations  and sum over ``the most relevant''
states, both for equal time
\cite{rmk,Damerau07,Boos08,Kormos10,Trippe10,Dugave13,Dugave14} and
dynamical correlators
\cite{muss,Saleur00,AKT,EK08,EK:finiteT,PT10,PC14,deNP15,deNP16,GKKKS17,DNP18,CCP19}.  
These summations can again be approached either numerically or analytically.

The numerical approach focuses on finite systems of about a hundred
particles in the case of the Bose gas and works in momentum space,
i.e. considers the dynamical structure factor as a function of
frequency and momentum \cite{PC14}. It then sums the dominant
contributions to the dynamical structure factor in the sense that the
f-sum rule is satisfied to a very high accuracy.

To make analytical progress it is essential to identify the classes of
states that give the dominant contributions in a given range of
frequencies and momenta or space and time
\cite{kitanineetcformfactor}. Known results suggest that 
in interacting theories this generally requires the summation over an
infinite number of states. Firstly, the large space and time asymptotics of
zero temperature dynamical correlators in interacting models has been
shown to be determined by an arbitrary number of (soft) particle-hole 
excitations over the ground state around the Fermi points and the
saddle points of the dispersions of elementary excitations
\cite{kitanineetcformfactor,kozlowski1}. Secondly, it has been shown
that the asymptotic behaviours of dynamical correlations of semi-local
operators in thermal and other finite entropy states involves an
arbitrary number of (soft) particle-hole excitations \cite{GFE20} over
the macro state of interest. Truncating this sum to a finite number of
particle-hole excitations leads to a result that diverges in time.
In the zero temperature case it has been shown that it is possible to
take the thermodynamic limit of (partial) spectral sums and obtain a
representation in terms of (dressed) excitations in the thermodynamic
limit \cite{kozlowski1,kitanineetcformfactor}. An analogous result for
the finite temperature/entropy case would be highly desirable, but is
not known at present. In Refs \cite{deNP16,DNP18,CCP19} such an
expansion in terms of thermodynamic particle-hole excitations was
conjectured. It is based an phenomenological assumptions on how
partial sums over states in the finite volume combine into
thermodynamic form factors. It also exhibits singularities, whose
regularization is not presently known.

Given this state of affairs it is highly desirable to obtain explicit
results through \textit{ab initio} calculations that do not require
any assumptions, i.e. carrying out the spectral sum in a finite volume
and then taking the thermodynamic limit exactly. In order to make
progress in this direction we consider the spectral sum in the
framework of an expansion in the inverse interaction strength $c^{-1}$
around the impenetrable limit. Strong coupling expansions have previously been used at
zero temperature and for static correlators at finite temperatures
\cite{Creamer81,jimbomiwa,Creamer86,korepin84}. More recently the
$1/c$ contribution for the finite temperature dynamical
density-density correlation function was determined in
\cite{deNP15}. This contribution has a particularly simple structure
similar to that of the impenetrable limit, that does not carry over to
the next orders, and as a consequence until now it has been unclear
how to determine higher orders in this expansion. In the following we
develop a method for calculating the 
higher orders of this expansion and apply it to obtain the
contribution to the dynamical density-density correlator at order
$c^{-2}$. The general idea of the $1/c$ expansion, and more generally
of strong coupling expansions in integrable models, is as follows. A
consequence of integrability is that $N$-particle energy eigenstates in
a finite volume can be labelled by $N$ rapidity variables
\be
|\boldsymbol{\lambda}\rangle
=|\lambda_1,\dots,\lambda_N\rangle.
\ee
These rapidities are in a one-to-one correspondence with sets of
(half-odd) integers $\{I_j\}$ through the quantization conditions in the
finite volume
\be
\{\lambda_1,\dots,\lambda_N\} \leftrightarrow \{I_1,\dots, I_N\}.
\ee
The energy and momentum of these states are given by
\begin{align}
E(\boldsymbol{\lambda})&=\sum_{j=1}^N\epsilon(\lambda_j)\ ,\quad
P(\boldsymbol{\lambda})=\sum_{j=1}^Np(\lambda_j)\ ,
\end{align}
where $\epsilon(\lambda)$ and $p(\lambda)$ parametrize the energy and
momentum of a single-particle excitation over the vacuum (reference) state.
For the Bose gas we have $\epsilon(\lambda)=\lambda^2$ and $p(\lambda)=\lambda$.
Two-point correlation functions of a local operator ${\cal O}(x)$ in a
given energy eigenstate $|\boldsymbol{\lambda}\rangle$ thus have
spectral representations of the form
\be
\frac{\langle\boldsymbol{\lambda}|{\cal O}(x,t){\cal
    O}^\dagger(0,0)|\boldsymbol{\lambda}\rangle}{\langle\boldsymbol{\lambda}|\boldsymbol{\lambda}\rangle}= \sum_{M=0}^\infty\sum_{\{\mu_1,\dots,\mu_M\}}
\frac{|\langle\boldsymbol{\lambda}|{\cal
    O}(0,0)|\boldsymbol{\mu}\rangle|^2}
{\langle\boldsymbol{\lambda}|\boldsymbol{\lambda}\rangle\langle\boldsymbol{\mu}|\boldsymbol{\mu}\rangle} 
e^{i\big(E(\boldsymbol{\lambda})-E(\boldsymbol{\mu})\big)t-i
\big(P(\boldsymbol{\lambda})-P(\boldsymbol{\mu})\big)x}\ ,
\label{2pointfn}
\ee
where the first sum runs over the particle number and the second over
all $M$-particle energy eigenstates. The matrix elements
\be
F_{\cal O}(\boldsymbol{\lambda},\boldsymbol{\mu})=
\frac{\langle\boldsymbol{\lambda}|{\cal
    O}(0,0)|\boldsymbol{\mu}\rangle}
{\sqrt{\langle\boldsymbol{\lambda}|\boldsymbol{\lambda}\rangle\langle\boldsymbol{\mu}|\boldsymbol{\mu}\rangle}} 
\ee
are also known as \emph{form factors} and, as we will see, admit a
$1/c$-expansion
\be
F_{\cal O}(\boldsymbol{\lambda},\boldsymbol{\mu})=\sum_{n=0}^\infty
\frac{F_{{\cal O},n}(\boldsymbol{I},\boldsymbol{J})}{c^n},
\ee
where $\boldsymbol{I}=\{I_1,\dots,I_N\}$ and
$\boldsymbol{J}=\{J_1,\dots,J_M\}$ are the (half-odd) integers
corresponding to the rapidities $\lambda_1,\dots,\lambda_N$ and
$\mu_1,\dots,\mu_M$ respectively.
Similarly $E(\boldsymbol{\lambda})$ and
$P(\boldsymbol{\lambda})$ can be expanded in powers of $c^{-1}$
\be
E(\boldsymbol{\lambda})=\sum_{n=0}^\infty
\frac{E_n(\boldsymbol{I})}{c^n}\ ,\quad
P(\boldsymbol{\lambda})=\sum_{n=0}^\infty
\frac{P_n(\boldsymbol{I})}{c^n}.
\ee
Denoting the truncation of the sums to order ${\cal O}(c^{-j})$ by
$F^{(j)}_{\cal O}(\boldsymbol{I},\boldsymbol{J})$,
$E^{(j)}(\boldsymbol{I})$ and $P^{(j)}(\boldsymbol{I})$
respectively, the $1/c$-expansion at order ${\cal O}(c^{-j})$ is
defined as
\be
 \sum_{M=0}^\infty\sum_{\{\mu_1,\dots,\mu_M\}}
|F^{(j)}_{\cal O}(\boldsymbol{I},\boldsymbol{J})|^2
e^{i\big(E^{(j)}(\boldsymbol{I})-E^{(j)}(\boldsymbol{J})\big)t-i
\big(P^{(j)}(\boldsymbol{I})-P^{(j)}(\boldsymbol{J})\big)x}\ .
\label{1overcex}
\ee

We stress that the expansion sums certain $1/c$ contributions to all 
orders by virtue of the fact that although the (exactly known) energies and
momenta are expanded inside the exponentials, the exponentials are
\textit{not} expanded in $1/c$. In this sense the expansion 
is non-perturbative, and in fact rather different from more standard
(diagrammatic) approaches pursued in \cite{brand05}. As discussed in
detail below \fr{1overcex} is in fact both a $1/c$ expansion and an expansion \emph{in terms of
number of  particle-hole excitations}. At order $n$ in the expansion
(i) only excitations that involve at most $\lfloor
  \tfrac{n}{2} \rfloor +1$ particle-hole pairs contribute, and (ii)
  all terms up to ${\cal O}(c^{-n})$ contribute. Importantly, this ``mixed"
  expansion has a well-defined thermodynamic limit and is uniform in
  space and time. This is in contrast to both the bare $1/c$
  expansion that is non-uniform, and the bare expansion in the number of
  particle-hole excitations that is divergent in the thermodynamic
  limit.

Expectation values of the form \fr{2pointfn} are relevant in two
contexts.
\begin{enumerate}
\item{} By working in a micro-canonical ensemble dynamical response
functions at finite temperature can be cast in this form. In the
following we will use this to determine the finite temperature
dynamical structure factor in the Lieb-Liniger model.
\item{} At late times after quantum quenches local observables relax to
non-thermal stationary values
\cite{Polkovnikov2011,EFreview,CalabreseCardy2016,Vidmar2016}. 
It follows from the quench action approach \cite{CE13,caux16}
to quantum quenches that expectation values in the stationary state
in fact involve non-thermal energy eigenstates at finite energy
densities. This has been used to study the stationary behaviour of
certain one-point functions after (particular) quantum quenches
\cite{Kormos13,deNC14,denardis14,piroli16}. A natural extension is
then to consider linear response functions in such steady states
\cite{EEF12,FCG12}. These can be expressed in the form \fr{2pointfn}, where
$|\boldsymbol{\lambda}\rangle$ corresponds to the non-equilibrium steady
state relevant to the quench of interest.
\end{enumerate}
In the following we will consider both these cases and evaluate
\fr{2pointfn} for the density operator and general
$|\boldsymbol{\lambda}\rangle$. 

A brief summary of some of our key technical results is as follows. We
show that the $1/c$-expansion corresponds to an expansion in the number of
particle-hole excitations. This leads to a dramatic reduction in the
complexity of the spectral sum that needs to be carried
out. Interestingly, the contributions of one particle-hole and two
particle-hole excitations are individually \textit{divergent} in the
infinite volume limit $L\to\infty$. Moreover they individually 
depend on details of the ``averaging state''
$|\boldsymbol{\lambda}\rangle$ beyond the root distribution function in
the thermodynamic limit. Crucially, their sum is not divergent and is
independent of the choice of representative state $|\boldsymbol{\lambda}\rangle$, and is
well-defined. 

The manuscript is organized as follows. In Section \ref{sec1} we
introduce the Lieb-Liniger model and recall the key elements of its
Bethe Ansatz solution. In Section \ref{3bis} we report some important
intermediate results on the thermodynamic limit of expressions
computed within the Bethe Ansatz. In Section \ref{sec2} we discuss
the $1/c$-expansion up to and including ${\cal O}(c^{-2})$ of the Bethe
Ansatz equations, energy eigenvalues, form factors and the spectral
representation of the density-density correlation function. These
results are then used in Section \ref{sec3} to obtain a fully explicit
expression for the dynamical density-density correlator (and the
related dynamical structure factor) in the thermodynamic limit,
\emph{cf.} equations \eqref{corregenalt}, \eqref{chi2}, \eqref{DSF1ph}
and \eqref{DSF2ph2}. This 
constitutes the main result of our work. In Section \ref{sec6} we
obtain the asymptotic behaviour of the correlator and structure factor
in various regimes. In particular we perform non-trivial consistency
checks of our formulas, and recover known results from (nonlinear)
Luttinger liquid theory and generalized hydrodynamics (GHD)
\cite{CADY16,BCDF16,Doyon18}.

%\section {The delta-Bose gas\label{sec1}}

\section{\texorpdfstring{Lieb-Liniger model}{Lg}}
\label{sec1}
\subsection {Definition}

The Lieb-Liniger model \cite{LiebLiniger63a,Brezin64} is a non-relativistic
quantum field theory model with Hamiltonian
\begin{equation}
H=\int_0^L \D{x}\left[\psi^\dagger(x)
\Big(-\frac{\hbar^2}{2m}\frac{d^2}{dx^2}\Big)\psi(x)+c\psi^\dagger(x)\psi^\dagger(x)\psi(x)\psi(x)
\right]\,,
\label{HLL}
\end{equation}
where the canonical Bose field $\psi(x)$ satisfies equal-time
commutation relations
\begin{equation}
[\psi(x),\psi^\dagger(y)]=\delta(x-y)\,.
\end{equation}
In the following we set $\hbar=2m=1$ and impose periodic boundary
  conditions.
%There is a one-to-one 
%correspondance between the eigenstates and energies of the delta-Bose
%gas and those of the Lieb-Liniger model, that is a quantum mechanics
In first quantization \fr{HLL} corresponds to a quantum mechanical
system of $N$ particles with positions $0\leq x_1,...,x_N \leq L$ and
Hamiltonian
\begin{equation}
H=\sum_{k=1}^N-\left( \frac{\partial}{\partial x_k}\right)^2+2c \sum_{j<k}\delta(x_j-x_k)\,.
\end{equation}
For later convenience we define the density operator at position $x$
\begin{equation}
\sigma(x)=\psi^\dagger(x)\psi(x)\,,
\end{equation}
and its time-$t$ evolved version $\sigma(x,t)=e^{iHt}\sigma(x)e^{-iHt}$.
%The objective of the paper is to compute the infinite-size limit $L\to\infty$ of the two-point function $\langle \sigma(x,t)\sigma(0,0)\rangle$ averaged between \textit{any} eigenstates and for \textit{all} $x,t$ at order $1/c^2$ when $c\to\infty$.

\subsection {The Bethe ansatz solution}
\subsubsection{The spectrum}
The Lieb-Liniger model is solvable by the Bethe ansatz: the
energy $E$ and the momentum $P$ of an eigenstate
$|\pmb{\lambda}\rangle$ with $N$ bosons read 
\begin{equation}
E(\pmb{\lambda})=\sum_{i=1}^N \lambda_i^2\,,\qquad P(\pmb{\lambda})=\sum_{i=1}^N\lambda_i\,,
\end{equation}
where the rapidities $\pmb{\lambda}=\{\lambda_1,..,\lambda_N\}$
satisfy the following set of ``Bethe equations''
\begin{equation}
e^{iL\lambda_k}=\prod_{\substack{j=1\\j \neq k}}^N\frac{\lambda_k-\lambda_j+ic}{\lambda_k-\lambda_j-ic}\,,
\quad k=1,\dots, N.
\end{equation}
It is convenient to express them in logarithmic form
\begin{equation}
\label{belog}
\frac{\lambda_k}{2\pi}=\frac{I_k}{L}-\frac{1}{L}\sum_{j=1}^N \frac{1}{\pi}\arctan \frac{\lambda_k-\lambda_j}{c}\,,
\end{equation}
with $I_k$ an integer if $N$ is odd, a half-integer if $N$ is even. For $c>0$, which we will assume in this paper, all the solutions to this equation are real \cite{korepin}.

%%%%%%%%%%%%%%%%%%%%%%%%%%%%%%%%%%%%%%%%%%
\subsubsection{The density form factors}
%%%%%%%%%%%%%%%%%%%%%%%%%%%%%%%%%%%%%%%%%%
As set out in the introduction, our aim is to calculate the density-density correlation function in an eigenstate $|\pmb{\lambda}\rangle$
\begin{equation}
  \left\langle \sigma\left( x,t\right) \sigma \left( 0,0\right) \right\rangle =\frac {\left\langle \pmb{\lambda} \left| \sigma \left( x,t\right) \sigma \left( 0,0\right) \right| \pmb{\lambda} \right\rangle } {\left\langle \pmb{\lambda}  |\pmb{\lambda}  \right\rangle }\,.
\end{equation}
Our strategy is to use a Lehman representation in terms of energy eigenstates
$|\pmb{\mu}\rangle =|\mu_1,...,\mu_{N'}\rangle$, where
$\{\mu_1,\dots,\mu_{N'}\}$ are solutions to the Bethe equations \fr{belog}
\begin{equation}
\label{bigsumdensity}
\begin{aligned}
  \left\langle \sigma\left( x,t\right) \sigma \left( 0,0\right) \right\rangle &=\sum _{ \pmb{\mu}}\frac {\left| \left\langle \pmb{\lambda} |\sigma \left( 0\right) |\pmb{\mu}\right\rangle \right| ^{2}} {\left\langle \pmb{\lambda} \left| \pmb{\lambda} \right\rangle \left\langle \pmb{\mu}\right| \pmb{\mu}\right\rangle }e^{it\left( E\left( \pmb{\lambda}\right) -E\left( \pmb{\mu} \right) \right) +ix\left( P\left( \pmb{\mu}\right) -P\left( \pmb{\lambda}\right) \right) }\\
  &=\sum_{N'=0}^\infty \sum _{\mu_1<...<\mu_{N'}}\frac {\left| \left\langle \pmb{\lambda} |\sigma \left( 0\right) |\pmb{\mu}\right\rangle \right| ^{2}} {\left\langle \pmb{\lambda} \left| \pmb{\lambda} \right\rangle \left\langle \pmb{\mu}\right| \pmb{\mu}\right\rangle }e^{it\left( E\left( \pmb{\lambda}\right) -E\left( \pmb{\mu} \right) \right) +ix\left( P\left( \pmb{\mu}\right) -P\left( \pmb{\lambda}\right) \right) }\,.
\end{aligned}
\end{equation}
The (normalized) form factors of local operators between two Bethe
states have been derived in Refs ~\cite{Korepin82,Slavnov89,Slavnov90,KorepinSlavnov99,Oota04,KozlowskiForm11}.
In the case of the density operator $\sigma$,
the (square of the normalized) form factor between two eigenstates
$|\pmb{\lambda}\rangle,|\pmb{\mu}\rangle$ with respective numbers of Bethe roots
$N,N'$ reads 
\begin{equation}
\label{ffdensity}
\begin{aligned}
&\frac{|\langle \pmb{\lambda}|\sigma(0)|\pmb{\mu}\rangle|^2}{\langle\pmb{\lambda}|\pmb{\lambda}\rangle\langle\pmb{\mu}|\pmb{\mu}\rangle}=\delta_{N,N'}\frac{\left(\sum_{i=1}^N \mu_i-\lambda_i\right)^2}{L^{2N}\mathcal{N}_{\pmb{\lambda}}\mathcal{N}_{\pmb{\mu}}}\frac{\prod_{i\neq j}(\lambda_i-\lambda_j)(\mu_i-\mu_j)}{\prod_{i,j}(\lambda_i-\mu_j)^2}\prod_{i\neq j}\frac{\lambda_i-\lambda_j+ic}{\mu_i-\mu_j+ic}\\
&\times \left|\underset{i,j\neq p}{\det}\left[(V_i^+-V_i^-)\delta_{ij}+i(\mu_i-\lambda_i)\prod_{k\neq i}\frac{\mu_k-\lambda_i}{\lambda_k-\lambda_i}\left(\frac{2c}{(\lambda_i-\lambda_j)^2+c^2} -\frac{2c}{(\lambda_p-\lambda_j)^2+c^2}\right)\right] \right|^2\,.
\end{aligned}
\end{equation}
Here $p\in\{1,...,N\}$ can be freely chosen,
\begin{equation}
V_i^\pm=\prod_{k=1}^{N}\frac{\mu_k-\lambda_i\pm ic}{\lambda_k-\lambda_i\pm ic}\,,
\end{equation}
and $\mathcal{N}_{\pmb{\lambda}}$ is given by\cite{Gaudin71}
\begin{equation}
\label{norm}
\mathcal{N}_{\pmb{\lambda}}=\underset{i,j=1,...,N}{\det}\left[ \delta_{ij} \left(1+\frac{1}{L}\sum_{k=1}^N \frac{2c}{c^2+(\lambda_i-\lambda_k)^2}\right)-\frac{1}{L}\frac{2c}{c^2+(\lambda_i-\lambda_j)^2}\right]\,.
\end{equation}
\section{Thermodynamic description of eigenstates\label{3bis}}
In a finite system of size $L$ all eigenstates of the Hamiltonian are fully
characterized by a set of $N$ Bethe numbers $I_k$, or equivalently a
set of $N$ Bethe roots $\lambda_k$. The purpose of this section is to
explain how to turn this description into one based on (continuous)
distribution functions of these roots in the thermodynamic limit
$L\to\infty$ when $N$ scales like $L$. In particular, contrary to a
common misconception, we emphasize that the usual ``root densities" defined
below \textit{do not} fully characterize an eigenstate in the
thermodynamic limit; this observation turns out to be of crucial
importance in our calculation.

\subsection{Root density}
In the thermodynamic limit, any sum of a \textit{non-singular} (piece-wise continuous) function $f$ over the Bethe roots or Bethe numbers
\begin{equation}\label{1dsum}
S_L[f]=\frac{1}{L}\sum_{k}f(\lambda_k)\,,\qquad \tilde{S}_L[f]=\frac{1}{L}\sum_{k}f(\tfrac{I_k}{L})\,,
\end{equation}
is independent of the precise values taken individually by each $I_k$
or $\lambda_k$, and depends only on the \textit{number of Bethe roots
  or Bethe numbers in any given interval}. This information is encoded
in the so-called root density $\rho(\lambda)\geq 0$ and filling
function $0\leq \chi(\iota)\leq 1$. They are defined by the
requirement that in the large $L$ limit
\begin{align}
L\rho(\lambda)d\lambda &= \text{number of Bethe roots in }
[\lambda,\lambda+d\lambda]\ ,\nn
L\chi(\iota)d\iota&=  \text{number of Bethe numbers $I_k/L$ per length
in }[\iota,\iota+d\iota] .
\end{align}
In the thermodynamic limit the sums \fr{1dsum} can be turned into
  integrals over these functions
\begin{equation}
S_\infty[f]=\int_{-\infty}^\infty f(\lambda)\rho(\lambda)\D{\lambda}\,,\qquad \tilde{S}_\infty[f]=\int_{-\infty}^\infty f(\iota)\chi(\iota)\D{\iota}\,.
\end{equation}
The same holds for multidimensional sums of a multivariate
\textit{non-singular} function $f$, with 
\begin{equation}\label{2dsum2}
S_L[f]=\frac{1}{L^n}\sum_{k_1,\dots,k_n}f(\lambda_{k_1},\dots,\lambda_{k_n})\,,
\end{equation}
converging to
\begin{equation}\label{2dsum}
S_\infty[f]=\int_{-\infty}^\infty \dots\int_{-\infty}^\infty f(\lambda_1,...,\lambda_n)\rho(\lambda_1)\dots\rho(\lambda_n)\D{\lambda_1}\dots\D{\lambda_n} \,.
\end{equation}
As far as expressions of the form \eqref{1dsum} and \eqref{2dsum2} are
concerned, an eigenstate in the thermodynamic limit is entirely
characterized by the root density $\rho(\lambda)$, or equivalently the
filling function $\chi(\iota)$. To relate these two equivalent
quantities, we introduce the function $\vartheta(\lambda)$ as the
  $L\to\infty$ limit of a function $\vartheta(\lambda_k)\equiv
  \chi(\tfrac{I_k}{L})$ of the Bethe roots, where $I_k$ is the integer
  associated with $\lambda_k$. Using the Bethe equations
\eqref{belog} $\vartheta(\lambda)$ can be expressed in terms of $\chi$ and $\rho$ as 
\begin{equation}
\vartheta(\lambda)=\chi\left(\frac{\lambda}{2\pi}+\frac{1}{\pi}\int_{-\infty}^\infty \arctan \left(\tfrac{\lambda-\mu}{c}\right)\rho(\mu)\D{\mu} \right)\,.
\end{equation}
The filling function $\chi(\iota)$ and the root density are then
related through 
\begin{equation}
\label{vartheta}
\frac{\rho(\lambda)}{\vartheta(\lambda)}=\frac{1}{2\pi}+\frac{1}{2\pi}\int_{-\infty}^\infty \frac{2c}{c^2+(\lambda-\mu)^2}\rho(\mu)\D{\mu}\,.
\end{equation}
It is customary to introduce the so-called hole density $\rho_h(\lambda)$ defined by
\begin{equation}
\label{rhoh}
\frac{\rho(\lambda)}{\vartheta(\lambda)}=\rho(\lambda)+\rho_h(\lambda)\,,
\end{equation}
which again contains equivalent information to $\rho(\lambda)$ or
$\chi(\iota)$. When expressed in terms of the particle and hole
densities \fr{vartheta} is known as the \emph{thermodynamic limit of
the Bethe Ansatz equations} \cite{Takahashibook}. Finally, the particle
density is given by
\begin{equation}
\label{D}
D\equiv \int_{-\infty}^\infty \rho(x)\D{x} = \underset{L\to\infty}{\lim}\, \frac{N}{L}\,.
\end{equation}
We introduce the Fermi momentum $q_F$ defined by
\begin{equation}
q_F=\pi D\,.
\end{equation}
Although there is a simple relation between $D$ and $q_F$, we will in the following
sometimes use $D$ and sometimes $q_F$, depending on the physical
context at hand. We also denote (in the units where $\hbar=2m=1$)
\begin{equation}
\omega_F=q_F^2\,.
\end{equation}

%\subsection{Square-inverse pair densities \label{sipdsect}}
\subsection{\texorpdfstring{Pair distribution function}{Lg}
\label{sipdsect}}
\subsubsection{Definition}

Root densities entirely characterize the value of sums of the type
\eqref{1dsum} and \eqref{2dsum2} in the thermodynamic limit. However,
some functions of the Bethe roots \textit{cannot} be expressed
solely in terms of root densities in the thermodynamic limit, and as a
consequence can take \textit{different values in the thermodynamic
  limit} for states that have the same root density. An example is
provided by 
\begin{equation}\label{sipd}
\Sigma_L[g]=\frac{1}{L^3}\sum_{i\neq j}\frac{g(\lambda_i,\lambda_j)}{(\lambda_i-\lambda_j)^2}\,,
\end{equation}
that we will encounter below
\footnote{The summand does not need to be singular for this to happen:
Another example is $L\sum_i
g(\lambda_i)(\lambda_{i+1}-\lambda_i)^2$ if the Bethe roots are ordered $\lambda_1<\lambda_2<...<\lambda_N$.}.
The sum in \fr{sipd} by definition depends on the \emph{joint
distribution function} of pairs of roots separated by ${\cal O}(L^{-1})$, and
the latter clearly contains information beyond that contained in the
root density (which does not distinguish between roots separated
  by ${\cal O}(L^{-1})$). 
  
  We first note that if we impose 
the constraint $|\lambda_i-\lambda_j|>\epsilon$ for a $\epsilon>0$ then
$\Sigma_L[g]$ vanishes in the thermodynamic limit. Hence, it only
depends on $g(\lambda,\lambda)$ and its derivatives at
$\lambda$. Taylor expanding $g(\lambda_i,\lambda_j)$ for $\lambda_i$
close to $\lambda_j$ reduces the order of the pole and makes the
next terms vanish in the thermodynamic limit, so it depends only on
$g(\lambda,\lambda)$. Being a linear functional of $g$ it can be
written in the thermodynamic limit in the form 
\begin{equation}\label{sipddef}
\Sigma_\infty[g]=\int_{-\infty}^\infty g(\lambda,\lambda)\gamma_{-2}(\lambda)\D{\lambda}\,,
\end{equation}
where the function $\gamma_{-2}(\lambda)$ depends on the state. We call
$\gamma_{-2}(\lambda)$ a \emph{pair distribution function} as it
encodes information about the joint distribution of pairs of Bethe
roots. The index $-2$ relates to the fact that we are summing over the
inverse square of the difference between two Bethe roots.
%, that we will call 'square-inverse pair density'. The square-inverse
%pair density
The pair distribution function $\gamma_{-2}(\lambda)$ characterizes
certain properties of the thermodynamic limit of an eigenstate
and is \textit{unrelated} to the root density $\rho(\lambda)$. Two
states can have the same $\rho(\lambda)$ but different
$\gamma_{-2}(\lambda)$. 

The simplest example is that of
(translationally invariant) free fermions, where the Bethe roots reduce
to the single-particle momenta. Here we may construct two sequences of eigenstates
labelled by an integer $n$, with momenta
$\{\lambda_i=\tfrac{ni}{L}|i=1,\dots,N\}$ and
$\{\mu_{2i}=\tfrac{2ni}{L},\mu_{2i+1}=\tfrac{2ni+1}{L}|i=1,\dots,N/2\}$
respectively. In the thermodynamic limit both states are described
by a root density $\rho(\lambda)=1/n$, but the pair distribution
functions are different:
$\gamma_{-2}(\lambda)=\tfrac{\pi^2}{3n^2}$ for the first state and 
$\gamma_{-2}(\lambda)=1+\frac{\pi^2}{12n^2}+\sum_{m\neq 0}\frac{1}{(2nm+1)^2}$ for the second one.
%\footnote{The simplest 
%  counter-examples are $\lambda_i=\tfrac{ni}{L}$ and
%  $\mu_{2i}=\tfrac{2ni}{L},\mu_{2i+1}=\tfrac{2ni+1}{L}$ for an integer
%  $n>0$. They both have $\rho(\lambda)=\tfrac{1}{n}$, but
%  $\gamma(\lambda)=\tfrac{\pi^2}{3n^2}$ for the first one and
%  $\gamma(\lambda)=1+\tfrac{\pi^2}{12n^2}+\tfrac{1}{2n^2}\sum_{k\geq
%    1}\tfrac{1}{(k-1/(2n))^2}$ for the second one.}.\\ 

\subsubsection{(Generalized) micro-canonical ensemble and
representative states} 
The (generalized) micro-canonical ensemble average of a local operator
${\cal O}(x)$ is a priori defined as 
\bea
\frac{1}{C_L}\sum_{\boldsymbol{\nu}}\frac{\langle\boldsymbol{\nu}|{\cal O}(x)|\boldsymbol{\nu}\rangle}
{\langle\boldsymbol{\nu}|\boldsymbol{\nu}\rangle}\ ,
\label{GMC}
\eea
where the sum is over an appropriate ``shell'' of simultaneous
eigenstates of the Hamiltonian and the local conservation laws of the
theory. $C_L$ is the number of terms in the sum. In a large but finite volume this means that for thermal
averages we fix the energy within a window that contains an
exponential (in system size) number of eigenstates. In the case of
generalized micro-canonical ensembles we fix the eigenvalues of (some
or all) of the local conservation laws in an analogous fashion
\cite{Cassidy11,CE13}. It is believed that almost all states in the
sum in \fr{GMC} have identical local properties, and hence the sum
over states can be replaced by an expectation value with respect to a
single typical state $|\boldsymbol{\lambda}\rangle$ in the thermodynamic limit
\bea
\lim_{L\to\infty}\, \frac{1}{C_L}
\sum_{\boldsymbol{\nu}}\frac{\langle\boldsymbol{\nu}|{\cal O}(x)|\boldsymbol{\nu}\rangle}
    {\langle\boldsymbol{\nu}|\boldsymbol{\nu}\rangle}=
\lim_{L\to\infty}
\frac{\langle\boldsymbol{\lambda}|{\cal O}(x)|\boldsymbol{\lambda}\rangle}
    {\langle\boldsymbol{\lambda}|\boldsymbol{\lambda}\rangle}.
\eea
The state $|\boldsymbol{\lambda}\rangle$ is sometimes called a \emph{representative
  state} and we follow this terminology here. We note that in practice
there is a great deal of freedom in choosing a representative state in
a large, finite volume.

\subsubsection{Average over representative states}
As we have seen above, the thermodynamic limit of the sum \eqref{sipd}
cannot generally be expressed as an integral over the root density,
but depends on the choice of representative state in the
finite volume. The thermodynamic limit of these sums involves
the separate function $\gamma_{-2}(\lambda)$ defined in
\eqref{sipddef}. As we will see in the following, in our
calculations of the density-density correlation function the
dependence of certain intermediate quantities on $\gamma_{-2}(\lambda)$ 
eventually compensate and the end result depends only on the root
density. However, it is a priori possible that in other calculations
involving sums of form factors no such cancellations will occur
and the end result will indeed depend on the choice of representative state
through $\gamma_{-2}(\lambda)$ or an analogous quantity.

We now make the following observation. As we have discussed above,
averages with respect to a Bethe state $|\pmb{\lambda}\rangle$ often
emerge upon simplifying averages over exponentially (in system size)
many representative states corresponding to a given root density
$\rho(\lambda)$. By construction such averages will depend only
on the density. This then poses the question what value
\eqref{sipd} takes after averaging over all representative
states with same root density in the thermodynamic limit. We now
address this issue.

First, we need to define properly which states in a large finite
volume $L$ are acceptable representative states for a given root
density. 
%We say that a sequence of \textit{sets of states}
%$(\mathfrak{S}_L)_{L\in   \mathbb{N}}$ is \textit{complete} for the
We define a sequence of \textit{sets of states} to be
\textit{complete} for the root density $\rho$ if the corresponding
sequence of sets of solutions to the Bethe equations
$(\mathfrak{S}_L)_{L\in   \mathbb{N}}$ all give rise to the
density $\rho$ in the thermodynamic limit, and if the number of
elements of the set $\mathfrak{S}_L$ satisfies  
\begin{equation}
\log |\mathfrak{S}_L|= L S_{\rm YY}[\rho]+o(L)\,,
\end{equation}
where $S_{\rm YY}[\rho]$ is the Yang-Yang entropy \cite{YangYang69}
\begin{equation}
S_{\rm YY}[\rho]=\int
\Big[\big(\rho(\lambda)+\rho_h(\lambda)\big)
\log\big(\rho(\lambda)+\rho_h(\lambda)\big)-\rho(\lambda)\log \rho(\lambda)-\rho_h(\lambda) \log \rho_h(\lambda)\Big]\D{\lambda}\,.
\end{equation}
In order to build such a set in a large finite volume let us
consider a root density $\rho(\lambda)$ at a given particle
density $D=\int \rho(\lambda)\D{\lambda}$. Given the root density we
  may introduce a particle counting function by
\be
z(\lambda)=\int_{-\infty}^\lambda \rho(x)\D{x}\ .
\ee
Next we choose a ``coarsening function" $\epsilon_L$ with the
  property that $\epsilon_L\to 0$ and $L\epsilon_L\to\infty$ when
$L\to\infty$ --- for  example one can take $\epsilon_L=\tfrac{1}{\sqrt{L}}$.
We now split the real axis into $n_L$ ``bins"
$[x_{L,j},x_{L,j+1}]$ containing $\lfloor L\epsilon_L\rfloor$ Bethe roots by defining
$x_{L,1},...,x_{L,n_L+1}$ such that $z(x_{L,i})=i\epsilon_L$ for
$1\leq i\leq n_L+1=\lfloor D/\epsilon_L\rfloor$. 

Finally we define $\mathfrak{S}_L$ as the set containing all the
states in a finite volume $L$ that contain exactly $\lfloor
L\epsilon_L \rfloor$ Bethe roots in each of the $n_L$ bins
$[x_{L,i},x_{L,i+1}]$. All states in $\mathfrak{S}_L$ have
$N_L=(\lfloor D/\epsilon_L\rfloor-1)\lfloor L\epsilon_L \rfloor$ Bethe
roots, which for $L\to\infty$ by construction are
  distributed with density $\rho(\lambda)$. The number of
elements of $\mathfrak{S}_L$ will depend on the number of ``vacancies'' in
each of the bins, which in turn depend on the values of all the Bethe
roots since they interact via the Bethe equations. However,
asymptotically in $L$, we have $K_{L,i}=\lfloor L
(x_{L,i+1}-x_{L,i})(\rho(x_{L,i})+\rho_h(x_{L,i}))\rfloor$ vacancies in each
of the bins, so that
\begin{equation}
|\mathfrak{S}_L|=\prod_{i=1}^{n_L}\binom{K_{L,i}+{\cal O}(L^0)}{\lfloor L\epsilon_L \rfloor}\,.
\end{equation}
Using Stirling's formula in the large-$L$ limit one has
\begin{equation}
\log |\mathfrak{S}_L|= L S_{\rm YY}[\rho]+o(L)\,,
\end{equation}
which shows that $\mathfrak{S}_L$ is indeed a complete set of
representative states for a given root density $\rho(\lambda)$. 

We can now state our result for the average of \eqref{sipd} over all
representative states with root density $\rho(\lambda)$:
\begin{equation}\label{thmregula}
\underset{L\to\infty}{\lim}\, \frac{1}{|\mathfrak{S}_L|}\sum_{\{\lambda_i\}_i\in \mathfrak{S}_L}\frac{1}{L^3}\sum_{i\neq j} \frac{g(\lambda_i,\lambda_j)}{(\lambda_i-\lambda_j)^2}=\frac{\pi^2}{3}\int_{-\infty}^\infty g(\lambda,\lambda)(\rho(\lambda)+\rho_h(\lambda))\rho(\lambda)^2 \D{\lambda}\,.
\end{equation}
A proof of \fr{thmregula} is given in Appendix \ref{proofthm}.\\\hfill

We note that if we instead sum over rapidities distributed
regularly according to the inverse of the counting function $z^{-1}(\lambda)$
without imposing that the rapidities are solutions of the Bethe
equations, the sum takes a different value: 
\begin{equation}\label{regula}
\underset{L\to\infty}{\lim}\,  \frac{1}{L^3}\sum_{i\neq j} \frac{g(z^{-1}(i/L),z^{-1}(j/L))}{(z^{-1}(i/L)-z^{-1}(j/L))^2}=\frac{\pi^2}{3}\int_{-\infty}^\infty g(\lambda,\lambda)\rho(\lambda)^3 \D{\lambda}\,.
\end{equation}
If we sum over rapidities distributed regularly according to the
inverse of the counting function $z^{-1}(\lambda)$ and impose the
Bethe equations, the sum \eqref{sipd} is not easily expressed in
terms of $\rho$, but takes a value different from either
\eqref{regula} or \eqref{thmregula}. Hence formula \eqref{thmregula} 
is both non-trivial and non-intuitive.

\subsection{Principal values\label{principalvalue}}
\subsubsection{Single principal value}
The sums \eqref{2dsum2} can be expressed in terms of root densities in
the thermodynamic limit, provided $f$ is non-singular. We have seen in
the previous section that for functions $f$ with a quadratic singularity the
thermodynamic limit value of the sum cannot be expressed in terms of
the root density. We now turn to functions that are
singular but integrable in a principal value sense. This is
the case of the sum 
\begin{equation}\label{sipdpp}
\tilde{\Sigma}_L[g]=\frac{1}{L^2}\sum_{\substack{i,j\\i\neq j}}\frac{g(\lambda_i,\lambda_j)}{\lambda_i-\lambda_j}\,. 
\end{equation}
We will assume that $g$ and $\rho$ are continuous. Symmetrizing the sum, we have
\begin{equation}
\tilde{\Sigma}_L[g]=\frac{1}{2L^2}\sum_{\substack{i,j\\i\neq j}}\frac{g(\lambda_i,\lambda_j)-g(\lambda_j,\lambda_i)}{\lambda_i-\lambda_j}\,.
\end{equation}
The function $F(x,y)=\frac{g(x,y)-g(y,x)}{x-y}$ is regular, so that it
has the form of \eqref{2dsum2} and its thermodynamic limit can be
expressed in terms of $\rho$ according to 
\begin{equation}
\tilde{\Sigma}_\infty[g]=\frac{1}{2}\int_{-\infty}^\infty\int_{-\infty}^\infty \frac{g(\lambda,\mu)-g(\mu,\lambda)}{\lambda-\mu}\rho(\lambda)\rho(\mu)\D{\lambda} \D{\mu}\,.
\end{equation}
Since the integrand is finite, one can remove a small shell
$|\lambda-\mu|<\epsilon$ with an error of ${\cal O}(\epsilon)$, and then
un-symmetrize the sum. This yields 
\begin{equation}
\tilde{\Sigma}_\infty[g]=\dashint
\frac{g(\lambda,\mu)}{\lambda-\mu}\rho(\lambda)\rho(\mu)\D{\lambda}
\D{\mu}\,, 
\end{equation}
with the following usual definition of the principal value integral
\begin{equation}
\dashint \frac{F(\lambda)}{\lambda-\mu}\D{\lambda} =\underset{\epsilon\to 0}{\lim}\, \int_{|\lambda-\mu|>\epsilon} \frac{F(\lambda)}{\lambda-\mu}\D{\lambda}\,.
\end{equation}
Hence, sums of type \eqref{sipdpp} can indeed be expressed in
terms of root densities.\hfill

In contrast partial sums like
\begin{equation}
\frac{1}{L}\sum_{\substack{i\\i\neq j}}\frac{g(\lambda_i,\lambda_j)}{\lambda_i-\lambda_j}\,,
\end{equation}
at fixed $j$ cannot be expressed in terms of the root density in the
thermodynamic limit.

\subsubsection{Double principal values}
Higher-dimensional sums of the form
\begin{equation}\label{sipdpp3}
\tilde{\Sigma}_L[g]=\frac{1}{L^3}\sum_{\substack{i,j,k\\i\neq j\\ j\neq k}}\frac{g(\lambda_i,\lambda_j,\lambda_k)}{(\lambda_i-\lambda_j)(\lambda_j-\lambda_k)}\,,
\end{equation}
can be treated likewise, but with subtleties hiding in the fact that
$i$ can be equal to $k$. Separating out the term with $i=k$ and
symmetrizing the remaining sum gives
\begin{equation}
\begin{aligned}
&\tilde{\Sigma}_L[g]=\frac{1}{6L^3}\sum_{\substack{i\neq j\\ j\neq k\\ i\neq k}}\sum_{\sigma\in\mathfrak{S}_3}\frac{g(\lambda_{\sigma(i)},\lambda_{\sigma(j)},\lambda_{\sigma(k)})(\lambda_{\sigma(k)}-\lambda_{\sigma(i)})\sign(\sigma)}{(\lambda_i-\lambda_j)(\lambda_j-\lambda_k)(\lambda_k-\lambda_i)}-\frac{1}{L^3}\sum_{\substack{i\neq j}}\frac{g(\lambda_i,\lambda_j,\lambda_i)}{(\lambda_i-\lambda_j)^2}\,.
\end{aligned}
\end{equation}
The first term is regular so that \eqref{2dsum} can be used,
while the second term is of the type \eqref{sipd} and can be
expressed in terms of $\gamma_{-2}(\lambda)$. In the first
term we can remove the region where
$|\lambda-\mu|<\epsilon\text{ or }|\lambda-\nu|<\epsilon\text{ or }|\nu-\mu|<\epsilon$ with an error that is
${\cal O}(\epsilon)$, and then un-symmetrize the integral. One obtains
\begin{equation}\label{eq1double}
\tilde{\Sigma}_\infty[g]=\ddashint\frac{g(\lambda,\mu,\nu)\rho(\lambda)\rho(\mu)\rho(\nu)}{(\lambda-\mu)(\mu-\nu)}\D{\lambda} \D{\mu} \D{\nu}-\int_{-\infty}^\infty g(\lambda,\lambda,\lambda)\gamma_{-2}(\lambda)\D{\lambda}\,,
\end{equation}
where the simultaneous principal value in the triple-integral is
defined as
\begin{equation}
\ddashint\frac{F(\lambda,\mu,\nu)}{(\lambda-\mu)(\mu-\nu)}\D{\lambda}
\D{\mu} \D{\nu}=
\underset{\epsilon\to 0}{\lim}\, \int_{\substack{|\lambda-\mu|>\epsilon\\ |\mu-\nu|>\epsilon\\|\lambda-\nu|>\epsilon}}  \frac{F(\lambda,\mu,\nu)}{(\lambda-\mu)(\mu-\nu)}\D{\lambda} \D{\mu} \D{\nu}\,.
\end{equation}
As shown in Appendix \ref{proofPB} this can be expressed in terms
of the successive principal value triple-integral according to a
Poincar\'e-Bertrand-like formula 
\begin{equation}\label{PB}
\begin{aligned}
\ddashint\frac{F(\lambda,\mu,\nu)}{(\lambda-\mu)(\mu-\nu)}\D{\lambda}
\D{\mu} \D{\nu}=\dashint\frac{F(\lambda,\mu,\nu)}{(\lambda-\mu)(\mu-\nu)}\D{\lambda}\D{\mu} \D{\nu}+\frac{\pi^2}{3}\int_{-\infty}^\infty F(\lambda,\lambda,\lambda)\D{\lambda}\,.
\end{aligned}
\end{equation}
where we defined
\begin{equation}\label{dashint}
\begin{aligned}
\dashint\frac{F(\lambda,\mu,\nu)}{(\lambda-\mu)(\mu-\nu)}\D{\lambda}
\D{\mu} \D{\nu}&=\int \D{\mu} \dashint \D{\nu} \frac{1}{\mu-\nu} \dashint \D{\lambda} \frac{F(\lambda,\mu,\nu)}{\lambda-\mu}\\
&= \int \D{\mu} \,\underset{\epsilon\to 0}{\lim}\, \int_{|\nu-\mu|>\epsilon} \D{\nu} \frac{1}{\mu-\nu} \underset{\epsilon'\to 0}{\lim}\,\int_{|\mu-\lambda|>\epsilon'} \D{\lambda} \frac{F(\lambda,\mu,\nu)}{\lambda-\mu}\,.
\end{aligned}
\end{equation}
It can also be expressed as
\begin{equation}\label{dashint22}
\begin{aligned}
\dashint\frac{F(\lambda,\mu,\nu)}{(\lambda-\mu)(\mu-\nu)}\D{\lambda}\D{\mu} \D{\nu}&= \int \D{\nu}\dashint \D{\mu} \frac{1}{\mu-\nu} \dashint \D{\lambda} \frac{F(\lambda,\mu,\nu)}{\lambda-\mu}\\
&=\int \D{\lambda} \dashint \D{\mu} \frac{1}{\lambda-\mu} \dashint \D{\nu} \frac{F(\lambda,\mu,\nu)}{\mu-\nu}\\
&=\int \D{\mu}\dashint \D{\lambda}  \frac{1}{\lambda-\mu} \dashint \D{\nu} \frac{F(\lambda,\mu,\nu)}{\mu-\nu}\,,
\end{aligned}
\end{equation}
and
\begin{equation}\label{dashint33}
\dashint\frac{F(\lambda,\mu,\nu)}{(\lambda-\mu)(\mu-\nu)}\D{\lambda}\D{\mu} \D{\nu}=\underset{\epsilon,\epsilon'\to 0}{\lim}\,\int_{\substack{|\lambda-\mu|>\epsilon\\|\mu-\nu|>\epsilon'}}\frac{F(\lambda,\mu,\nu)}{(\lambda-\mu)(\mu-\nu)}\D{\lambda}\D{\mu} \D{\nu}\,,
\end{equation}
as shown in Appendices \ref{dashint22p}.
Using these principal value integral identities we can rewrite \eqref{eq1double}
in the form
\begin{equation}\label{simpledouble}
\tilde{\Sigma}_\infty[g]=\dashint\frac{g(\lambda,\mu,\nu)\rho(\lambda)\rho(\mu)\rho(\nu)}{(\lambda-\mu)(\mu-\nu)}\D{\lambda}\D{\mu} \D{\nu}+\int_{-\infty}^\infty g(\lambda,\lambda,\lambda)\left[\frac{\pi^2}{3}\rho(\lambda)^3-\gamma_{-2}(\lambda)\right]\D{\lambda}\,.
\end{equation}

%
%
%We report in Appendices \ref{proofthm} and \ref{principalvalue} two important additional discussions. In Appendix \ref{proofthm} is calculated the sum \eqref{sipd} when averaged over all the representative states that have the same root density in the thermodynamic limit, in which case it can be expressed in terms of the root density indeed. In Appendix  \ref{principalvalue} are reported important comments on singular sums over Bethe roots, but whose singularity is still integrable in a principal value sense. These aspects will turn out to be crucial in the calculations below.

\subsection{Examples of root densities\label{exrootg}}
The calculations presented in this paper
hold for a generic piece-wise continuous root density $\rho(\lambda)$.
Two applications we have in mind is to thermal states and
non-equilibrium steady states after quantum quenches, and we now
discuss specific root densities that arise in these contexts.
%%%%%%%%%%%%%%%%%%%%%%%%%%%%%%%%%%%%%%%%%%%%%%
\subsubsection{Thermal states \label{exroot}}
%%%%%%%%%%%%%%%%%%%%%%%%%%%%%%%%%%%%%%%%%%%%%%
Thermal states are characterized by root densities
that maximise the Yang-Yang entropy at inverse temperature
$\beta$ \cite{YangYang69}. Defining the so-called dressed energy 
$\varepsilon_{\rm dr}(\lambda)$ by 
\begin{equation}
\vartheta(\lambda)=\frac{1}{1+e^{\beta\varepsilon_{\rm dr}(\lambda)}}\,,
\label{dresseden}
\end{equation}
the filling function $\vartheta(\lambda)$ of a thermal state is such that
\begin{equation}
\varepsilon_{\rm dr}(\lambda)=\lambda^2-h-\frac{1}{2\pi\beta}\int_{-\infty}^\infty \frac{2c}{c^2+(\lambda-\mu)^2}\log(1+e^{-\beta\varepsilon_{\rm dr}(\mu)})\D{\mu}\,.
\label{dresseden2}
\end{equation}
Here $h$ is a chemical potential that is used to fix the desired
  particle density $D$.
In practice one first solves the nonlinear integral equation
\fr{dresseden2} and then uses \fr{dresseden} to determine $\rho(\lambda)$
from the linear integral equation \fr{vartheta}.

A particular case of thermal states is the zero temperature ground
state, obtained in the limit $\beta\to\infty$. Its root density satisfies 
\begin{equation}
\label{0t}
\rho(\lambda)=\frac{1}{2\pi}+\frac{c}{\pi}\int_{-Q}^Q \frac{\rho(\mu)}{c^2+(\lambda-\mu)^2}\D{\mu}\,,
\end{equation}
with $Q$ defined such that
\begin{equation}
\int_{-Q}^Q \rho(\lambda)\D{\lambda}= D\,.
\end{equation}

\subsubsection{Non-equilibrium steady states \label{exroot2}}
Refs \cite{Kormos13,denardis14} considered a particular interaction
quench in the Lieb-Liniger model, where the system is initially in the
ground state of \fr{HLL} for $c=0$, and is subsequently time-evolved
with the Lieb-Liniger Hamiltonian at a finite value of $c$. The root
density characterizing the steady state reached at late times was
determined in \cite{denardis14} and remarkably allows for a closed form
solution
\begin{align}\label{rhoss}
\rho_{\rm
  ss}(\lambda)&=\frac{\tau}{4\pi\big(1+a(\lambda/c)\big)}\frac{\D{
  a(\lambda/c)}}{\D{\tau}} \ ,
\end{align}
where
$\tau=\frac{1}{c}\int_{-\infty}^{\infty} \rho_{\rm ss}(x)\D{x}$ and 
\begin{align}
a(x)&=\frac{2\pi\tau}{x\sinh(2\pi x)}I_{1-2ix}(4\sqrt{\tau})
I_{1+2ix}(4\sqrt{\tau})\,,
\end{align}
with $I$ the modified Bessel function.
%%%%%%%%%%%%%%%%%%%%%%%%%%%%%%%%%%%%%%%%%%
\section{\texorpdfstring{$1/c$}{Lg} expansion of the Lieb-Liniger model\label{sec2}}
%%%%%%%%%%%%%%%%%%%%%%%%%%%%%%%%%%%%%%%%%%
In this section we perform an expansion around the limit
$c\to\infty$ at order $1/c^2$ of the energy levels and form factors in
the Lieb-Liniger model, at fixed $L$ and fixed Bethe numbers. We then expose the consequences it has on the spectral sum \eqref{bigsumdensity} in Section \ref{suminterm}.

\subsection{The Bethe equations}
The Bethe equations \eqref{belog} admit a regular $1/c$ expansion at
large $c$. In the following, in order to expand the form factor at
order $1/c^2$ we will need the value of the Bethe roots at order
$1/c^3$. The Bethe equations \eqref{belog} at order $1/c^3$ read 
\begin{equation}
\lambda_i=\frac{2\pi
  I_i}{L}-\frac{2}{cL}\sum_{k=1}^N(\lambda_i-\lambda_k)+\frac{2}{3c^3L}\sum_{k=1}^N(\lambda_i-\lambda_k)^3+{\cal
  O}(c^{-5})\,.
\end{equation}
This gives the following expression for the Bethe roots in terms of
the Bethe numbers at order $1/c^3$ 
\begin{equation}
\label{besolve}
\lambda_i=\frac{2\pi}{1+\tfrac{2D}{c}}\frac{I_i}{L}+\frac{4\pi
}{c(1+\tfrac{2D}{c})}\frac{1}{L}\sum_{j=1}^N
\frac{I_j}{L}+\frac{1}{3\pi
  c^3}\left(\frac{2\pi}{1+\tfrac{2D}{c}}\right)^4\frac{1}{L}\sum_{j=1}^N
\left(\frac{I_i-I_j}{L}\right)^3+{\cal O}(c^{-4})\,.
\end{equation}
The alert reader will have noticed that some of the terms
contain higher powers of $1/c$ than the order at which
we are working, that is $1/c^3$. We find it useful throughout
the manuscript to retain certain ``resummed" expressions of $1/c$ as
they appear in calculations, both for clarity and convenience since
they often happen to compensate each other. In any case, keeping these
resummed expressions in $1/c$ does not affect the validity of the 
equations at the order considered.  
%
%We obtain then the value of the momentum and energy
%\begin{equation}
%\label{EP}
%\begin{aligned}
%E(\pmb{\lambda})&=\left(\frac{2\pi}{1+\tfrac{2D}{c}}\right)^2\sum_{i=1}^N \left(\frac{I_i}{L}\right)^2+\frac{1+\tfrac{D}{c}}{(1+\tfrac{2D}{c})^2} \frac{16\pi^2}{cL}\left(\sum_{i=1}^N \frac{I_i}{L}\right)^2+O(c^{-3})\\
%P(\pmb{\lambda})&=2\pi\sum_{i=1}^N \frac{I_i}{L}
%\end{aligned}
%\end{equation}

\subsection{The form factors}
\subsubsection{\sfix{Leading order in $1/c$ of the form factor between two generic states}}
The behaviour of the $1/c$ expansion of a form factor
\eqref{ffdensity} between states $|\pmb{\lambda}\rangle$ and
$|\pmb{\mu}\rangle$ depends on the ``relative positions" of the
Bethe numbers of one state to the other. To see this, let us determine
the leading order in $1/c$ of the form factor \eqref{ffdensity}
without making any assumptions on the eigenstates
$|\pmb{\lambda}\rangle$ and $|\pmb{\mu}\rangle$.
It is then straightforward to see that when $c\to\infty$
\begin{equation}
\begin{aligned}
V_j^+-V_j^-&=\frac{2}{ic}\sum_{k=1}^N (\mu_k-\lambda_k)+{\cal O}(c^{-2})\\
\mathcal{N}_{\pmb{\lambda}}&=1+{\cal O}(c^{-1})\,,
\end{aligned}
\end{equation}
while the non-diagonal term in the determinant appearing in the
form factor is of order ${\cal O}(c^{-3})$. We conclude that
\begin{equation}
\label{ffc0}
\frac{|\langle \pmb{\lambda}|\sigma(0)|\pmb{\mu}\rangle|^2}{\langle\pmb{\lambda}|\pmb{\lambda}\rangle\langle\pmb{\mu}|\pmb{\mu}\rangle}=\frac{\left(\sum_i \mu_i-\lambda_i\right)^{2N}}{L^{2N}}\left(\frac{2}{c}\right)^{2N-2}\frac{\prod_{i\neq j}(\lambda_i-\lambda_j)(\mu_i-\mu_j)}{\prod_{i,j}(\lambda_i-\mu_j)^2}(1+{\cal O}(c^{-1}))\,.
\end{equation}
We see that the order in $1/c$ of this expression entirely
depends on the roots $\lambda_k$ and $\mu_k$. To be specific, let us
now denote by $I_k$ and $J_k$ the Bethe numbers of $\pmb{\lambda}$ and
$\pmb{\mu}$ respectively, and define  
\begin{equation}
\nu=N-|\{I_k\} \cap \{J_k\}|\,,
\end{equation}
the number of Bethe numbers present in $\pmb{\lambda}$ and absent from
$\pmb{\mu}$.
If $\lambda_i$ and $\mu_j$ have different Bethe numbers, then from
\eqref{besolve} we have $\lambda_i-\mu_j={\cal O}(c^0)$, whereas if they have
the same Bethe number then at least $\lambda_i-\mu_j={\cal O}(c^{-1})$. It follows that 
\begin{equation}
\frac{|\langle \pmb{\lambda}|\sigma(0)|\pmb{\mu}\rangle|^2}{\langle\pmb{\lambda}|\pmb{\lambda}\rangle\langle\pmb{\mu}|\pmb{\mu}\rangle}={\cal O}(c^{-2(\nu-1)})\,.
\end{equation}
Hence expanding in $1/c$ naturally orders the Lehman representation
\eqref{bigsumdensity} into an expansion in terms of number of
\textit{particle-hole excitations} of $\pmb{\mu}$ above
$\pmb{\lambda}$, i.e. of the number of changes in the Bethe numbers of
$\pmb{\mu}$ compared to those of $\pmb{\lambda}$. This means that if
one considers \eqref{bigsumdensity} at order $c^{-m}$, then only
intermediate states $\pmb{\mu}$ with $\nu\leq \tfrac{m}{2}+1$
contribute to the sum. We note however that the converse is not true:
restricting \eqref{bigsumdensity} to e.g. one-particle-hole
excitations would still involve arbitrarily high orders in $1/c$. 

Since our goal is to compute correlations at order $1/c^2$, we only
need to investigate the restriction of \eqref{ffc0} to one- and
two-particle-hole excitations.

\subsubsection{\sfix{Order $1/c^2$ of form factors involving a single
particle-hole excitation}}
In this section we consider one-particle-hole excitations of the state
$|\pmb{\mu}\rangle$ above $|\pmb{\lambda}\rangle$. Up to reordering
the roots, we can assume that the Bethe numbers $I_k$ of
$\pmb{\lambda}$ differ from those $J_k$ of $\pmb{\mu}$ only at a
  single position $a$:
\begin{equation}
\forall i\neq a\quad I_i=J_i\,,\qquad J_a-I_a\equiv n\neq 0\,.
\end{equation}
Since the excited particle cannot coincide with an already existing
particle, we also have the constraint 
\begin{equation}
\forall i\neq a\quad I_a+n\neq I_i\,.
\end{equation}
This has the following consequences at order $1/c^3$ on the value of
the Bethe roots. Using \eqref{besolve} we have 
\begin{equation}
\label{corrmodif}
\mu_a=\lambda_a+\frac{2\pi n}{L(1+2D/c)}\left(1+\frac{2}{cL}\right)+{\cal O}(c^{-3})\,,
\end{equation}
while for $i\neq a$, we obtain
\begin{equation}
\label{corrnonmodif}
\mu_i=\lambda_i+\frac{4\pi n}{cL^2(1+2D/c)}\left(1-\frac{(\lambda_i-\lambda_a)^2}{c^2}+\frac{2\pi n}{c^2L}(\lambda_i-\lambda_a)-\frac{1}{3c^2}\left(\frac{2\pi n}{L}\right)^2\right)+{\cal O}(c^{-4})\,.
\end{equation}
Using \fr{corrmodif} and \fr{corrnonmodif} we can determine the
various terms entering the expression of the form factor at order $c^{-2}$

\begin{align}
\prod_{i\neq
  j}\frac{\lambda_i-\lambda_j+ic}{\mu_i-\mu_j+ic}&=1-\frac{1}{c^2}\left(\frac{2\pi
  n}{L}\right)^2\sum_{i\neq a}1+\frac{4\pi n}{L c^2}\sum_{i\neq
  a}\lambda_i-\lambda_a\ ,\nn
V_i^+-V_i^-&=\frac{4\pi n}{icL}\left(1-\frac{(\lambda_i-\lambda_a)^2}{c^2}\right),\nn
\prod_{\substack{i\neq j\\ i\neq a\\ j\neq
    a}}\frac{(\lambda_i-\lambda_j)(\mu_i-\mu_j)}{(\lambda_i-\mu_j)^2}&=1+\left(\frac{4\pi
  n}{c L^2}\right)^2\sum_{\substack{i\neq j\\ i\neq a\\ j\neq
    a}}\frac{1}{(\lambda_i-\lambda_j)^2}\ ,\nn
\prod_{i\neq  a}\frac{(\lambda_i-\lambda_a)^2}{(\mu_i-\lambda_a)^2}&=1-\frac{8\pi
  n}{cL^2(1+2D/c)}\sum_{i\neq a}\frac{1}{\lambda_i-\lambda_a}\nn
&\quad+\Big(\frac{4\pi n}{cL^2}\Big)^2\bigg[2\bigg(\sum_{i\neq
  a}\frac{1}{\lambda_i-\lambda_a}\bigg)^2
  +\sum_{i\neq a}\frac{1}{(\lambda_i-\lambda_a)^2}\bigg],
\end{align}
\begin{align}
\prod_{i\neq a}\frac{(\mu_i-\mu_a)^2}{(\lambda_i-\mu_a)^2}&=1+\frac{8\pi n}{cL^2(1+\frac{2D}{c})}\sum_{i\neq a}\frac{1}{\lambda_i-\lambda_a-\tfrac{2\pi n}{L(1+\frac{2D}{c})}}\nn
&+\Big(\frac{4\pi n}{cL^2}\Big)^2\bigg[2\Big(\sum_{i\neq a}\frac{1}{\lambda_i-\lambda_a-\tfrac{2\pi n}{L(1+\frac{2D}{c})}}\Big)^2
+\sum_{i\neq a}\frac{1}{(\lambda_i-\lambda_a-\tfrac{2\pi n}{L(1+\frac{2D}{c})})^2}\bigg],\nn
\prod_{i=1}^N\frac{1}{(\lambda_i-\mu_i)^2}&=\frac{4(1+\frac{2D}{c})^{2N}}{(1+\frac{2}{cL})^2}\frac{c^{2N-2}L^{4N-2}}{n^{2N}(4\pi)^{2N}}\nn
&\quad\times\biggl[1+\frac{2}{c^2}\sum_{i\neq
    a}\biggl((\lambda_i-\lambda_a)^2-\frac{2\pi
    n}{L}(\lambda_i-\lambda_a)+\frac{1}{3}\left(\frac{2\pi
    n}{L}\right)^2\biggr)\biggr],\nn
\mathcal{N}_{\pmb{\lambda}}&=\mathcal{N}_{\pmb{\mu}}=\bigg(1+\frac{2D}{c}\bigg)^{N-1}\,,
\end{align}
\begin{align}
i(\mu_l-\lambda_l)\prod_{k\neq l}\frac{\mu_k-\lambda_l}{\lambda_k-\lambda_l}&\left(\frac{2c}{(\lambda_l-\lambda_j)^2+c^2} -\frac{2c}{(\lambda_p-\lambda_j)^2+c^2}\right)={\cal O}(c^{-4}).
\end{align}

Putting everything together we have at order $c^{-2}$

\begin{align}
\label{ffcm2}
&\frac{|\langle \pmb{\lambda}|\sigma(0)|\pmb{\mu}\rangle|^2}{\langle\pmb{\lambda}|\pmb{\lambda}\rangle\langle\pmb{\mu}|\pmb{\mu}\rangle}=
\frac{(1+\tfrac{2D}{c})^2}{(1+\frac{2}{cL})^2}\frac{1}{L^2}\Bigg[1+\frac{4}{cL(1+\tfrac{2D}{c})}\frac{2\pi n}{L}\sum_{i\neq a}\Big(\frac{1}{\lambda_i-\lambda_a-\tfrac{2\pi n}{L(1+\frac{2D}{c})}}-\frac{1}{\lambda_i-\lambda_a}\Big)\nn
&+\frac{4}{c^2L^2}\Big(\frac{2\pi n}{L}\Big)^2\Bigg(-\frac{L^2}{12}\sum_{i\neq a}1+\sum_{\substack{i\neq j\\j\neq a}}\frac{1}{(\lambda_i-\lambda_j)^2}+2\Big(\sum_{i\neq a}\frac{1}{\lambda_i-\lambda_a-\tfrac{2\pi n}{L(1+\frac{2D}{c})}}-\frac{1}{\lambda_i-\lambda_a}\Big)^2\nn
&\qquad\qquad\qquad +\sum_{i\neq a}\frac{1}{(\lambda_i-\lambda_a-\tfrac{2\pi n}{L(1+\frac{2D}{c})})^2}
\Bigg)\Bigg]+{\cal O}(c^{-3})\ .
\end{align}

%This expression can be checked numerically with infinite precision, since it holds at fixed finite $L$ and since $c$ can be made as large as desired.

%\subsubsection{\sfix{Order $1/c^2$ of the two-particle-hole excitations form factors}}
\subsubsection{\sfix{Order $1/c^2$ of form factors involving two
    particle-hole excitations}}
We now consider two particle-hole excitations. Up to re-ordering the
roots of $\pmb{\mu}$, we can assume its Bethe numbers differ from
those of $\pmb{\lambda}$ only at positions $a$ and $b\neq a$, and
thus assume
\begin{equation}
\forall i\neq a,b\quad I_i=J_i\,,\qquad J_a-I_a\equiv n\neq 0\,,\qquad J_b-I_b\equiv m\neq 0\,.
\end{equation}
Since the excited particles cannot coincide with an already existing particle, we also have the constraints
\begin{equation}
\forall i\neq a,b\quad I_a+n\neq I_i\,,\qquad \forall i\neq a,b\quad I_a+m\neq I_i\,.
\end{equation}
Moreover we must also exclude the case where one of the excited
particles fill the hole left by the other, since this reduces to a
single particle-hole excitation and is therefore already
covered. The corresponding constraint is  
\begin{equation}
I_a+n\neq I_b\,,\qquad I_b+m\neq I_a\,.
\end{equation}
Finally we have to exclude the case where the two excited particles coincide
\begin{equation}
I_a+n\neq I_b+m\,.
\end{equation}
%Since the form factor is already at order $1/c^2$, we only \bl{require} the
%leading corrections in $1/c$ to the roots $\mu_k$ compared to
%$\lambda_k$. From \eqref{besolve} we obtain 
%\begin{equation}
%\label{mu2ph}
%\mu_i=\begin{cases}
%\lambda_i+\frac{4\pi (n+m)}{cL^2}+O(c^{-2})\quad\text{if }i\neq a,b\\
%\lambda_a+\frac{2\pi n}{L}+O(c^{-1})\quad\text{if }i= a\\
%\lambda_b+\frac{2\pi m}{L}+O(c^{-1})\quad\text{if }i= b\,.
%\end{cases}
%\end{equation}
From \eqref{besolve} we obtain 
\begin{equation}
\label{mu2ph}
\mu_i=\begin{cases}
\lambda_i+\frac{4\pi (n+m)}{c\big(1+\frac{2D}{c}\big)L^2}+{\cal O}(c^{-3})\quad\text{if }i\neq a,b\\
\lambda_a+\frac{2\pi n}{L\big(1+\frac{2D}{c}\big)}+\frac{4\pi (n+m)}{c\big(1+\frac{2D}{c}\big)L^2}+{\cal O}(c^{-3})\quad\text{if }i= a\\
\lambda_b+\frac{2\pi m}{L\big(1+\frac{2D}{c}\big)}+\frac{4\pi (n+m)}{c\big(1+\frac{2D}{c}\big)L^2}+{\cal O}(c^{-3})\quad\text{if }i= b\,.
\end{cases}
\end{equation}
%Since the form factor is already at order $1/c^2$, we only require the
%leading corrections in $1/c$ to the roots $\mu_k$ compared to $\lambda_k$.

We can now investigate the form taken by \eqref{ffc0} for
these values of roots. At leading order in $1/c$ we have 
\begin{equation}
\frac{\prod_{i\neq
    j}(\lambda_i-\lambda_j)(\mu_i-\mu_j)}{\prod_{i,j}(\lambda_i-\mu_j)^2}=\frac{(\lambda_a-\lambda_b)^2(\mu_a-\mu_b)^2}{(\lambda_a-\mu_b)^2(\lambda_b-\mu_a)^2}\frac{1}{\prod_i(\lambda_i-\mu_i)^2}(1+{\cal
  O}(c^{-1}))\,,
\end{equation}
which, when substituted in \eqref{ffc0} yields the following
leading order expression of the form factor for two-particle-hole
excitations 
\begin{equation}
\label{ffc4}
\frac{|\langle
  \pmb{\lambda}|\sigma(0)|\pmb{\mu}\rangle|^2}{\langle\pmb{\lambda}|\pmb{\lambda}\rangle\langle\pmb{\mu}|\pmb{\mu}\rangle}=\frac{4}{c^2L^4}\frac{(n+m)^4}{n^2m^2}\frac{(\lambda_a-\lambda_b)^2(\lambda_a-\lambda_b+\tfrac{2\pi(n-m)}{L(1+2D/c)})^2}{(\lambda_a-\lambda_b+\tfrac{2\pi n}{L(1+2D/c)})^2(\lambda_a-\lambda_b-\tfrac{2\pi m}{L(1+2D/c)})^2}+{\cal
  O}(c^{-3})\,.
\end{equation}

%Again, this expression can be checked numerically with infinite precision.

\subsection{The Lehmann representation}
We can now write the Lehmann representation \eqref{bigsumdensity} for
the density-density correlation functions at order $1/c^2$. As
explained in the previous section, only one and two particle-hole
excitations contribute to \eqref{bigsumdensity} at order $1/c^2$, and
the corresponding form factors were computed at this order in the
previous subsections. This leaves us with working out the
phases in the corresponding terms in \eqref{bigsumdensity} at
order $1/c^2$.   

\subsubsection{\sfix{The phase for a single particle-hole excitation}}
For excitations with one particle and one hole, it follows from
\eqref{corrmodif} and $\eqref{corrnonmodif}$ that
\begin{align}
x\big(P(\pmb{\mu})-P(\pmb{\lambda})\big)&
+t\big(E(\pmb{\lambda})-E(\pmb{\mu})\big)=x \frac{2\pi n}{L}
-t\biggl[\frac{8\pi n}{cL^2(1+\tfrac{2D}{c})}\sum_{i}\lambda_i\nn
&\qquad+\Big(\frac{2\pi n}{L(1+\tfrac{2D}{c})}\Big)^2(1+\tfrac{4}{cL}+\tfrac{4D}{c^2 L})+\frac{4\pi n}{L(1+\tfrac{2D}{c})}\lambda_a\biggr]
+{\cal O}(c^{-3})\,.
\end{align}
It will be convenient to perform the following change of variable $x'$ defined as
\begin{equation}
x'=x(1+\tfrac{2D}{c})-\frac{4\delta_L}{c}t\,,
\end{equation}
where
\begin{equation}
\delta_L=\frac{1}{L}\sum_{i=1}^N \lambda_i\,.
\end{equation}
Then the phase becomes

\begin{align}
x(P(\pmb{\mu})-P(\pmb{\lambda}))+t(E(\pmb{\lambda})-E(\pmb{\mu}))&=
-t \frac{2\pi n}{L(1+\tfrac{2D}{c})}\left[\frac{2\pi
    n}{L(1+\tfrac{2D}{c})}+2\lambda_a +{\cal O}(L^{-1})\right]\nn
&\qquad+x' \frac{2\pi n}{L(1+\tfrac{2D}{c})}+{\cal O}(c^{-3})\,.
\end{align}

For later convenience we define
\begin{equation}\label{delta}
\delta\equiv \underset{L\to\infty}{\lim}\, \delta_L=\int_{-\infty}^\infty x\rho(x)\D{x}\,.
\end{equation}

\subsubsection{The phase for two particle-hole excitations}
%Form factors involving two particle-hole excitations are already at
%order $1/c^2$, so that \bl{in order to achieve} a precision of $1/c^2$
%in the end we only need to compute the phase at order $c^0$. We have
Using \eqref{mu2ph} 
\begin{equation}
\begin{aligned}
&x(P(\pmb{\mu})-P(\pmb{\lambda}))+t(E(\pmb{\lambda})-E(\pmb{\mu}))=x \frac{2\pi (n+m)}{L}
+t\Biggl[\lambda_a^2-\Big(\lambda_a+\frac{2\pi n}
  {L\big(1+\frac{2D}{c}\big)}\Big)^2\nn
&\qquad+\lambda_b^2-\Big(\lambda_b+\frac{2\pi
    m}{L\big(1+\frac{2D}{c}\big)} \Big)^2
-\frac{8\pi(n+m)}{cL^2\big(1+\frac{2D}{c}\big)}\sum_j\lambda_j\\
%-\frac{1}{cL}\Big(\frac{4\pi(n+m)}{L\big(1+\frac{2D}{c}\big)}\Big)^2(1+\frac{D}{c})
&\qquad-\frac{16\pi^2}{L^4(1+\tfrac{2D}{c})^2c^2}(n+m)^2\sum_i 1-\frac{16\pi^2}{L^3(1+\tfrac{2D}{c})^2c}(n+m)^2
\Biggr] +{\cal O}(c^{-3})\,.
\end{aligned}
\end{equation}

We can express this in terms of $x'$ as well
%\begin{equation}
%\begin{aligned}
%x(P(\pmb{\mu})-P(\pmb{\lambda}))+t(E(\pmb{\lambda})-E(\pmb{\mu}))&=x' \frac{2\pi (n+m)}{L}%\\
%&+t\left[\lambda_a^2-\left(\lambda_a+\frac{2\pi n}{L} \right)^2+\lambda_b^2-\left(\lambda_%b+\frac{2\pi m}{L} \right)^2 \right]\\
%&+O(c^{-1})\,.
%\end{aligned}
%\end{equation}
\begin{align}
&x(P(\pmb{\mu})-P(\pmb{\lambda}))+t(E(\pmb{\lambda})-E(\pmb{\mu}))=x'
\frac{2\pi (n+m)}{L\big(1+\frac{2D}{c}\big)}
+t\Biggl[\lambda_a^2-\Big(\lambda_a+\frac{2\pi n}
  {L\big(1+\frac{2D}{c}\big)}\Big)^2\nn
&\qquad+\lambda_b^2-\Big(\lambda_b+\frac{2\pi
    m}{L\big(1+\frac{2D}{c}\big)} \Big)^2
+{\cal O}(L^{-1})\Biggr] +{\cal O}(c^{-3})\,.
\end{align}

\subsubsection{The sum over intermediate states\label{suminterm}}
So far we have expanded all the terms arising in
\eqref{bigsumdensity} at order $1/c^2$, at a fixed $L$ for arbitrary
eigenstates $|\pmb{\lambda}\rangle$ and $|\pmb{\mu}\rangle$ with fixed Bethe numbers. We
have shown that the sum truncates to one- and two-particle-hole
excitations, and that the resulting terms are well-defined functions
of the excitation parameters $n$ and $m$. 

However, as the Lieb Liniger model is a field theory and not a
lattice model it features an \textit{infinite} number of
particle-hole states even if $L$ is finite, so that
\eqref{bigsumdensity} is still an infinite sum even if it involves 
only one- and two-particle-hole excitations. This creates two notable
problems. The first one is that we encounter infinite sums of the type
$\sum_{k=-\infty}^\infty k^n e^{ik^2t+ikx}$ for $n=0,1,2$ which are
\textit{ill-defined as functions of $x,t$} (except if $n=0$ and $t\neq 0$). The explanation for this
behaviour is that $\left\langle \sigma\left( x,t\right) \sigma \left(
0,0\right) \right\rangle$, similarly to the propagator of a quantum
particle, should be understood as a \textit{probability amplitude}
that is meant to be integrated against a smooth and localized function
of $x$ and $t$, or, stated differently, that it must be understood as
a distribution in $x,t$. The second problem is that the $1/c$
expansion of a form factor $\langle
\pmb{\lambda}|\sigma(0)|\pmb{\mu}\rangle$ has been performed for fixed 
Bethe numbers, whereas in the
spectral sum at fixed $c$ there are always  excited states with Bethe roots
larger than $c$. This poses a potential problem of commuting two
limits.

%{\color{red} I rewrote the following discussion, as the argument that
%the asymptotic expansion is carried out at fixed $L$ then a priori
%constrains $c$ to be much larger than $L$, and we then want to commute
%the two limits (which I think will not be correct for arbitrary
%states). This is exactly the issue J.-S.Caux was stressing.} 

In order to address these problems we are going to impose that all the rapidities involved in the spectral sum \eqref{bigsumdensity} are smaller than a certain \textit{cut-off} $\Lambda$, that can be taken as large as desired. Firstly, this imposes a restriction
of the state $|\pmb{\lambda}\rangle$, in which we are calculating our
expectation value. We require that for all roots $|\lambda_j|<\Lambda$, i.e. that the density $\rho(\lambda)$ vanishes for $|\lambda|>\Lambda$; this is a mild restriction in the following sense. In practice we are interested in the dynamical response in macro
states characterized by root distributions $\rho(\lambda)$ that decay
faster than $|\lambda|^{-2}$ for $\lambda\to\infty$, which is a
necessary condition for the energy density of the state to be
finite (for example in the thermal state the decay is Gaussian). We therefore can
always approximate $\rho(\lambda)$ to any given accuracy by a root
density $\rho_\Lambda(\lambda)$, which vanishes outside the interval
$[-\Lambda,\Lambda]$. Moreover this truncation can be done in an infinitely differentiable way, so it does not affect the regularity of the root density $\rho(\lambda)$. Secondly, this cut-off also restrains the sum \eqref{bigsumdensity} to excited states
$|\pmb{\mu}\rangle$ such that the $|\mu_i|<\Lambda$, which removes the problem of possible excited rapidities becoming larger than $c$. Hence we define a $\Lambda$-regularized correlation
function $\left\langle \sigma\left( x,t\right) \sigma \left(
0,0\right) \right\rangle_\Lambda$ as 
\begin{equation}
\label{bigsumdensitylambda}
\begin{aligned}
  \left\langle \sigma\left( x,t\right) \sigma \left( 0,0\right) \right\rangle_\Lambda &=\sum _{\ontop{\pmb{\mu}}{\forall i,\,|\mu_i|<\Lambda}}\frac {\left| \left\langle \pmb{\lambda} |\sigma \left( 0\right) |\pmb{\mu}\right\rangle \right| ^{2}} {\left\langle \pmb{\lambda} \left| \pmb{\lambda} \right\rangle \left\langle \pmb{\mu}\right| \pmb{\mu}\right\rangle }e^{it\left( E\left( \pmb{\lambda}\right) -E\left( \pmb{\mu} \right) \right) +ix\left( P\left( \pmb{\mu}\right) -P\left( \pmb{\lambda}\right) \right) }\,.
\end{aligned}
\end{equation}

The correlator $\left\langle \sigma\left( x,t\right) \sigma \left(
0,0\right) \right\rangle_\Lambda$ defined in this way and expanded in $1/c$ has a
regular thermodynamic limit $L\to\infty$, as we will see below. Now,
in order to recover the true correlation functions \eqref{bigsumdensity},
one would like to then take the limit $\Lambda\to\infty$. It turns out
that such a limit of $\left\langle \sigma\left( x,t\right) \sigma
\left( 0,0\right) \right\rangle_\Lambda$ seen as a function of $x,t$
does not exist. To be more specific one encounters problematic terms
of the form
\begin{equation}
\label{intdiverge}
I_n(\Lambda|t,x)=\int_{-\Lambda}^\Lambda  \mu^n
e^{-it\mu^2+ix\mu}\D{\mu}\,,\quad
n=0,1,2,
\end{equation}
for which the limit $\Lambda\to\infty$ does not
exist (except for $n=0$ if $t\neq 0$). However, the
limit exists in a distribution sense, i.e. the integral of $I_n(\Lambda|t,x)$ over any
smooth localized function of $x,t$ has a well-defined limit when
$\Lambda\to\infty$. This is all we require, since the
correlation function is in any case meant to be integrated with a
smooth localized function of $x,t$.   

To take the limit we perform an integration by part and obtain
\begin{equation}
I_n(\Lambda|t,x)=\frac{xI_{n-1}(\Lambda|t,x)}{2t}
+\frac{(n-1)I_{n-2}(\Lambda|t,x)}{2it}
+\frac{e^{-ix\Lambda}(-1)^{n-1}-e^{ix\Lambda}}{2it}e^{-it\Lambda^2}\Lambda^{n-1}\,.
\end{equation}
In particular we have
\begin{equation}
\label{reg}
\begin{aligned}
I_1(\Lambda|t,x)&=\frac{x}{2t}I_0(\Lambda|t,x)+\frac{e^{-ix\Lambda}-e^{ix\Lambda}}{2it}e^{-it\Lambda^2}\\
I_2(\Lambda|t,x)&=\Big(\left(\frac{x}{2t}\right)^2+\frac{1}{2it}\Big)I_0(\Lambda|t,x)-\frac{e^{-ix\Lambda}+e^{ix\Lambda}}{2it}e^{-it\Lambda^2}\Lambda+x\frac{e^{-ix\Lambda}-e^{ix\Lambda}}{4it^2}e^{-it\Lambda^2}\,,
\end{aligned}
\end{equation}
where
\begin{align}
\underset{\Lambda\to\infty}{\lim} I_0(\Lambda|t,x)&
= \int_{-\infty}^\infty e^{-it\mu^2+i\mu x}\D{\mu}\,,\qquad \text{if
}t\neq0\ ,\nn
I_0(\Lambda|t,x)&=\frac{e^{ix\Lambda}-e^{-ix\Lambda}}{ix}\,,\qquad\qquad
\text{if }t=0.
\end{align}
Terms like $e^{-it\Lambda^2 \mp ix \Lambda}\Lambda^n$ and $e^{ix\Lambda}$
do not have limits when $\Lambda\to\infty$ as a function of $x,t$. In
a distribution sense however, they vanish when $\Lambda\to\infty$ in
the sense that their integral with any smooth localized function of
$x,t$ vanishes when $\Lambda\to\infty$. Hence we obtain that when
$\Lambda\to\infty$ $I_n(\Lambda|t,x)$ tends to $I_n(t,x)$ with
$I_n(0,x)=0$ and
\begin{equation}
\label{reg2}
\begin{aligned}
I_1(t,x)&=\frac{x}{2t}I_0(t,x)\ ,\\
I_2(t,x)&=\bigg(\Big(\frac{x}{2t}\Big)^2+\frac{1}{2it}\bigg)I_0(t,x)\ ,\\
I_0(t,x)&=\int_{-\infty}^\infty e^{-it\mu^2+i\mu x}\D{\mu}\ ,\qquad t\neq0.
\end{aligned}
\end{equation}
One notices that these limits are exactly those obtained by introducing a small imaginary part in time and taking $\Lambda\to\infty$ 
\begin{equation}\label{epsto0}
I_n(t,x)=\underset{\epsilon\to 0^+}{\lim}\,\int_{-\infty}^\infty  \mu^ne^{-i(t-i\epsilon)\mu^2+ix\mu}\D{\mu}\,.
\end{equation}
However, such a small imaginary part cannot be incorporated from the beginning in \eqref{bigsumdensitylambda}, since $E(\pmb{\lambda})-E(\pmb{\mu})$ can take both signs when $|\pmb{\lambda}\rangle$ is not the ground state.

These limits will be useful in the
following sections in order to take the limit $\Lambda\to\infty$ of
the $\Lambda$-regularized correlation functions.\\

At order $1/c^2$ we therefore have the following decomposition 
\begin{equation}
\label{omegacorr}
\left\langle \sigma\left( x,t\right) \sigma \left( 0,0\right)
\right\rangle_\Lambda=D^2+\mathcal{C}^\Lambda_1(x,t)+\mathcal{C}^\Lambda_2(x,t)+{\cal
  O}(c^{-3})\ ,
\end{equation}
where $\mathcal{C}^\Lambda_{1,2}(x,t)$ are defined in the following. Introducing the convenient notations
%} {\color{red} I re-typeset the following equation to make it
%easier to digest -- this required introducing the $\lambda_{a,n}$. I
%also dropped ${\cal O}(1/L)$ terms in the exponentials.}
\be
\lambda_{a,n}\equiv\lambda_a+\frac{2\pi n}{L(1+\frac{2D}{c})},
\ee
and
\be
\lp=L\gam,
\ee
we have the following contribution at order $c^{-2}$ of the one-particle-hole excitations
\begin{equation}
\label{s1}
\begin{aligned}
&\mathcal{C}^\Lambda_1(x,t)=\frac{(1+\tfrac{2D}{c})^2}{L^2}\sum_{a=1}^N \sum_{\substack{n\\ \forall k,\, \lambda_{a,n}\neq\lambda_k\\ |\lambda_{a,n}|<\Lambda}}\Bigg[1+\frac{4}{cL}\frac{2\pi n}{\lp}\sum_{i\neq a}\bigg(\frac{1}{\lambda_i-\lambda_{a}-\tfrac{2\pi n}{\lp}}-\frac{1}{\lambda_i-\lambda_a}\Bigg)\\
&+\frac{4}{c^2L^2}\Big(\frac{2\pi n}{\lp}\Big)^2\bigg(-\frac{L^2}{12}\sum_{i\neq a}1+\sum_{\substack{i\neq j\\j\neq a}}\frac{1}{(\lambda_i-\lambda_j)^2}+2\Big(\sum_{i\neq
    a}\frac{1}{\lambda_i-\lambda_{a}-\tfrac{2\pi n}{\lp}}-\frac{1}{\lambda_i-\lambda_a}\Big)^2\nn
& +\sum_{i\neq a}\frac{1}{(\lambda_i-\lambda_{a}-\tfrac{2\pi n}{\lp})^2}
\bigg)\Bigg]\ 
\exp\bigg(ix' (\lambda_{a,n}-\lambda_{a})+it \left(\lambda_a^2-\lambda_{a,n}^2 \right) \bigg)\,.
\end{aligned}
\end{equation}
We already neglected a global factor $(1+\frac{2}{cL})^2$ that is $1$ in the thermodynamic limit, as well as a ${\cal O}(L^{-1})$ contribution in the exponential. We also used $\frac{1}{L^2 c^2}=\frac{1}{\lp^2 c^2}+{\cal O}(c^{-3})$ at order $c^{-2}$. 

\begin{figure}[H]
\begin{center}
\begin{tikzpicture}[scale=1]
\draw[->]     (3.5,0.25) arc (180: 0:1.5);
\node at (-1,0) {.};
\draw[black] (-0.5,0) circle (3pt);
\node at (0,0) {.};
\draw[black] (0.5,0) circle (3pt);
\node at (1,0) {.};
\draw[black] (1.5,0) circle (3pt);
\draw[black] (2,0) circle (3pt);
\node at (2.5,0) {.};
\node at (3,0) {.};
\draw[black] (3.5,0) circle (3pt);
\draw[black] (4,0) circle (3pt);
\draw[black] (4.5,0) circle (3pt);
\draw[black] (5,0) circle (3pt);
\node at (5.5,0) {.};
\draw[black] (6,0) circle (3pt);
\node at (6.5,0) {.};
\node at (7,0) {.};
\draw[black] (7.5,0) circle (3pt);
\node at (8,0) {.};
\node at (8.5,0) {.};
\draw[black] (9,0) circle (3pt);
\node at (9.5,0) {.};
\draw[black] (10,0) circle (3pt);
\draw[black] (10.5,0) circle (3pt);
\draw[black] (11,0) circle (3pt);
\node at (11.5,0) {.};
\draw[black] (12,0) circle (3pt);
\node at (12.5,0) {.};
\filldraw[black] (-0.5,0) circle (2pt);
\filldraw[black] (0.5,0) circle (2pt);
\filldraw[black] (1.5,0) circle (2pt);
\filldraw[black] (2,0) circle (2pt);
\filldraw[black] (6.5,0) circle (2pt);
\filldraw[black] (4,0) circle (2pt);
\filldraw[black] (4.5,0) circle (2pt);
\filldraw[black] (5,0) circle (2pt);
\filldraw[black] (6,0) circle (2pt);
\filldraw[black] (9,0) circle (2pt);
\filldraw[black] (7.5,0) circle (2pt);
\filldraw[black] (10,0) circle (2pt);
\filldraw[black] (10.5,0) circle (2pt);
\filldraw[black] (11,0) circle (2pt);
\filldraw[black] (12,0) circle (2pt);
\end{tikzpicture}
\end{center}
\caption{Sketch of a one-particle-hole excitation: positions of the
momenta of the representative state (empty circles) and the
intermediate state (filled circles) respectively, and position of 
the holes (dots).} 
\label{exfnp1k0}
\end{figure}
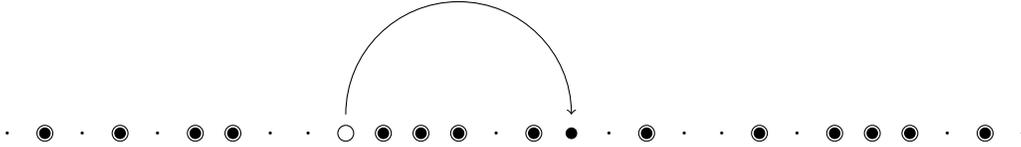

In Figure~\ref{exfnp1k0} we show the distribution of Bethe numbers for
the particle-hole excitations that are summed over in \fr{s1}. Compared to
the representative state we have changed a single integer.

For the two particle-hole excitations the sum in \eqref{bigsumdensity}
is over the set $\{a,b\}$ and over $n,m$ with the constraint
$\mu_a<\mu_b$. Since the form factor is symmetric upon swapping $a,b$
and $n,m$ simultaneously, this constraint can be taken into
account with a factor $1/2$ and with imposing $\mu_a\neq\mu_b$. The sum over
$\{a,b\}$ as a set can be transformed into a sum over $a\neq b$ as a
couple with a factor $1/2$ as well. Hence we have the leading
contribution of the two particle-hole excitations 
\begin{align}
\label{s2}
\mathcal{C}^\Lambda_2(x,t)&=\frac{1}{c^2L^4}\sum_{a\neq
    b}\sum_{\substack{n\\\forall i, \,
      \lambda_{a,n}\neq\lambda_i\\ |\lambda_{a,n}|<\Lambda}}\quad
  \sum_{\substack{m\\\forall i, \, \lambda_{b,m} \neq \lambda_i\\ \lambda_{b,m}\neq \lambda_{a,n}\\ |\lambda_{b,m}|<\Lambda}}
\frac{(n+m)^4}{n^2m^2}\frac{(\lambda_a-\lambda_b)^2(\lambda_a-\lambda_b+\tfrac{2\pi}{\lp}(n-m))^2}{(\lambda_a-\lambda_b+\tfrac{2\pi}{\lp}n)^2(\lambda_a-\lambda_b-\tfrac{2\pi}{\lp}m)^2}\nn
&\qquad\qquad\times
\exp\bigg(it\left[\lambda_a^2-\lambda_{a,n}^2+\lambda_b^2-\lambda_{b,m}^2\right]+ix'
[\lambda_{a,n}-\lambda_a+\lambda_{b,m}-\lambda_b] \bigg)\,. 
\end{align}

%\begin{equation}
%\label{s2}
%\begin{aligned}
%&\mathcal{C}^\Lambda_2(x,t)=\frac{1}{c^2L^4}\sum_{a\neq b}\sum_{\substack{n\\\forall i, \, n \neq \tfrac{L(\lambda_i-\lambda_a)}{2\pi}\\ |\lambda_a+\tfrac{2\pi n}{L}|<\Lambda}}\qquad \sum_{\substack{m\\\forall i, %\, m \neq \tfrac{L(\lambda_i-\lambda_b)}{2\pi}\\m\neq n+\tfrac{L(\lambda_a-\lambda_b)}{2\pi}\\ |\lambda_b+\tfrac{2\%pi m}{L}|<\Lambda}}\\
%&\frac{(n+m)^4}{n^2m^2}\frac{(\lambda_a-\lambda_b)^2(\lambda_a-\lambda_b+\tfrac{2\pi}{L}(n-m%))^2}{(\lambda_a-\lambda_b+\tfrac{2\pi}{L}n)^2(\lambda_a-\lambda_b-\tfrac{2\pi}{L}m)^2}\\
%&\times \exp\left[it\left[\lambda_a^2-(\lambda_a+\tfrac{2\pi n}{L})^2+\lambda_b^2-(\lambda_b%+\tfrac{2\pi m}{L})^2\right]+ix' \frac{2\pi(n+m)}{L} \right]\,.
%\end{aligned}
%\end{equation}

\begin{figure}[H]
\begin{center}
\begin{tikzpicture}[scale=1]
\draw[->]     (7.5,0.25) arc (180: 0:0.5);
\draw[->]     (3.5,0.25) arc (0: 180:1.25);
\node at (-1,0) {.};
\draw[black] (-0.5,0) circle (3pt);
\node at (0,0) {.};
\draw[black] (0.5,0) circle (3pt);
\node at (1,0) {.};
\draw[black] (1.5,0) circle (3pt);
\draw[black] (2,0) circle (3pt);
\node at (2.5,0) {.};
\node at (3,0) {.};
\draw[black] (3.5,0) circle (3pt);
\draw[black] (4,0) circle (3pt);
\draw[black] (4.5,0) circle (3pt);
\draw[black] (5,0) circle (3pt);
\node at (5.5,0) {.};
\draw[black] (6,0) circle (3pt);
\node at (6.5,0) {.};
\node at (7,0) {.};
\draw[black] (7.5,0) circle (3pt);
\node at (8,0) {.};
\node at (8.5,0) {.};
\draw[black] (9,0) circle (3pt);
\node at (9.5,0) {.};
\draw[black] (10,0) circle (3pt);
\draw[black] (10.5,0) circle (3pt);
\draw[black] (11,0) circle (3pt);
\node at (11.5,0) {.};
\draw[black] (12,0) circle (3pt);
\node at (12.5,0) {.};
\filldraw[black] (-0.5,0) circle (2pt);
\filldraw[black] (0.5,0) circle (2pt);
\filldraw[black] (1.5,0) circle (2pt);
\filldraw[black] (2,0) circle (2pt);
\filldraw[black] (1,0) circle (2pt);
\filldraw[black] (4,0) circle (2pt);
\filldraw[black] (4.5,0) circle (2pt);
\filldraw[black] (5,0) circle (2pt);
\filldraw[black] (6,0) circle (2pt);
\filldraw[black] (8.5,0) circle (2pt);
\filldraw[black] (9,0) circle (2pt);
\filldraw[black] (10,0) circle (2pt);
\filldraw[black] (10.5,0) circle (2pt);
\filldraw[black] (11,0) circle (2pt);
\filldraw[black] (12,0) circle (2pt);
\end{tikzpicture}
\end{center}
\caption{Sketch of a two particle-hole excitation: position of the
momenta of the representative state (empty circles) and the
intermediate state (filled circles) respectively, and position of 
the holes (dots).} 
\label{exfnp2k0}
\end{figure}
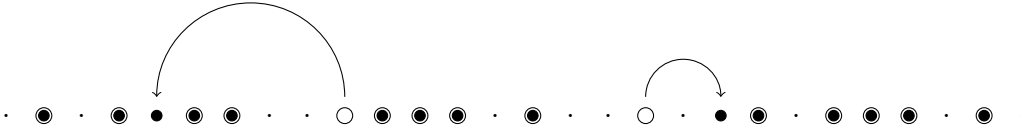

In Figure~\ref{exfnp2k0} we show the distribution of Bethe numbers for
the two particle-hole excitations that are summed over in \fr{s2}. Compared to
the representative state we have changed two integers.

%{\color{red}I would cut the following paragraph.}
%\bl{  
%We conclude this section by recalling the order of the three limits
%$c,L,\Lambda\to\infty$ involved in the calculation of the correlation
%functions from $\mathcal{C}^\Lambda_1(x,t)$ and
%$\mathcal{C}^\Lambda_2(x,t)$. We first take $c\to\infty$ by
%considering only the terms up to $c^{-2}$ in
%$\mathcal{C}^\Lambda_1(x,t)$ and $\mathcal{C}^\Lambda_2(x,t)$ at fixed
%$L$ and $\Lambda$. Then we take $L\to\infty$ in order to obtain the
%thermodynamic limit value of the $\Lambda$-correlation functions, and
%then we take the limit $\Lambda\to\infty$ to recover the full
%correlation functions. Although the calculation technically required
%to introduce a cutoff $\Lambda$ in order to carry out the expansion in
%$c^{-1}$, it can be sent to $\infty$ and the end result will not
%depend on any cutoff. 
%}
\subsection{Examples of root densities}
In this subsection we complete the $1/c$ expansion of the model by
determining the expansions of the root densities introduced in
Section \ref{exrootg}. 

\subsubsection{Hole density}
We introduced earlier the hole density $\rho_h(\lambda)$ in \eqref{rhoh} with $\vartheta(\lambda)$ given in terms of $\rho(\lambda)$ in \eqref{vartheta}. The hole density is a function of the root density $\rho(\lambda)$, and for a generic $\rho(\lambda)$ it reads at order $c^{-2}$
\begin{equation}
\label{rhohc}
\rho_h(\lambda)=\frac{1+\frac{2D}{c}}{2\pi}-\rho(\lambda)+{\cal O}(c^{-3})\,,
\end{equation}
where we recall that $D$ is defined in \eqref{D}. 
%%%%%%%%%%%%%%%%%%%%%%%%%%%%%%%%%%%%%%%%%%%%%%%
\subsubsection{Thermal states \label{thexp}}
%%%%%%%%%%%%%%%%%%%%%%%%%%%%%%%%%%%%%%%%%%%%%%%
Thermal states at finite inverse temperature $\beta<\infty$ are
defined in terms of the nonlinear integral equation for the dressed
energy \fr{dresseden2} and the thermodynamic limit of the Bethe Ansatz
equations \fr{vartheta}. These can be expanded in $1/c$ without
difficulty, and we obtain the following result for the particle
density at order $1/c^2$
\begin{equation}
\label{rhotemperature}
\rho(x)=\frac{1}{2\pi}\frac{A(c,\beta)}{1+e^{\beta x^2+B(c,\beta)}}\,,
\end{equation}
where
\begin{align}
A(c,\beta)&=1-\frac{\li_{1/2}(-e^{\beta h})}{\sqrt{\pi\beta}c}
+\frac{\li^2_{1/2}(-e^{\beta h})+\li_{-1/2}(-e^{\beta
    h})\li_{3/2}(-e^{\beta h})}{\pi\beta c^2}+{\cal O}(c^{-3}),\nn
B(c,\beta)&=-\beta h+\frac{\li_{3/2}(-e^{\beta
    h})}{\sqrt{\pi\beta}c}-\frac{\li_{1/2}(-e^{\beta
    h})\li_{3/2}(-e^{\beta h})}{\pi\beta c^2}+{\cal O}(c^{-3})\,.
\label{rhoT2}
\end{align}
%\begin{equation}
%\begin{aligned}
%A(c,\beta)&=1-\frac{1}{c\sqrt{\beta \pi}}\left(1-\frac{\li_{1/2}(-e^{\beta h})}{c\sqrt{\beta\pi}}\right)\li_{1/2}\left[-e^{-\tfrac{\li_{3/2}(-e^{\beta h})}{c\sqrt{\beta\pi}}+\beta h}\right]\\
%B(c,\beta)&=-\beta h+\frac{1}{c\sqrt{\beta\pi}} \li_{3/2}\left[-e^{-\tfrac{\li_{3/2}(-e^{\beta h})}{c\sqrt{\beta\pi}}+\beta h}\right]\,,
%\end{aligned}
%\end{equation}
We recall that $h$ is the chemical potential used to fix the particle
number $D$. In order to derive \fr{rhoT2}
we used the following relations
\begin{equation}
\begin{aligned}
\int_{-\infty}^\infty \frac{\D{x}}{1+e^{x^2+y}}&=-\sqrt{\pi}\li_{1/2}\left[-e^{-y}\right]\\
\int_{-\infty}^\infty \log(1+e^{-x^2-y})\D{x}&=-\sqrt{\pi}\li_{3/2}\left[-e^{-y}\right]\,.
\end{aligned}
\end{equation}

\subsubsection{Zero temperature ground state}
Equation \eqref{0t} for the ground state root density can be expanded in $1/c$ to yield
\begin{align}
\label{rho0t}
\rho(\lambda)&=\frac{1+\frac{2D}{c}}{2\pi}\1_{|\lambda|<Q}+{\cal O}(c^{-3})\,,\nn
Q&=\frac{q_F}{1+\frac{2D}{c}}+{\cal O}(c^{-3})\,.
\end{align}
Here $\1_\mathcal{P}$ is the indicator function, equal to $1$ if the affirmation $\mathcal{P}$ is true and $0$ if it is false. The Luttinger parameter $K=(2\pi\rho(Q))^2$ \cite{korepin} is
\begin{equation}
\label{K}
K=1+\frac{4D}{c}+\frac{4D^2}{c^2}+{\cal O}(c^{-3})\,.
\end{equation}

%%%%%%%%%%%%%%%%%%%%%%%%%%%%%%%%%%%%%%%%%
\section {The thermodynamic limit of correlation functions\label{sec3}}
%%%%%%%%%%%%%%%%%%%%%%%%%%%%%%%%%%%%%%%%%
In this section we perform explicitly the sum over intermediate states
in \eqref{omegacorr} at order $1/c^2$ in the infinite volume limit
$L\to\infty$.  
\subsection{One particle-hole excitations}
Our starting point is $\mathcal{C}^\Lambda_1(x,t)$ as
defined in \eqref{s1}. In the following we consider the different
orders in the $1/c$-expansion and derive integral representations of
the corresponding contributions to $\mathcal{C}^\Lambda_1(x,t)$ in the
thermodynamic limit. As we have noted before, we retain certain
resummed expressions in this expansion for convenience, an example
being the factor $(1+\frac{2D}{c})$. When we refer to a given order of
the $1/c$-expansion this should be understood modulo such factors.
%%%%%%%%%%%%%%%%%%%%%%%%%%%%
\subsubsection{\sfix{Order $c^0$}}
%%%%%%%%%%%%%%%%%%%%%%%%%%%%
Let us focus first on the leading order ${\cal O}(c^{0})$, namely
\begin{equation}
\mathcal{A}_0=\frac{(1+\tfrac{2D}{c})^2}{L^2}
\sum_{a} \sum_{\substack{n\\ \forall k,\, \lambda_{a,n}\neq\lambda_k\\ |\lambda_{a,n}|<\Lambda}} e^{ix' (\lambda_{a,n}-\lambda_{a})+it \left(\lambda_a^2-\lambda_{a,n}^2 \right)}\,.
\end{equation}
We rewrite this as
\begin{align}
\mathcal{A}_0&=\frac{(1+\tfrac{2D}{c})^2}{L^2}\sum_{a} \sum_{\substack{n\\ |\lambda_{a,n}|<\Lambda}}e^{ix' (\lambda_{a,n}-\lambda_{a})+it \left(\lambda_a^2-\lambda_{a,n}^2 \right)}\nn
&-\frac{(1+\tfrac{2D}{c})^2}{L^2}\sum_{a} \sum_{\substack{k\\ |\lambda_k|<\Lambda}}e^{ix' (\lambda_{k}-\lambda_{a})+it \left(\lambda_a^2-\lambda_{k}^2 \right)}\,.
\end{align}
Using \eqref{2dsum} the sums over $a$ and $k$ can be turned into
integrals over the root density $\rho(\lambda)$, and the sum
over $n$ into an integral with density $\tfrac{1+\frac{2D}{c}}{2\pi}$.
Altogether we find
\begin{equation}
\mathcal{A}_0=(1+\tfrac{2D}{c})^2\int_{-\infty}^\infty
\D{\lambda}\ \rho(\lambda)\int_{-\Lambda}^\Lambda \D{\mu}\ \rho_h(\mu)
e^{it(\lambda^2-\mu^2)+ix'(\mu-\lambda)}+{\cal O}(L^{-1})\,,
\end{equation}
where we used the expression \eqref{rhohc} for the hole density
$\rho_h$ at order $c^{-2}$.
%%%%%%%%%%%%%%%%%%%%%%%%%%%%%%
\subsubsection{\sfix{Order $c^{-1}$}}
%%%%%%%%%%%%%%%%%%%%%%%%%%%%%%
We next turn to the $c^{-1}$ term
\begin{align}
\mathcal{A}_1=4\frac{(1+\tfrac{2D}{c})^2}{cL^3}\sum_{a}
&\sum_{\substack{n\\ \forall k,\, \lambda_{a,n}\neq \lambda_k\\ |\lambda_{a,n}|<\Lambda}}\frac{2\pi n}{\lp}\sum_{i\neq a}\Bigg(\frac{1}{\lambda_i-\lambda_{a}-\frac{2\pi n}{\lp}}-\frac{1}{\lambda_i-\lambda_a}\Bigg)\nn
&\times e^{ix' (\lambda_{a,n}-\lambda_{a})+it \left(\lambda_a^2-\lambda_{a,n}^2 \right)}\,.
\end{align}
We rewrite this as
\begin{align}
\mathcal{A}_1=4\frac{(1+\tfrac{2D}{c})^2}{cL^3}\sum_{a}
\sum_{i\neq a}&
\sum_{\substack{n\\ \lambda_{a,n}\neq \lambda_i\\  |\lambda_{a,n}|<\Lambda}}\frac{2\pi n}{\lp}\bigg(\frac{1}{\lambda_i-\lambda_a-\tfrac{2\pi n}{\lp}}-\frac{1}{\lambda_i-\lambda_a}\bigg)\nn
&\times e^{ix' (\lambda_{a,n}-\lambda_{a})+it \left(\lambda_a^2-\lambda_{a,n}^2 \right)}\nn
\quad-4\frac{(1+\tfrac{2D}{c})^2}{cL^3}\sum_{a} \sum_{i\neq a}&\sum_{\substack{k\\k\neq i\\  |\lambda_k|<\Lambda}}(\lambda_k-\lambda_a)\bigg(\frac{1}{\lambda_i-\lambda_k}-\frac{1}{\lambda_i-\lambda_a}\bigg)\nn
&\times e^{ix' (\lambda_{k}-\lambda_{a})+it \left(\lambda_a^2-\lambda_{k}^2 \right)}\,.
\end{align}
This term involves either a sum over regularly spaced integers $n$
that becomes an integral with density $\tfrac{1+2D/c}{2\pi}$ in the
thermodynamic limit, or sums of the type \eqref{sipdpp} that can be
expressed as principal part integrals over the root density. We obtain  
\begin{align}
\mathcal{A}_1=\frac{4(1+\tfrac{2D}{c})^2}{c}\int_{-\infty}^\infty
\D{\lambda}\ \rho(\lambda)&\int_{-\Lambda}^\Lambda \D{\mu}
(\mu-\lambda)\left[\dashint \frac{\rho(u)}{u-\mu}\D{u}-\dashint
\frac{\rho(u)}{u-\lambda}\D{u}\right]\nn
&\times e^{it(\lambda^2-\mu^2)+ix'(\mu-\lambda)}\bigg(
\frac{1+\frac{2D}{c}}{2\pi}-\rho(\mu)\bigg)  
+{\cal O}(L^{-1}) \,.
\end{align}

Introducing the Hilbert transform $\tilde{\rho}$ of $\rho$ by
\begin{equation}
\label{rhotilde}
\tilde{\rho}(\lambda)=\dashint \frac{\rho(u)}{\lambda-u}\D{u}\,,
\end{equation}
permits us to rewrite this contribution in the form
\begin{align}
\mathcal{A}_1&=-\frac{4(1+\tfrac{2D}{c})^2}{c}\int_{-\infty}^\infty
\D{\lambda} \rho(\lambda)\int_{-\Lambda}^\Lambda \D{\mu}
\rho_h(\mu)(\mu-\lambda)
\big[\tilde{\rho}(\mu)-\tilde{\rho}(\lambda)\big]
e^{it(\lambda^2-\mu^2)+ix'(\mu-\lambda)}\nn
&\qquad+{\cal O}(L^{-1}) \,.
\end{align}
%%%%%%%%%%%%%%%%%%%%%%%%%%%%%%%%%
\subsubsection{\sfix{Order $c^{-2}$: first contribution}}
%%%%%%%%%%%%%%%%%%%%%%%%%%%%%%%%%
We now consider contributions involving the factor
\begin{align}
\label{bigterm}
&\bigg(\sum_{i\neq a}\frac{1}{\lambda_i-\lambda_{a}-\frac{2\pi n}{\lp}}-\frac{1}{\lambda_i-\lambda_a}\bigg)^2=\sum_{\substack{i\neq a\\j\neq a}}\frac{1}{(\lambda_i-\lambda_a)(\lambda_j-\lambda_a)}\nn
&\qquad\qquad+\sum_{\substack{i\neq a\\j\neq
    a}}\frac{1}{(\lambda_i-\lambda_{a}-\frac{2\pi n}{\lp})(\lambda_j-\lambda_{a}-\frac{2\pi n}{\lp})}-2\sum_{\substack{i\neq a\\j\neq a}}\frac{1}{(\lambda_i-\lambda_{a}-\frac{2\pi n}{\lp})(\lambda_j-\lambda_a)}\,,
\end{align}
which are more delicate. The first term on the right hand side
gives rise to a contribution
\begin{align}
\mathcal{A}_2=8\frac{(1+\tfrac{2D}{c})^2}{c^2L^4}\sum_{a}
&\sum_{\substack{n\\ \forall k,\, \lambda_{a,n}\neq
      \lambda_k\\ |\lambda_{a,n}|<\Lambda}}
\left(\frac{2\pi n}{\lp}\right)^2\sum_{\substack{i,j\\i\neq a\\j\neq a}}\frac{1}{(\lambda_i-\lambda_a)(\lambda_j-\lambda_a)}\nn
&\times e^{ix' (\lambda_{a,n}-\lambda_{a})+it \left(\lambda_a^2-\lambda_{a,n}^2 \right)}\,.
\end{align}
We rewrite this as
\begin{align}
\mathcal{A}_2=&8\frac{(1+\tfrac{2D}{c})^2}{c^2L^4}\sum_{a}
\sum_{\substack{n\\ |\lambda_{a,n}|<\Lambda}}\left(\frac{2\pi n}{\lp}\right)^2\sum_{\substack{i,j\\i\neq a\\j\neq a}}\frac{e^{ix' (\lambda_{a,n}-\lambda_{a})+it \left(\lambda_a^2-\lambda_{a,n}^2 \right)}}{(\lambda_i-\lambda_a)(\lambda_j-\lambda_a)} \nn
&-8\frac{(1+\tfrac{2D}{c})^2}{c^2L^4}\sum_{a}
\sum_{\substack{k\\ |\lambda_k|<\Lambda}}\left(\lambda_k-\lambda_a\right)^2\sum_{\substack{i,j\\i\neq a\\j\neq a}}\frac{e^{ix' (\lambda_{k}-\lambda_{a})+it \left(\lambda_a^2-\lambda_{k}^2 \right)}}{(\lambda_i-\lambda_a)(\lambda_j-\lambda_a)} \,.
\end{align}
The two terms are of the form \eqref{sipdpp3} and we apply
\eqref{simpledouble} to express them in terms of the root density
$\rho(\lambda)$ and the pair distribution function
$\gamma_{-2}(\lambda)$ defined in \eqref{sipddef}, with a triple
integral with successive principal values defined in
\eqref{dashint}. This yields  
\begin{align}
\mathcal{A}_2=&-\frac{8}{c^2}\int_{-\Lambda}^\Lambda \D{\mu} \rho_h(\mu)e^{-it\mu^2+ix'\mu}\dashint \frac{(\mu-\lambda)^2 \rho(\lambda)\rho(u)\rho(v)}{(u-\lambda)(\lambda-v)}e^{it\lambda^2-ix'\lambda}\D{\lambda} \D{u} \D{v}\nn
&-\frac{8}{c^2}\int_{-\infty}^\infty \D{\lambda}\int_{-\Lambda}^\Lambda \D{\mu} (\mu-\lambda)^2\rho_h(\mu)\left[\tfrac{\pi^2}{3}\rho(\lambda)^3-\gamma_{-2}(\lambda) \right]e^{it(\lambda^2-\mu^2)+ix'(\mu-\lambda)}\D{\lambda} \D{\mu}\nn
&+{\cal O}(c^{-3})+{\cal O}(L^{-1})\,.
\end{align}
The definition of the successive principal value integral allows us to rewrite it in terms of $\tilde{\rho}$, to give
\begin{align}
\mathcal{A}_2=&\frac{8}{c^2}\int_{-\infty}^\infty \D{\lambda}\rho(\lambda)\int_{-\Lambda}^\Lambda \D{\mu} \rho_h(\mu)(\mu-\lambda)^2\tilde{\rho}(\lambda)^2e^{it(\lambda^2-\mu^2)+ix'(\mu-\lambda)}\nn
&-\frac{8}{c^2}\int_{-\infty}^\infty \D{\lambda}\int_{-\Lambda}^\Lambda \D{\mu} (\mu-\lambda)^2\rho_h(\mu)\left[\tfrac{\pi^2}{3}\rho(\lambda)^3-\gamma_{-2}(\lambda) \right]e^{it(\lambda^2-\mu^2)+ix'(\mu-\lambda)}\D{\lambda} \D{\mu}\nn
&+{\cal O}(c^{-3})+{\cal O}(L^{-1})\,.
\end{align}
%%%%%%%%%%%%%%%%%%%%%%%%%%%%%%%%%
\subsubsection{\sfix{Order $c^{-2}$: second contribution}}
%%%%%%%%%%%%%%%%%%%%%%%%%%%%%%%%%
The second term on the right hand side of \eqref{bigterm} is
particularly cumbersome to deal with. We first treat the
case $i=j$ separately, and for all terms with $i\neq j$ we apply a
partial fraction decomposition with respect to $n$, so that we have only one $n$
appearing in the denominator. Finally we split the sum over $n$
as the difference of sums over vacancies and
particles. Specifically, we have for
$f(u)=u^2e^{ix'u+it(\lambda_a^2-(\lambda_a+u)^2))}$  
\begin{align}
\label{bigcemposi}
&\sum_{\substack{i,j,n\\ \forall k,\,\lambda_{a,n}\neq\lambda_k\\|\lambda_{a,n}|<\Lambda \\ i,j\neq a}}\frac{f(\tfrac{2\pi  n}{\lp})}{(\lambda_i-\lambda_{a}-\frac{2\pi n}{\lp}  )(\lambda_j-\lambda_{a}-\frac{2\pi n}{\lp})}=
\sum_{\substack{i,n\\ \lambda_{a,n}\neq\lambda_i\\|\lambda_{a,n}|<\Lambda \\i\neq a}}\frac{f(\tfrac{2\pi n}{\lp})}{(\lambda_i-\lambda_{a}-\frac{2\pi n}{\lp})^2}-\sum_{\substack{i,k\\ i\neq k\\ |\lambda_k|<\Lambda \\i\neq a}}\frac{f(\lambda_k-\lambda_a)}{(\lambda_i-\lambda_k)^2}\nn
&\qquad\qquad+\sum_{\substack{i,j,n\\ \lambda_{a,n}\neq
    \lambda_i\\  |\lambda_{a,n}|<\Lambda \\i,j\neq a,\ i\neq j}}\frac{f(\tfrac{2\pi
    n}{\lp})}{(\lambda_j-\lambda_i)(\lambda_i-\lambda_{a}-\frac{2\pi n}{\lp})}-\sum_{\substack{i,j,n\\ \lambda_{a,n}\neq
    \lambda_j\\ |\lambda_{a,n}|<\Lambda\\i,j\neq a,\ i\neq j\\ }}\frac{f(\tfrac{2\pi n}{\lp})}{(\lambda_j-\lambda_i)(\lambda_j-\lambda_{a}-\frac{2\pi n}{\lp})}\nn
&\qquad\qquad
-\sum_{\substack{i,j,k\\ i\neq j,\ i\neq k\\ |\lambda_k|<\Lambda\\i,j\neq a}}\frac{f(\lambda_k-\lambda_a)}{(\lambda_j-\lambda_i)(\lambda_i-\lambda_k)}+\sum_{\substack{i,j,k\\ i\neq j,\ j\neq k\\ |\lambda_k|<\Lambda\\i,j\neq a }}\frac{f(\lambda_k-\lambda_a)}{(\lambda_j-\lambda_i)(\lambda_j-\lambda_k)}\,.
\end{align}

In all these terms, the conditions $i,j\neq a$ only give rise to
subleading contributions in $L$, so that they can be
discarded. The first term on the right hand side of
\fr{bigcemposi} gives rise to a contribution to $\mathcal{C}_1^\Lambda(x,t)$ of
the form
\begin{equation}
\begin{aligned}
\mathcal{A}_3=8\frac{(1+\tfrac{2D}{c})^2}{c^2L^4}\sum_{a} &\sum_{\substack{n\\ \lambda_{a,n}\neq\lambda_i\\ |\lambda_{a,n}|<\Lambda}}\left(\frac{2\pi n}{\lp}\right)^2\sum_{\substack{i\\i\neq a}}\frac{e^{ix' (\lambda_{a,n}-\lambda_{a})+it \left(\lambda_a^2-\lambda_{a,n}^2 \right)}}{(\lambda_i-\lambda_{a}-\frac{2\pi n}{\lp})^2}\,.
\end{aligned}
\end{equation}
As this is proportional to $L^{-4}$ and only involves
three sums the dominant contribution arises from the double pole. Using $\sum_{n\neq
  0}\tfrac{1}{n^2}=\tfrac{\pi^2}{3}$ for the sum over $n$, we obtain 
\begin{equation}
\begin{aligned}
\mathcal{A}_3=\frac{8}{c^2}\int_{-\infty}^\infty
(\lambda-\mu)^2\rho(\lambda)\frac{\pi^2}{3}\frac{\rho(\mu)}{(2\pi)^2}e^{it(\lambda^2-\mu^2)+ix'(\mu-\lambda)}\D{\lambda}
\D{\mu}+{\cal O}(c^{-3})+{\cal O}(L^{-1})\,.
\end{aligned}
\end{equation}
%%%%%%%%%%%%%%%%%%%%%%%%%%%%%%%%%%
\subsubsection{\sfix{Order $c^{-2}$: third contribution}}
%%%%%%%%%%%%%%%%%%%%%%%%%%%%%%%%%%
The second term on the right hand side of \eqref{bigcemposi}
gives rise to a contribution
\be
\mathcal{A}_4=-8\frac{(1+\tfrac{2D}{c})^2}{c^2L^4}\sum_{a} \sum_{\substack{i,k\\i,k\neq a\\ i\neq k}}\frac{\left(\lambda_k-\lambda_a\right)^2}{(\lambda_i-\lambda_k)^2}
e^{ix' (\lambda_{k}-\lambda_{a})+it \left(\lambda_a^2-\lambda_{k}^2 \right)}\,.
\ee
The sum is of the form \eqref{sipd} and according to
\eqref{sipddef} in the thermodynamic limit gives rise to integrals
over the pair distribution function
\begin{equation}
\mathcal{A}_4=-\frac{8}{c^2}\int_{-\infty}^\infty
(\lambda-\mu)^2\rho(\lambda)\gamma_{-2}(\mu)e^{it(\lambda^2-\mu^2)+ix'(\mu-\lambda)}\D{\lambda}
\D{\mu}+{\cal O}(c^{-3})+{\cal O}(L^{-1})\,.
\end{equation}
%%%%%%%%%%%%%%%%%%%%%%%%%%%%%%%%%%%
\subsubsection{\sfix{Order $c^{-2}$: fourth contribution}}
%%%%%%%%%%%%%%%%%%%%%%%%%%%%%%%%%%%
The third and fourth terms in \eqref{bigcemposi} are ``hybrid" terms
mixing sums over $\lambda_i$'s and sums over regularly distributed
$n$'s. They give rise to a contribution to ${\cal C}_1^\Lambda(x,t)$ of
the form
\begin{align}
\mathcal{A}_5=16\frac{(1+\tfrac{2D}{c})^2}{c^2L^4}\sum_{a} &\sum_{\substack{i,j\\i,j\neq a\\i\neq j}}\sum_{\substack{n\\\lambda_{a,n}\neq\lambda_i\\ |\lambda_{a,n}|<\Lambda}}\left(\frac{2\pi n}{\lp}\right)^2\frac{e^{ix' (\lambda_{a,n}-\lambda_{a})+it \left(\lambda_a^2-\lambda_{a,n}^2 \right)}}{(\lambda_i-\lambda_{a}-\frac{2\pi n}{\lp})(\lambda_j-\lambda_i)} \,.
\end{align}
By symmetrizing over $i,j$, one obtains a sole pole in $n$, but since
$n$ is regularly distributed and avoids only the pole one can
convert the sum into a principal value integral. This leads to
integrals with two successive principal values
\begin{align}
\mathcal{A}_5=&\frac{16}{c^2}\int_{-\infty}^\infty \D{\lambda}\rho(\lambda)\int_{-\Lambda}^\Lambda \D{\mu} \frac{1}{2\pi}(\mu-\lambda)^2e^{it(\lambda^2-\mu^2)+ix'(\mu-\lambda)}\dashint \D{u} \frac{\rho(u)}{u-\mu}\dashint \D{v} \frac{\rho(v)}{v-u}\nn
&+{\cal O}(c^{-3})+{\cal O}(L^{-1})\,.
\end{align}
This can be simplified further by expressing the rightmost double
integral in terms of $\tilde{\rho}(\lambda)$. To that end, let us
consider the integral of this term as a function of $\mu$ with an
arbitrary continuous function  $\varphi(\mu)$. Using \eqref{PB} we have  
\begin{equation}
\int \D{\mu} \varphi(\mu)\dashint \D{u} \frac{\rho(u)}{\mu-u}\dashint \D{v} \frac{\rho(v)}{u-v}=\ddashint \frac{\varphi(\mu)\rho(u)\rho(v)}{(\mu-u)(u-v)}\D{\mu} \D{u} \D{v}-\frac{\pi^2}{3}\int \varphi(\mu)\rho(\mu)^2\D{\mu}\,.
\end{equation}

Under the simultaneous principal value triple integral it is
legitimate to decompose $\tfrac{1}{(\mu-u)(u-v)} =
\tfrac{1}{\mu-v}(\tfrac{1}{\mu-u} + \tfrac{1}{u-v})$
and split the integral into two since $|\mu-v|>\epsilon$:
\begin{equation}
\ddashint \frac{\varphi(\mu)\rho(u)\rho(v)}{(\mu-u)(u-v)}\D{\mu} \D{u} \D{v}=\ddashint \frac{\varphi(\mu)\rho(u)\rho(v)}{(\mu-v)(\mu-u)}\D{\mu} \D{u} \D{v}+\ddashint \frac{\varphi(\mu)\rho(u)\rho(v)}{(\mu-v)(u-v)}\D{\mu} \D{u} \D{v}\,.
\end{equation}
We then use \eqref{PB} to express the two simultaneous principal
value triple integrals in terms of successive principal value integrals
\begin{align}
\int \D{\mu} \varphi(\mu)\dashint \D{u} \frac{\rho(u)}{\mu-u}\dashint \D{v} \frac{\rho(v)}{u-v}=&
\int \D{\mu} \varphi(\mu)\dashint \D{v} \frac{\rho(v)}{\mu-v}\dashint \D{u}
\frac{\rho(u)}{\mu-u}-\pi^2\int \varphi(\mu)\rho(\mu)^2\D{\mu}\nn
&+\int \D{\mu} \varphi(\mu)\dashint \D{v} \frac{\rho(v)}{\mu-v}\dashint \D{u}
\frac{\rho(u)}{u-v}\ .
\end{align}

The first integral on the right hand side is $\int
\varphi(\mu)\tilde{\rho}(\mu)^2\D{\mu}$ while the third equals minus
the left hand side. Using that this identity holds for any
continuous function $\varphi(\mu)$ we conclude that  
\begin{equation}\label{rhorhotilda}
\dashint  \D{\lambda} \frac{\rho(\lambda)}{\mu-\lambda}\dashint \D{u} \frac{\rho(u)}{\lambda-u}=\frac{1}{2}\tilde{\rho}(\mu)^2-\frac{\pi^2}{2}\rho(\mu)^2\,.
\end{equation}
Putting everything together we obtain
\begin{align}
\mathcal{A}_5=&\frac{8}{c^2}\int_{-\infty}^\infty \D{\lambda}\rho(\lambda)\int_{-\Lambda}^\Lambda \D{\mu} \frac{1}{2\pi}(\mu-\lambda)^2\tilde{\rho}(\mu)^2e^{it(\lambda^2-\mu^2)+ix'(\mu-\lambda)}\nn
&-\frac{8\pi^2}{c^2}\int_{-\infty}^\infty
\D{\lambda}\rho(\lambda)\int_{-\Lambda}^\Lambda \D{\mu}
\frac{1}{2\pi}(\mu-\lambda)^2\rho(\mu)^2e^{it(\lambda^2-\mu^2)+ix'(\mu-\lambda)}+{\cal
  O}(c^{-3})+{\cal O}(L^{-1})\,.
\end{align}
%%%%%%%%%%%%%%%%%%%%%%%%%%%%%%%%%%%
\subsubsection{\sfix{Order $c^{-2}$: fifth contribution}}
%%%%%%%%%%%%%%%%%%%%%%%%%%%%%%%%%%%
The fifth and fourth terms on the right hand side of
\eqref{bigcemposi} are of the form \eqref{sipdpp3} and give rise to a
contribution
\begin{align}
\mathcal{A}_6=-16\frac{(1+\tfrac{2D}{c})^2}{c^2L^4}\sum_{a}
&\sum_{\substack{i,j,k\\i,j,k\neq a\\i\neq j, \ i\neq k}}\frac{(\lambda_k-\lambda_a)^2}{(\lambda_j-\lambda_i)(\lambda_i-\lambda_k)} e^{ix' (\lambda_{k}-\lambda_{a})+it \left(\lambda_a^2-\lambda_{k}^2 \right)}\,.
\end{align}
Applying \eqref{simpledouble} with successive principal values
and then using \eqref{rhorhotilda} we find
\begin{align}
\mathcal{A}_6=&-\frac{8}{c^2}\int_{-\infty}^\infty \D{\lambda}\rho(\lambda)\int_{-\Lambda}^\Lambda \D{\mu} \rho(\mu)(\mu-\lambda)^2\tilde{\rho}(\mu)^2e^{it(\lambda^2-\mu^2)+ix'(\mu-\lambda)}\nn
&+\frac{8\pi^2}{c^2}\int_{-\infty}^\infty \D{\lambda}\rho(\lambda)\int_{-\Lambda}^\Lambda \D{\mu} (\mu-\lambda)^2\rho(\mu)^3e^{it(\lambda^2-\mu^2)+ix'(\mu-\lambda)}\nn
&-\frac{16}{c^2}\int_{-\infty}^\infty\int_{-\infty}^\infty (\lambda-\mu)^2\rho(\lambda)\left[\tfrac{\pi^2}{3}\rho(\mu)^3-\gamma_{-2}(\mu) \right]e^{it(\lambda^2-\mu^2)+ix'(\mu-\lambda)}\D{\lambda} \D{\mu}\nn
&+{\cal O}(c^{-3})+{\cal O}(L^{-1})\,.
\end{align}

%Under this form, one can take the thermodynamic limit $L\to\infty$ and express the result in terms of successive-principal-value triple integrals.  It yields
%\begin{equation}
%\begin{aligned}
%&\frac{8}{c^2}\int_{-\infty}^\infty (\lambda-\mu)^2\rho(\lambda)\left[\tfrac{\pi^2}{3}\frac{\rho(\mu)}{(2\pi)^2}-\gamma(\mu) \right]e^{it(\lambda^2-\mu^2)+ix'(\mu-\lambda)}d\lambda d\mu\\
%&+\frac{8}{c^2}\int_{-\infty}^\infty d\lambda\rho(\lambda)\int_{-\Lambda}^\Lambda d\mu \rho_h(\mu)(\mu-\lambda)^2\tilde{\rho}(\mu)^2e^{it(\lambda^2-\mu^2)+ix'(\mu-\lambda)}\\
%&-\frac{16}{c^2}\int_{-\infty}^\infty\int_{-\infty}^\infty (\lambda-\mu)^2\rho(\lambda)\left[\tfrac{\pi^2}{3}\rho(\mu)^3-\gamma(\mu) \right]e^{it(\lambda^2-\mu^2)+ix'(\mu-\lambda)}d\lambda d\mu\\
%&+O(\Lambda^{-1}L^0)+O(L^{-1})\,.
%\end{aligned}
%\end{equation}
%%%%%%%%%%%%%%%%%%%%%%%%%%%%%%%%%%%%%
\subsubsection{\sfix{Order $c^{-2}$: sixth contribution}}
%%%%%%%%%%%%%%%%%%%%%%%%%%%%%%%%%%%%%
Finally, the last term in \eqref{bigterm} gives rise to a contribution
\begin{align}
\mathcal{A}_7=-16\frac{(1+\tfrac{2D}{c})^2}{c^2L^4}&\sum_{a} \sum_{\substack{n\\ \forall k,\, \lambda_{a,n}\neq\lambda_k\\ |\lambda_{a,n}|<\Lambda}}\left(\frac{2\pi n}{\lp}\right)^2\sum_{\substack{i,j\\i\neq a\\j\neq a}}\frac{e^{ix' (\lambda_{a,n}-\lambda_{a})+it \left(\lambda_a^2-\lambda_{a,n}^2 \right)}}{(\lambda_i-\lambda_{a}-\frac{2\pi n}{\lp})(\lambda_j-\lambda_a)}\,.
\end{align}
By again decomposing the sum over $n$ as a sum over vacancies
minus a sum over particles we find
\begin{align}
\mathcal{A}_7=&-\frac{16}{c^2}\int_{-\infty}^\infty
\D{\lambda}\rho(\lambda)\int_{-\Lambda}^\Lambda \D{\mu}
\rho_h(\mu)(\mu-\lambda)^2\tilde{\rho}(\lambda)\tilde{\rho}(\mu)e^{it(\lambda^2-\mu^2)+ix'(\mu-\lambda)}\nn
&+{\cal O}(c^{-3})+{\cal O}(L^{-1})\,.
\end{align}
%

%%%%%%%%%%%%%%%%%%%%%%%%%%%%%%%%%%%%%%%%%%%%%%%%%%%%%%%%%%%%%
\subsubsection{\sfix{Result for the contribution of} one
    particle-hole excitations} 
%%%%%%%%%%%%%%%%%%%%%%%%%%%%%%%%%%%%%%%%%%%%%%%%%%%%%%%%%%%%%
We leave the remaining contributions to $\mathcal{C}^\Lambda_1(x,t)$
untouched, i.e. in sum form, since they will be cancelled by
contributions from two particle-hole excitations to the
  correlator. Our final result for $\mathcal{C}^\Lambda_1(x,t)$
is thus given by

\begin{align}
\label{cm2oneph}
&\mathcal{C}^\Lambda_1(x,t)=\Omega_1^\Lambda
+(1+\tfrac{2D}{c})^2\int_{-\infty}^\infty
\D{\lambda}\ \rho(\lambda)\int_{-\Lambda}^\Lambda \D{\mu}\ \rho_h(\mu)\Bigg[
1-\frac{4}{c}(\mu-\lambda)(\tilde{\rho}(\mu)-\tilde{\rho}(\lambda))\nn
&\qquad\qquad\qquad+\frac{8}{c^2}(\mu-\lambda)^2(\tilde{\rho}(\mu)-\tilde{\rho}(\lambda))^2-\frac{8\pi^2}{c^2}(\mu-\lambda)^2[\rho(\mu)]^2\Bigg]e^{it(\lambda^2-\mu^2)+ix'(\mu-\lambda)}\nn
%&-\frac{4(1+\tfrac{2D}{c})^2}{c}\int_{-\infty}^\infty d\lambda\ \rho(\lambda)\int_{-\Lambda}^\Lambda d\mu\ \rho_h(\mu)(\mu-\lambda)\big[\tilde{\rho}(\mu)-\tilde{\rho}(\lambda)\big] e^{it(\lambda^2-\mu^2)+ix'(\mu-\lambda)} \nn
%&+\frac{8}{c^2}\int_{-\infty}^\infty
%d\lambda\ \rho(\lambda)\int_{-\Lambda}^\Lambda d\mu\
%\rho_h(\mu)(\mu-\lambda)^2\big[\tilde{\rho}(\mu)-\tilde{\rho}(\lambda)\big]^2e^{it(\lambda^2-\mu^2)+ix'(\mu-\lambda)}\nn
%&-\frac{8\pi^2}{c^2}\int_{-\infty}^\infty
%d\lambda\ \rho(\lambda)\int_{-\Lambda}^\Lambda d\mu\ \rho_h(\mu)
%(\mu-\lambda)^2
%\big(\rho(\mu)\big)^2e^{it(\lambda^2-\mu^2)+ix'(\mu-\lambda)}\nn
&-\frac{8}{c^2}\int_{-\infty}^\infty \D{\lambda} \int_{-\Lambda}^\Lambda \D{\mu}(\lambda-\mu)^2\rho_h(\mu)\left[\frac{\pi^2}{3}[\rho(\lambda)]^3-\gamma_{-2}(\lambda) \right]e^{it(\lambda^2-\mu^2)+ix'(\mu-\lambda)}\nn
&+\frac{8}{c^2}\int_{-\infty}^\infty \D{\lambda}\int_{-\infty}^\infty \D{\mu}(\lambda-\mu)^2\rho(\lambda)\left[\frac{\rho(\mu)}{12}-\frac{2\pi^2}{3}[\rho(\mu)]^3+\gamma_{-2}(\mu) \right]e^{it(\lambda^2-\mu^2)+ix'(\mu-\lambda)}\nn
&+{\cal O}(c^{-3})+{\cal O}(L^{-1})\, ,
\end{align}
where we have defined
\begin{align}
\label{omega1l}  
\Omega_1^\Lambda&= -\frac{1}{c^2L^2}\sum_{a}\sum_{\substack{\forall k,\,\lambda_{a,n}\neq\lambda_k\\ |\lambda_{a,n}|<\Lambda}}
\bigg[ \frac{1}{3}\sum_{\substack{i\\ i\neq a}}1
  -\frac{4}{L^2}\sum_{\substack{i,j\\i\neq j\\ j\neq
      a}}\frac{1}{(\lambda_i-\lambda_j)^2}-\frac{4}{L^2}\sum_{\substack{i\\ i\neq
      a}}\frac{1}{(\lambda_i-\lambda_a-\frac{2\pi n}{\lp})^2}\bigg]\nn
&\qquad\qquad\qquad\qquad\qquad\times\Big(\frac{2\pi
  n}{\lp}\Big)^2\ e^{it[\lambda_a^2-\lan^2]+ix' [\lan-\lambda_a]}\ .
\end{align}

%%%%%%%%%%%%%%%%%%%%%%%%%%%%%%%%%%%%%%%%%%%%%
\subsection{Two-particle-hole excitations}
%%%%%%%%%%%%%%%%%%%%%%%%%%%%%%%%%%%%%%%%%%%%%

%%%%%%%%%%%%%%%%%%%%%%%%%%%%%%%%%%%%%%%%%%%%%
\subsubsection{A partial fraction decomposition}
%%%%%%%%%%%%%%%%%%%%%%%%%%%%%%%%%%%%%%%%%%%%%
The computation of $\mathcal{C}^\Lambda_{2}(x,t)$ defined in
\eqref{s2} is slightly different. In order to proceed we decompose 
the form factor into partial fractions with respect to $n$, and then
$m$: 
%\begin{align}
%&\frac{(n+m)^4}{n^2m^2}\frac{(\lambda_a-\lambda_b)^2(\lan-\lbm)^2}{(\lan-\lambda_b)^2(\la%mbda_a-\lbm)^2}=\nn
%&\bigg[\frac{2\pi n}{L\gam}\bigg]^2\left[\frac{1}{(\tfrac{2\pi m}{L\gam})^2}+\frac{2}{(\l%ambda_a-\lambda_b)\frac{2\pi m}{L\gam}}+\frac{1}{(\lambda_a-\lambda_{b,m})^2}+\frac{2}{(\%lambda_a-\lambda_b)(\lambda_a-\lambda_{b,m})} \right]\nn
%&+\frac{2\pi n}{L\gam}\left[\frac{2}{\frac{2\pi m}{L\gam}}+\frac{2(\lambda_a-\lambda_b)}{%(\lambda_a-\lbm)^2}+\frac{2}{\lambda_a-\lbm}\right]\nn
%&+\left[\frac{2(\lambda_a-\lambda_b)}{\frac{2\pi m}{L\gam}}+\frac{(\lambda_a-\lambda_b)^2%}{(\lambda_a-\lbm)^2}+\frac{2(\lambda_a-\lambda_b)}{\lambda_a-\lbm}\right]\nn
%&+\bigg[\frac{2\pi n}{L\gam}\bigg]^{-1}\left[-2(\lambda_a-\lambda_b)+2\frac{2\pi m}{L\gam%}-\frac{2\big(\frac{2\pi m}{L\gam}\big)^2}{\lambda_a-\lambda_b}+\frac{2(\lambda_a-\lambda%_b)^2}{\lambda_a-\lbm}\right]\nn
%&+\left(\frac{2\pi n}{L}\right)^{-2}\left(\frac{2\pi m}{L}\right)^{2}\nn
%&+\left(\lan-\lambda_b\right)^{-1}\left[2(\lambda_a-\lambda_b)-\frac{2(\lambda_a-\lambda_%b)^2}{\frac{2\pi m}{L\gam}}-2\frac{2\pi m}{L\gam}+\frac{2(\frac{2\pi m}{L\gam})^2}{\lambd%a_a-\lambda_b}\right]\nn
%&+\left(\lan-\lambda_b\right)^{-2}\left[(\lambda_a-\lambda_b)^2-2(\lambda_a-\lambda_b)\fr%ac{2\pi m}{L\gam}+\left(\frac{2\pi m}{L\gam}\right)^2\right]\,.
%\end{align}
\begin{align}
&\frac{(n+m)^4}{n^2m^2}\frac{(\lambda_a-\lambda_b)^2(\lambda_a-\lambda_b+\tfrac{2\pi(n-m)}{L\gam})^2}{(\lambda_a-\lambda_b+\tfrac{2\pi
    n}{L\gam})^2(\lambda_a-\lambda_b-\tfrac{2\pi m}{L\gam})^2}=\nn
&\left(\frac{2\pi n}{\lp}\right)^2\left[\frac{1}{(\tfrac{2\pi m}{\lp})^2}+\frac{2}{(\lambda_a-\lambda_b)\frac{2\pi m}{\lp}}+\frac{1}{(\lambda_a-\lambda_b-\frac{2\pi m}{\lp})^2}+\frac{2}{(\lambda_a-\lambda_b)(\lambda_a-\lambda_b-\frac{2\pi m}{\lp})} \right]\nn
&+\frac{2\pi n}{\lp}\left[\frac{2}{\frac{2\pi m}{\lp}}+\frac{2(\lambda_a-\lambda_b)}{(\lambda_a-\lambda_b-\frac{2\pi m}{\lp})^2}+\frac{2}{\lambda_a-\lambda_b-\frac{2\pi m}{\lp}}\right]\nn
&+\left[\frac{2(\lambda_a-\lambda_b)}{\frac{2\pi m}{\lp}}+\frac{(\lambda_a-\lambda_b)^2}{(\lambda_a-\lambda_b-\frac{2\pi m}{\lp})^2}+\frac{2(\lambda_a-\lambda_b)}{\lambda_a-\lambda_b-\frac{2\pi m}{\lp}}\right]\nn
&+\Big(\frac{2\pi n}{\lp}\Big)^{-1}\left[-2(\lambda_a-\lambda_b)+2\frac{2\pi m}{\lp}-\frac{2(\frac{2\pi m}{\lp})^2}{\lambda_a-\lambda_b}+\frac{2(\lambda_a-\lambda_b)^2}{\lambda_a-\lambda_b-\frac{2\pi m}{\lp}}\right]\nn
&+\Big(\frac{2\pi n}{\lp}\Big)^{-2}\Big(\frac{2\pi m}{\lp}\Big)^{2}\nn
&+\Big(\lambda_a-\lambda_b+\frac{2\pi n}{\lp}\Big)^{-1}\left[2(\lambda_a-\lambda_b)-\frac{2(\lambda_a-\lambda_b)^2}{\frac{2\pi m}{\lp}}-2\frac{2\pi m}{\lp}+\frac{2(\frac{2\pi m}{\lp})^2}{\lambda_a-\lambda_b}\right]\nn
&+\Big(\lambda_a-\lambda_b+\frac{2\pi n}{\lp}\Big)^{-2}\left[(\lambda_a-\lambda_b)^2-2(\lambda_a-\lambda_b)\frac{2\pi m}{\lp}+\left(\frac{2\pi m}{\lp}\right)^2\right]\,.
\end{align}
%Here, and throughout the following analysis, we use that as the
%form factors are already of order ${\cal O}(c^{-2})$.
 
We now use that the sum is invariant under the simultaneous
reparametrisations $n'=m-\tfrac{\lp(\lambda_a-\lambda_b)}{2\pi}$ and
$m'=n+\tfrac{\lp(\lambda_a-\lambda_b)}{2\pi}$ (which corresponds to
swapping the position of the two excited particles) to bring all the
poles into poles in $n$ or $m$, the only exceptions being
$[\tfrac{2\pi n}{\lp}(\lambda_a-\lambda_b-\frac{2\pi m}{\lp})]^{-1}$
and $[\tfrac{2\pi m}{\lp}(\lambda_a-\lambda_b+\frac{2\pi n}{\lp})]^{-1}$ which cannot be
transformed further. Next we use the 
invariance under swapping $n,m$ and $a,b$ simultaneously (which
corresponds to renaming dummy variables) to bring all the poles into poles in $m$
only, with the exception of $[\tfrac{2\pi m}{\lp}(\lambda_a-\lambda_b+\frac{2\pi n}{\lp})]^{-1}$. We obtain 

\begin{align}
\mathcal{C}^\Lambda_{2}(x,t)&=\frac{4}{c^2L^4}\sum_{a\neq b}
\sum_{\substack{n\\\forall i, \, \lan\neq\lambda_i\\ |\lambda_{a,n}|<\Lambda}}\
\sum_{\substack{m\\\forall j,\, \lbm \neq \lambda_j\\\lbm\neq \lan \\ |\lambda_{b,m}|<\Lambda}}
%e^{-it\Big[\frac{2\pi n}{L\gam}\big(\frac{2\pi n}{L\gam}+2\lambda_a \big)+\frac{2\pi m}{L\gam}\big(\frac{2\pi m}{L\gam}+2\lambda_b \big)\Big]}\nn
%&\quad\times e^{ix' \frac{2\pi (n+m)}{L}}
e^{ix'[\lan-\lambda_a+\lbm-\lambda_b]+it\big[\lambda_a^2-\lambda_{a,n}^2+\lambda_b^2-\lambda_{b,m}^2\big]}\nn
&\times\biggl[\frac{n^2}{m^2}+2\frac{(\tfrac{2\pi
      n}{\lp})^2}{(\lambda_a-\lambda_b)\tfrac{2\pi
      m}{\lp}}+2\frac{n}{m}
+2\frac{\lambda_a-\lambda_b}{\tfrac{2\pi
    m}{\lp}}-\frac{(\lambda_a-\lambda_b)^2}{\tfrac{2\pi
    m}{\lp}(\lambda_a-\lambda_b+\frac{2\pi n}{\lp})}\biggr].
\end{align}

%This expression is supposed to match exactly \eqref{s2} in finite
%$L$; we can check numerically that the two sums are indeed exactly
%equal.
%\bl{We recall that in the exponential factor we retain terms to order
%${\cal O}(c^{-2})$ and concomitantly require the Bethe roots at that
%  order.}
We now carry out the sums over $m$ and $b$ in order to bring this
to a form similar to the contribution from one particle hole excitations. We will
denote the resulting five terms by $\Sigma_i$ for $i=1,\dots,5$ and
treat them one at a time.
%%%%%%%%%%%%%%%%%%%%%%%%%%%%%%%%%%%%%%%%%%%%%
\subsubsection{\sfix{First term $\Sigma_1$}}
\label{ssec:sigma1}
%%%%%%%%%%%%%%%%%%%%%%%%%%%%%%%%%%%%%%%%%%%%%
In this subsection we take the thermodynamic limit of
\begin{align}
&\Sigma_1=\frac{4}{c^2L^4}\sum_{a\neq b}\sum_{\substack{n\\\forall k,
      \, \lan\neq\lambda_k\\ |\lambda_{a,n}|<\Lambda}}\
\sum_{\substack{m\\\forall i, \, \lbm
    \neq\lambda_i\\\lbm\neq\lan\\ |\lambda_{b,m}|<\Lambda}}\
\frac{n^2}{m^2}\ 
e^{it\left[\lambda_a^2-\lambda_{a,n}^2+\lambda_b^2-\lambda_{b,m}^2\right]+ix'
[\lan-\lambda_a+\lbm-\lambda_b]}\ .
\end{align}
We begin by splitting the exponential factor
\begin{align}
e^{it\left[\lambda_b^2-\lambda_{b,m}^2\right]+ix'[\lbm-\lambda_b]}=&
\left(e^{it\left[\lambda_b^2-\lambda_{b,m}^2\right]+ix'[\lbm-\lambda_b]}-1 \right)+1.
\label{splitexp}
\end{align}
Performing the sum over $m$ for the second term in
  \fr{splitexp} gives 
\begin{align}
\Sigma_1&=\tilde{\Sigma}_1+\frac{1}{c^2L^2}\sum_{a}
\sum_{\substack{n\\\forall k, \, \lan \neq
    \lambda_k\\|\lambda_{a,n}|<\Lambda}}\Big(\frac{2\pi n}{\lp}\Big)^2
e^{it[\lambda_a^2-\lambda_{a,n}^2]+i x' [\lan-\lambda_a]}\nn
&\qquad\qquad\times \bigg[ \frac{1}{3}\sum_{\substack{i\\i\neq a}}1
-\frac{4}{L^2}\Big\{\sum_{\substack{i,j\\i\neq j,\ j\neq
    a\\|\lambda_j|<\Lambda}}\frac{1}{(\lambda_i-\lambda_j)^2}
+\sum_{\substack{j\\j\neq a}}\frac{1}{(\lambda_a-\lambda_j+\frac{2\pi n}{\lp})^2}\Big\}\bigg],
\end{align}
where
\be
\tilde{\Sigma}_1=\frac{4}{c^2L^4}\sum_{a\neq b}\sum_{\substack{n\\\forall k,
      \, \lan\neq\lambda_k\\ |\lambda_{a,n}|<\Lambda}}\
\sum_{\substack{m\\\forall i, \, \lbm
    \neq\lambda_i\\\lbm\neq\lan\\ |\lambda_{b,m}|<\Lambda}}\
\frac{n^2}{m^2}\ e^{it\left[\lambda_a^2-\lambda_{a,n}^2\right]+ix'[\lan-\lambda_a]}
\bigg[e^{it\left[\lambda_b^2-\lambda_{b,m}^2\right]+ix'[\lbm-\lambda_b]}-1 \bigg].
\ee
The advantage of this representation is that the pole in $m$ is now
only of order $1$. Writing the sum over $m$ as a sum over $m\neq 0$
minus sums over particles one obtains 

\begin{align}
\tilde{\Sigma}_1&=\frac{4}{c^2L^4}\sum_{a\neq b}
\sum_{\substack{n\\ \forall k, \, \lan \neq
\lambda_k\\ |\lambda_{a,n}|<\Lambda}}\sum_{\substack{m\neq 0\\ |\lambda_{b,m}|<\Lambda}}
\frac{n^2}{m^2}  \Big[e^{it\left[\lambda_b^2-\lambda_{b,m}^2\right]+ix'[\lbm-\lambda_b]}-1 \Big]
e^{it\left[\lambda_a^2-\lambda_{a,n}^2\right]+ix'[\lan-\lambda_a]}\nn
&-\frac{4}{c^2L^4}\sum_{a\neq b}\sum_{\substack{n\\ \forall k, \, \lan
    \neq \lambda_k\\ |\lambda_{a,n}|<\Lambda}}\sum_{\substack{i\\i\neq
    b\\ |\lambda_i|<\Lambda}}\frac{\Big(\frac{2\pi n}{\lp}\Big)^2}{(\lambda_b-\lambda_i)^2}
\Big[e^{it\left[\lambda_b^2-\lambda_i^2\right]+ix'[\lambda_i-\lambda_b]}-1 \Big]
e^{it\left[\lambda_a^2-\lambda_{a,n}^2\right]+ix'[\lan-\lambda_a]}\nn
&-\frac{4}{c^2L^4}\sum_{a\neq b}\sum_{\substack{n\\ \forall k, \, \lan
    \neq \lambda_k\\ |\lambda_{a,n}|<\Lambda}}\frac{\Big(\frac{2\pi
    n}{\lp}\Big)^2}{(\lambda_b-\lambda_a-\frac{2\pi n}{\lp})^2}
\Big[e^{it\left[\lambda_b^2-\lambda_{a,n}^2\right]+ix'
    [\lambda_{a,n}-\lambda_b]}-1 \Big]\nn
&\qquad\qquad\qquad\qquad\qquad\times\ e^{it\left[\lambda_a^2-\lambda_{a,n}^2\right]+ix'[\lan-\lambda_a]}\ .
\end{align}

The first term is a sum over regularly spaced integers $m$ with
only a simple pole. In the thermodynamic limit it can therefore be
expressed in terms of a principal value integral with a constant density
$\tfrac{1+\frac{2D}{c}}{2\pi}$. The second term is of type
\eqref{sipdpp} and gives rise to an integral over the root
density $\rho(\lambda)$ in the thermodynamic limit. The last term
is negligible in $L$. We find

\begin{align}
\label{sigma1}
\Sigma_1=&\Omega^\Lambda_2+\frac{4A_{x',t}}{c^2}\int_{-\infty}^\infty \D{\lambda}
\rho(\lambda)\int_{-\Lambda}^\Lambda \D{\mu}
\rho_h(\mu)(\mu-\lambda)^2e^{it(\lambda^2-\mu^2)+ix'(\mu-\lambda)}\nn
&+{\cal  O}(\Lambda^{-1}L^0)+{\cal  O}(L^{-1})\,,
\end{align}
where we defined
\begin{equation}
\label{aintrep}
A_{x,t}=\int_{-\infty}^\infty \D{u} \dashint_{-\infty}^{\infty}\D{v}\frac{\rho(u)\rho_h(v)}{(u-v)^2}(e^{it(u^2-v^2)+ix(v-u)}-1)\,,
\end{equation}
and
\begin{align}
\label{omega2l}
\Omega_2^\Lambda=&
  \frac{1}{c^2L^2}\sum_{a}\sum_{\substack{n\\\forall k, \, \lan \neq \lambda_k\\|\lan|<\Lambda}}\left(\frac{2\pi n}{L}\right)^2e^{it[\lambda_a^2-\lan^2]+ix'[\lan-\lambda_a]}\nn
&\qquad\qquad\times \bigg[ \frac{1}{3}\sum_{\substack{i\\i\neq a}}1
-\frac{4}{\lp^2}\Big\{\sum_{\substack{i,j\\i\neq j,\ j\neq
    a\\|\lambda_j|<\Lambda}}\frac{1}{(\lambda_i-\lambda_j)^2}
+\sum_{\substack{j\\j\neq a}}\frac{1}{(\lambda_a-\lambda_j+\frac{2\pi
    n}{\lp})^2}\Big\}\bigg].
\end{align}

%%%%%%%%%%%%%%%%%%%%%%%%%%%%%%%%%%%%%%%%%%%%%
\subsubsection{\sfix{Second term $\Sigma_2$}}
\label{ssec:sigma2}
%%%%%%%%%%%%%%%%%%%%%%%%%%%%%%%%%%%%%%%%%%%%%
The next contribution is
\begin{align}
&\Sigma_2=\frac{4}{c^2L^4}\sum_{a\neq b}\suman \sumbm
\frac{2(\tfrac{2\pi n}{\lp})^2}{(\lambda_a-\lambda_b)\tfrac{2\pi m}{\lp}} \
e^{it\left[\lambda_a^2-\lambda_{a,n}^2+\lambda_b^2-\lambda_{b,m}^2\right]+ix'
[\lan-\lambda_a+\lbm-\lambda_b]}\,.
\end{align}
Writing the sum over $m$ again as sums over vacancies minus
particles we have

\begin{align}
&\Sigma_2=\frac{8}{c^2L^4}\sum_{a\neq b}\suman
\sum_{\substack{m\neq 0\\|\lambda_{b,m}|<\Lambda}}\frac{(\tfrac{2\pi
    n}{\lp})^2}{(\lambda_a-\lambda_b)\tfrac{2\pi
    m}{\lp}}e^{it\left[\lambda_a^2-\lambda_{a,n}^2+\lambda_b^2-\lambda_{b,m}^2\right]+ix'
[\lan-\lambda_a+\lbm-\lambda_b]}\nn
&-\frac{8}{c^2L^4}\sum_{a\neq b}\suman\
\sum_{\substack{i\\i\neq b\\|\lambda_i|<\Lambda}}\frac{(\tfrac{2\pi
    n}{\lp})^2}{(\lambda_a-\lambda_b)(\lambda_i-\lambda_b)}e^{it\left[\lambda_a^2-\lan^2+\lambda_b^2-\lambda_i^2\right]+ix'[\lan-\lambda_a+\lambda_i-\lambda_b]}\nn
&-\frac{8}{c^2L^4}\sum_{a\neq b}\suman
\frac{(\tfrac{2\pi n}{\lp})^2}{(\lambda_a-\lambda_b)(\tfrac{2\pi n}{\lp}+\lambda_a-\lambda_b)}e^{it\left[\lambda_a^2-2\lan^2+\lambda_b^2\right]+ix'[2\lan-\lambda_a-\lambda_b]}\,.
\label{sigma2_a}
\end{align}

In the  first two lines of \fr{sigma2_a} the sums over $n$
are regular. The first line involves only sums of the form
\eqref{sipdpp}, while the second line is of the form
\eqref{sipdpp3} and the thermodynamic limit can be worked out using
 \eqref{simpledouble}. The third term, after
splitting the sum over $n$ as sums over vacancies minus
particles is seen to be negligible in $L$. Hence we obtain 
\begin{align}
\Sigma_2=&\frac{8}{c^2}\int_{-\infty}^\infty \D{\lambda} \rho(\lambda)\int_{-\Lambda}^\Lambda \D{\mu} \rho_h(\mu)(\mu-\lambda)^2 B_{x',t}(\lambda)e^{it(\lambda^2-\mu^2)+ix'(\mu-\lambda)}\nn
&+\frac{8}{c^2}\int_{-\infty}^\infty \D{\lambda} \rho(\lambda)\int_{-\Lambda}^\Lambda \D{\mu} \rho_h(\mu)(\mu-\lambda)^2 \left[\frac{\pi^2}{3}\rho(\lambda)^3-\gamma_{-2}(\lambda) \right]e^{it(\lambda^2-\mu^2)+ix'(\mu-\lambda)}\nn
&+{\cal O}(\Lambda^{-1}L^0)+{\cal O}(L^{-1})\,,
\end{align}
where $B_{x,t}(\lambda)$ is defined in terms of  principal
value integrals by
\begin{equation}\label{bintrep}
B_{x,t}(\lambda)= \dashint_{-\infty}^{\infty} \D{u} \dashint_{-\infty}^{\infty} \D{v}
\frac{\rho(u)\rho_h(v)}{(v-u)(\lambda-u)}e^{it(u^2-v^2)+ix(v-u)}\, .
\end{equation}
%%%%%%%%%%%%%%%%%%%%%%%%%%%%%%%%%%%%%%%%%%%%%
\subsubsection{\sfix{Third term $\Sigma_3$}}
\label{ssec:sigma3}
%%%%%%%%%%%%%%%%%%%%%%%%%%%%%%%%%%%%%%%%%%%%%
In this subsection we take the thermodynamic limit of
\begin{align}
&\Sigma_3=\frac{4}{c^2L^4}\sum_{a\neq b}\suman\ \sumbm
2\frac{n}{m}\
e^{it\left[\lambda_a^2-\lambda_{a,n}^2+\lambda_b^2-\lambda_{b,m}^2\right]+ix'
[\lan-\lambda_a+\lbm-\lambda_b]}\ .
\end{align}
Expressing the sum over $m$ as the difference of sums over
vacancies and particles $\Sigma_3$ reduces to terms of the form
\eqref{sipdpp} that can be readily expressed as integrals over root densities.
We obtain
\begin{align}
\Sigma_3=&\frac{8C_{x',t}}{c^2}\int_{-\infty}^\infty \D{\lambda} \rho(\lambda)\int_{-\Lambda}^\Lambda \D{\mu} \rho_h(\mu)(\mu-\lambda)e^{it(\lambda^2-\mu^2)+ix'(\mu-\lambda)}
+{\cal O}(\Lambda^{-1}L^0)+{\cal O}(L^{-1})\,,
\end{align}
where we have defined
\begin{equation}
\label{cintrep}
C_{x,t}=\int_{-\infty}^\infty \D{u} \dashint_{-\infty}^{\infty} \D{v} \frac{\rho(u)\rho_h(v)}{v-u}e^{it(u^2-v^2)+ix(v-u)}\,.
\end{equation}

%%%%%%%%%%%%%%%%%%%%%%%%%%%%%%%%%%%%%%%%%%%%%
\subsubsection{\sfix{Fourth term $\Sigma_4$}}
\label{ssec:sigma4}
%%%%%%%%%%%%%%%%%%%%%%%%%%%%%%%%%%%%%%%%%%%%%
The next contribution is given by
\begin{align}
\Sigma_4&=\frac{4}{c^2L^4}\sum_{a\neq b}
\suman\ \sumbm
 2\frac{\lambda_a-\lambda_b}{\tfrac{2\pi m}{L'}}
e^{it\left[\lambda_a^2-\lambda_{a,n}^2+\lambda_b^2-\lambda_{b,m}^2\right]+ix'
[\lan-\lambda_a+\lbm-\lambda_b]}\ ,
\end{align}
and can be treated in complete analogy with $\Sigma_3$. We obtain
\be
\Sigma_4=\frac{8}{c^2}\int_{-\infty}^\infty \D{\lambda} \rho(\lambda)\int_{-\Lambda}^\Lambda \D{\mu} \rho_h(\mu)(\lambda C_{x',t}-D_{x',t})e^{it(\lambda^2-\mu^2)+ix'(\mu-\lambda)}
+{\cal O}(\Lambda^{-1}L^0)+{\cal O}(L^{-1})\,,
\ee
where we have defined
\begin{equation}
\label{dintrep}
D_{x,t}=\int_{-\infty}^\infty \D{u} \dashint_{-\infty}^{\infty}\D{v}\,  u\frac{\rho(u)\rho_h(v)}{v-u}e^{it(u^2-v^2)+ix(v-u)}\,.
\end{equation}

%%%%%%%%%%%%%%%%%%%%%%%%%%%%%%%%%%%%%%%%%%%%%
\subsubsection{\sfix{Fifth term $\Sigma_5$}}
\label{ssec:sigma5}
%%%%%%%%%%%%%%%%%%%%%%%%%%%%%%%%%%%%%%%%%%%%%
The final contribution to $\mathcal{C}_2^\Lambda(x,t)$ is
\begin{align}
\Sigma_5&=-\frac{4}{c^2L^4}\sum_{a\neq b}\suman\ \sumbm
\frac{(\lambda_a-\lambda_b)^2}{\tfrac{2\pi m}{\lp}(\lambda_a-\lambda_b+\tfrac{2\pi n}{\lp})}\nn
&\qquad\qquad\qquad\qquad\qquad\times
e^{it\left[\lambda_a^2-\lambda_{a,n}^2+\lambda_b^2-\lambda_{b,m}^2\right]+ix'
[\lan-\lambda_a+\lbm-\lambda_b]}\ .
\end{align}
In order to take the thermodynamic limit we rewrite this as
\begin{align}
\Sigma_5&=-\frac{4}{c^2L^4}\sum_{a\neq b}\suman
\sum_{\substack{m\neq 0\\|\lambda_{b,m}|<\Lambda}}
\frac{(\lambda_a-\lambda_b)^2\ e^{it\left[\lambda_a^2-\lambda_{a,n}^2+\lambda_b^2-\lambda_{b,m}^2\right]+ix'
  [\lan-\lambda_a+\lbm-\lambda_b]}}{\tfrac{2\pi
    m}{\lp}(\lambda_a-\lambda_b+\tfrac{2\pi n}{\lp})}\nn
&+\frac{4}{c^2L^4}\sum_{a\neq b}\sum_{\substack{n\\ \lambda_{a,n}\neq\lambda_b\\|\lambda_{a,n}|<\Lambda}}\sum_{\substack{i\\ i\neq
    b\\|\lambda_{i}|<\Lambda}}\frac{(\lambda_a-\lambda_b)^2\ e^{it\left[\lambda_a^2-\lan^2+\lambda_b^2-\lambda_i^2\right]+ix'[\lan-\lambda_a+\lambda_i-\lambda_b]}}{(\lambda_i-\lambda_b)(\lambda_a-\lambda_b+\tfrac{2\pi n}{\lp})}\nn
&-\frac{4}{c^2L^4}\sum_{a\neq b}\sum_{\substack{k\\k\neq
    b\\|\lambda_k|<\Lambda}}\sum_{\substack{i\\i\neq
    b\\|\lambda_i|<\Lambda}}\frac{(\lambda_a-\lambda_b)^2\ e^{it\left[\lambda_a^2-\lambda_k^2+\lambda_b^2-\lambda_i^2\right]+ix'[\lambda_i+\lambda_k-\lambda_a-\lambda_b]}}{(\lambda_i-\lambda_b)(\lambda_k-\lambda_b)}\nn
&+\frac{4}{c^2L^4}\sum_{a\neq b}\sum_{\substack{n\\\lan\neq
    \lambda_b\\|\lambda_{a,n}|<\Lambda}}
\frac{(\lambda_a-\lambda_b)^2}{(\lambda_a-\lambda_b+\tfrac{2\pi
    n}{\lp})^2}\ e^{it\left[\lambda_a^2-2\lan^2+\lambda_b^2\right]+2ix'[2\lan-\lambda_a-\lambda_b]}\nn
&-\frac{4}{c^2L^4}\sum_{a\neq b}\sum_{\substack{k\\k\neq
    b\\|\lambda_k|<\Lambda}}
\frac{(\lambda_a-\lambda_b)^2}{(\lambda_k-\lambda_b)^2}\ e^{it\left[\lambda_a^2-2\lambda_k^2+\lambda_b^2\right]+ix'[2\lambda_k-\lambda_a-\lambda_b]}\,.
\end{align}
The first two lines are of type \eqref{sipdpp} while the third
and fifth lines are of types \eqref{sipdpp3} and \eqref{sipd}
respectively. Finally, in the fourth line we use that $\sum_{n\neq
  0}\tfrac{1}{n^2}=\tfrac{\pi^2}{3}$ to arrive at
\begin{align}
\Sigma_5&=\frac{4}{c^2}\int_{-\infty}^\infty \D{\lambda}
\rho(\lambda)\int_{-\Lambda}^\Lambda \D{\mu}
\rho_h(\mu)\Big[(\mu-2\lambda)C_{x',t}+D_{x',t}-(\lambda-\mu)^2B_{x',t}(\mu)\Big]\nn
&\hskip 4cm\times e^{it(\lambda^2-\mu^2)+ix'(\mu-\lambda)}\nn
&+\frac{4}{c^2}\int_{-\infty}^\infty \D{\lambda}\rho(\lambda)\int_{-\infty}^\infty \D{\mu}(\lambda-\mu)^2\left[\frac{\pi^2}{3}\rho(\mu)^3+\frac{\pi^2}{3}\frac{\rho(\mu)}{(2\pi)^2}-2\gamma_{-2}(\mu) \right]e^{it(\lambda^2-\mu^2)+ix'(\mu-\lambda)}\nn
&+{\cal O}(\Lambda^{-1}L^0)+{\cal O}(L^{-1})\,.
\end{align}
%%%%%%%%%%%%%%%%%%%%%%%%%%%%%%%%%%%%%%%%%%%%%
\subsubsection{\sfix{Result for the contribution from two particle-hole
  excitations} }
%%%%%%%%%%%%%%%%%%%%%%%%%%%%%%%%%%%%%%%%%%%%%
Combining the results of sections \ref{ssec:sigma1}-\ref{ssec:sigma5}
we arrive at the following expression for the two particle-hole
contribution to the density-density correlation function
\begin{align}
&\mathcal{C}_{2}^\Lambda(x,t)=\Omega_2^\Lambda
%\frac{1}{c^2L^2}\sum_{a}\sum_{\substack{n\\\forall k, \, \lan \neq \lambda_k\\|\lan|<\Lam%bda}}\left(\frac{2\pi n}{L}\right)^2e^{it[\lambda_a^2-\lan^2]+ix'[\lan-\lambda_a]}\nn
%&\qquad\qquad\times \bigg[ \frac{1}{3}\sum_{\substack{i\\i\neq a}}1
%-\frac{4}{\lp^2}\Big\{\sum_{\substack{i,j\\i\neq j,\ j\neq
%    a\\|\lambda_j|<\Lambda}}\frac{1}{(\lambda_i-\lambda_j)^2}
%+\sum_{\substack{j\\j\neq a}}\frac{1}{(\lambda_a-\lambda_j+\frac{2\pi n}{\lp})^2}\Big\}\bigg],\nn
+\frac{4}{c^2}\int_{-\infty}^\infty \D{\lambda}  \rho(\lambda)\int_{-\Lambda}^\Lambda \D{\mu}
\rho_h(\mu)\Big[(\mu-\lambda)^2[A_{x',t}+2B_{x',t}(\lambda)-B_{x',t}(\mu)]\nn
&\hskip5cm    +(3\mu-2\lambda)C_{x',t}-D_{x',t} \Big]e^{it(\lambda^2-\mu^2)+ix'(\mu-\lambda)}\nn
&+\frac{8}{c^2}\int_{-\infty}^\infty \D{\lambda} \rho(\lambda)\int_{-\Lambda}^\Lambda \D{\mu} \rho_h(\mu)(\lambda-\mu)^2 \left[\frac{\pi^2}{3}\rho(\lambda)^3-\gamma_{-2}(\lambda) \right]e^{it(\lambda^2-\mu^2)+ix'(\mu-\lambda)}\nn
&+\frac{4}{c^2}\int_{-\infty}^\infty \D{\lambda}\rho(\lambda)\int_{-\infty}^\infty \D{\mu}(\lambda-\mu)^2\left[\frac{\pi^2}{3}\rho(\mu)^3+\frac{\rho(\mu)}{12}-2\gamma_{-2}(\mu) \right]e^{it(\lambda^2-\mu^2)+ix'(\mu-\lambda)}\nn
&+{\cal O}(\Lambda^{-1}L^0)+{\cal O}(L^{-1})\ ,
\end{align}
where $\Omega_2^\Lambda$ has been defined in \fr{omega2l}.

%%%%%%%%%%%%%%%%%%%%%%%%%%%%%%%%%%%%%%%%%%%%%%%%%%%%%%%%%%%%%%%
\subsection{\sfix{Density-density correlations in arbitrary macro
states for all $x$ and $t$ at order ${\cal O}(c^{-2})$}}
%%%%%%%%%%%%%%%%%%%%%%%%%%%%%%%%%%%%%%%%%%%%%%%%%%%%%%%%%%%%%%%

%%%%%%%%%%%%%%%%%%%%%%%%%%%%%%%%%%%%%%%%%%%%%%%%%%%%%%%%%%%%%%%
\subsubsection{Compensation of divergent parts}
%%%%%%%%%%%%%%%%%%%%%%%%%%%%%%%%%%%%%%%%%%%%%%%%%%%%%%%%%%%%%%%
As explained above, the ${\cal O}(c^{-2})$ contributions due to one-
and two particle-hole excitations are individually divergent in
  the thermodynamic limit. The divergent parts are given in
\fr{omega1l} and \fr{omega2l} respectively.
Their difference is

\be
\Omega_1^\Lambda-\Omega_2^\Lambda=\frac{4}{c^2 L^4}\sum_{a}\suman
\sum_{\substack{i,j\\i\neq j,\ j\neq a\\|\lambda_j|>\Lambda}}\frac{1}{(\lambda_i-\lambda_j)^2}\
e^{it[\lambda_a^2-\lan^2]+ix'[\lan-\lambda_a]}\,.
\ee
%
%and behaves as
%\bl{
%\begin{equation}
%\Omega_1^\Lambda-\Omega_2^\Lambda=C(x,t)\frac{L}{\Lambda^2 c^2}+{\cal
%  O}(L^0)+{\cal O}(c^{-3})\, .
%\end{equation}
%Here the factor $C(x,t)$} depends on $x,t$ and $\rho(\pm \Lambda)$
%but is of order ${\cal O}(\Lambda^0)$ and ${\cal
%  O}(L^0)$.
%{\color{red} I have rewritten the following in light of the changes we made to
%  section \ref{ssec:lehmann}.}
Crucially this vanishes for the class of root densities
we use in our $\Lambda$-regularization discussed in Section \ref{suminterm}, i.e.
$\rho(\lambda)=0$ for $|\lambda|>\Lambda$. Indeed, the second sum is
zero whenever all the roots satisfy $|\lambda_j|<\Lambda$. We conclude
that within our regularization scheme all divergences cancel at order ${\cal O}(c^{-2})$,
but they do so in a non-trivial fashion: divergent contributions from
intermediate states with one particle-hole excitation precisely cancel
those arising from intermediate states with two particle-hole excitations.

%%%%%%%%%%%%%%%%%%%%%%%%%%%%%%%%%%%%%%%%%%%%%%%%%%%%%%%%%%%%%%%
\subsubsection{Compensation of contributions that depend on the choice
of representative state}
%%%%%%%%%%%%%%%%%%%%%%%%%%%%%%%%%%%%%%%%%%%%%%%%%%%%%%%%%%%%%%%
As we have seen above, the contributions from both one- and two-particle-hole
excitations in the thermodynamic limit individually depend on the
choice of the representative state through the pair distribution function
$\gamma_{-2}(\lambda)$. Importantly, these contributions exactly
cancel one another and the full correlation function \textit{does
not} depend on the representative state. 

%%%%%%%%%%%%%%%%%%%%%%%%%%%%%%%%%%%%%%%%%%%%%%%%
\subsubsection{\sfix{$\Lambda$-regularized} correlation function}
%%%%%%%%%%%%%%%%%%%%%%%%%%%%%%%%%%%%%%%%%%%%%%%%
Combining the results for the one and two particle-hole excitations we
obtain the following result for the dynamical density-density
correlator in the $\Lambda$-regularization

\begin{align}
\label{corregen}
\langle \sigma(x,t)\sigma(0,0)\rangle_\Lambda=&D^2+\int_{-\infty}^\infty \D{\lambda} \int_{-\Lambda}^\Lambda  \D{\mu} f_{x',t}(\lambda,\mu)e^{it(\lambda^2-\mu^2)+ix'(\mu-\lambda)}\nn
&+{\cal O}(\Lambda^{-1}L^0)+{\cal O}(L^{-1})\,,
\end{align}
where the integrand is given by
\begin{equation}
f_{x,t}(\lambda,\mu)=\chi_{x,t}^{(1)}(\lambda,\mu)+\frac{\chi_{x,t}^{(2)}(\lambda,\mu)}{c^2}+{\cal
  O}(c^{-3})\ . 
\end{equation}
Here the contributions due to one and two particle-hole
excitations are respectively
%{\color{red} While I like boxed equations I don't think journals allow them.}

\begin{align}
\chi^{(1)}_{x,t}(\lambda,\mu)=&(1+\tfrac{2D}{c})^2\rho(\lambda)\rho_h(\mu)\bigg[1-\frac{4}{c}(\mu-\lambda)(\tilde{\rho}(\mu)-\tilde{\rho}(\lambda))+\frac{8}{c^2}(\mu-\lambda)^2(\tilde{\rho}(\mu)-\tilde{\rho}(\lambda))^2\nn
&\qquad\qquad\qquad+\frac{4\pi^2}{c^2}(\mu-\lambda)^2\rho(\mu)\rho_h(\mu)\bigg],\nn
\chi^{(2)}_{x,t}(\lambda,\mu)=&4\rho(\lambda)\rho_h(\mu)\Big[(\mu-\lambda)^2[A_{x,t}+2B_{x,t}(\lambda)-B_{x,t}(\mu)]+(3\mu-2\lambda)C_{x,t}-D_{x,t} \Big]\,.
\label{chi2}
\end{align}
The function $\tilde{\rho}(\lambda)$ is defined in \eqref{rhotilde}
and the four functions $A_{x,t}$, $B_{x,t}(\lambda)$, $C_{x,t}$ and
$D_{x,t}$ are given in \eqref{aintrep}, \eqref{bintrep},
\eqref{cintrep} and \eqref{dintrep} respectively.

Some comments on the term
$\frac{4\pi^2}{c^2}(\mu-\lambda)^2\rho(\lambda)\rho(\mu)\rho_h(\mu)^2$ 
are in order. This term arises from the sum of the
contributions involving the pair distribution function $\gamma_{-2}(\lambda)$
in both $\mathcal{C}^\Lambda_1(x,t)$ and $\mathcal{C}^\Lambda_2(x,t)$. Strictly
speaking it therefore involves two particle-hole excitations 
as well one particle-hole excitations. Since it does not involve
double integrals, as is the case for the other contributions from
$\mathcal{C}^\Lambda_2(x,t)$, we have chosen to include it entirely in
$\chi^{(1)}_{x,t}(\lambda,\mu)$. It can be interpreted as a ``dressing" of
contributions arising from one particle-hole excitations by
two particle-hole excitations.  
%%%%%%%%%%%%%%%%%%%%%%%%%%%%%%%%%%%%%%%%
\subsubsection{Dynamical correlations}
%%%%%%%%%%%%%%%%%%%%%%%%%%%%%%%%%%%%%%%%
The result \eqref{corregen} gives the thermodynamic limit of the
$\Lambda$-regularized correlation function. We now remove the
cutoff dependence by taking the limit $\Lambda\to\infty$. The
resulting ill-defined integrals \eqref{intdiverge} are to be
understood as distributions following \eqref{reg2}. To express the
limit $\Lambda\to\infty$ in terms of well-defined integrals we
consider the expansion of $f_{x,t}(\lambda,\mu)$ around $\mu\to\infty$  
\begin{equation}
f_{x,t}(\lambda,\mu)=\mu^2 \varphi^{(2)}_{x,t}(\lambda)+\mu
\varphi^{(1)}_{x,t}(\lambda)+\varphi^{(0)}_{x,t}(\lambda)+o(\mu^0)\ .
\end{equation}
%\bl{Here} the functions $\varphi^{(i)}_{x,t}(\lambda)$ in general depend on the
%behaviour of $\rho(\lambda)$ at infinity. 
Defining
\begin{equation}
\tilde{f}_{x,t}(\lambda,\mu)=\left[\left(\tfrac{x}{2t}\right)^2+\tfrac{1}{2it}-\mu^2\right]
\varphi^{(2)}_{x,t}(\lambda)+\left[\tfrac{x}{2t}-\mu\right]
\varphi^{(1)}_{x,t}(\lambda)+f_{x,t}(\lambda,\mu)\, ,
\end{equation}
it follows from \fr{reg2} that we can express the limit
$\Lambda\to\infty$ of \fr{corregen}
as a function of $x$ and $t\neq 0$
\begin{equation}
\label{corregent}
 \left\langle \sigma\left( x,t\right) \sigma \left( 0,0\right)
 \right\rangle =D^2+\int_{-\infty}^\infty\int_{-\infty}^\infty
 \tilde{f}_{x',t}(\lambda,\mu)e^{it(\lambda^2-\mu^2)+ix'(\mu-\lambda)}\D{\lambda}
 \D{\mu}+{\cal O}(L^{-1})\ .
\end{equation}
For the energy of a macro state to be well-defined we need
$\rho(\mu)=o(\mu^{-2})$ for $\mu\to\infty$. From this we have 

\begin{align}
\varphi^{(2)}_{x,t}(\lambda)&=4\frac{\gam^3}{2\pi c^2}\rho(\lambda)
\Big[2\tilde{\rho}(\lambda)^2+A_{x',t}+2B_{x',t}(\lambda) \Big]\ ,\nn
\varphi^{(1)}_{x,t}(\lambda)&=4\frac{\gam^3}{2\pi c}\rho(\lambda)
\bigg[\tilde{\rho}(\lambda)
-\frac{4\lambda\tilde{\rho}(\lambda)^2+4D
    \tilde{\rho}(\lambda)+2\lambda A_{x',t}
    +4\lambda B_{x',t}(\lambda)-2C_{x',t}}{c}
    \bigg],\nn
  %  +\frac{1}{c}\Big[-4\lambda\tilde{\rho}(\lambda)^2-4D
%    \tilde{\rho}(\lambda)-2\lambda A_{x',t}\nn
%&\hskip5cm-4\lambda B_{x',t}(\lambda)+2C_{x',t} \Big]\bigg],\nn
\varphi^{(0)}_{x,t}(\lambda)&=4\frac{\gam^3}{2\pi c}\rho(\lambda)
\bigg[-D-\lambda\tilde{\rho}(\lambda)+\frac{1}{c}\Big[2D^2+8D\lambda\tilde{\rho}(\lambda)+2\lambda^2\tilde{\rho}(\lambda)^2\nn
&\hskip4cm +\lambda^2A_{x',t}+2\lambda^2 B_{x',t}(\lambda)-2D_{x',t}-4\lambda C_{x',t}\Big] \bigg]\,.
\end{align}
Alternatively, one can also write, using \eqref{epsto0}
\begin{align}\label{corregenalt}
\langle \sigma(x,t)\sigma(0,0)\rangle=&D^2+\underset{\epsilon\to 0^+}{\lim}\,\int_{-\infty}^\infty  \int_{-\infty}^\infty f_{x',t}(\lambda,\mu)e^{it\lambda^2-i(t-i\epsilon)\mu^2+ix'(\mu-\lambda)}\D{\lambda}\D{\mu}+{\cal O}(L^{-1})\,.
\end{align}
%%%%%%%%%%%%%%%%%%%%%%%%%%%%%%%%%%%%
\subsubsection{Static correlations}
%%%%%%%%%%%%%%%%%%%%%%%%%%%%%%%%%%%%%
The result \eqref{corregent} is singular for $t\to 0$ since it behaves as
$\tfrac{1}{\sqrt{t}}e^{-x^2/t}$. However, in a distribution sense
we have $\tfrac{1}{\sqrt{t}}e^{-x^2/t} \to 0$ when $t\to 0$. Defining
\begin{equation}
\tilde{f}_{x,0}(\lambda,\mu)=\lim_{t\to 0}\left[
-\mu^2 \varphi^{(2)}_{x,t}(\lambda)-\mu\varphi^{(1)}_{x,t}(\lambda)-\varphi^{(0)}_{x,t}(\lambda)+f_{x,t}(\lambda,\mu)\right],
\end{equation}
we have the following representation of the static correlator as a
function of $x$ 
\begin{equation}
\label{corregenx}
 \left\langle \sigma\left( x,0\right) \sigma \left( 0,0\right)
 \right\rangle =D^2+\int_{-\infty}^\infty\int_{-\infty}^\infty
 \tilde{f}_{x',0}(\lambda,\mu)e^{ix'(\mu-\lambda)}\D{\lambda} \D{\mu}+
 {\cal O}(L^{-1})\,.
\end{equation}
Alternatively, one can also write
\begin{align}
\langle \sigma(x,0)\sigma(0,0)\rangle=&D^2+\underset{\epsilon\to 0^+}{\lim}\,\int_{-\infty}^\infty  \int_{-\infty}^\infty   f_{x',0}(\lambda,\mu)e^{-\epsilon\mu^2+ix'(\mu-\lambda)}\D{\lambda} \D{\mu}+{\cal O}(L^{-1})\,.
\end{align}

%%%%%%%%%%%%%%%%%%%%%%%%%%%%%%%%%%%%%%%%%%%%%%%%%%%%%%%%%%%%%%%%%%%
\subsection{\sfix{Dynamical structure factor in arbitrary macro states
    for all $\omega$, $q$ at order $c^{-2}$}}
%%%%%%%%%%%%%%%%%%%%%%%%%%%%%%%%%%%%%%%%%%%%%%%%%%%%%%%%%%%%%%%%%%%
Given the correlation function \fr{corregent} for all $x$ and $t$
we can determine the dynamical structure factor (DSF) $S(q,\omega)$ by
taking the Fourier transform 
\begin{equation}
\label{dsfdef}
S(q,\omega)=\int_{-\infty}^\infty\int_{-\infty}^\infty  \left[\underset{L\to\infty}{\lim}\,\langle\sigma(x,t) \sigma(0,0)\rangle\right] e^{i\omega t-iqx}\D{x}\D{t}\,.
\end{equation}
It is convenient to decompose $S(q,\omega)$ in terms of the
contributions of one and two particle-hole excitations, which we
denote by $S^{(1)}(q,\omega)$ and $S^{(2)}(q,\omega)$ respectively:
\begin{equation}
S(q,\omega)=D^2\delta(q)\delta(\omega)+S^{(1)}(q,\omega)+S^{(2)}(q,\omega)+{\cal
  O}(c^{-3})\,.
\end{equation}
In practice we determine the dynamical structure factor by first
computing the Fourier transform \eqref{dsfdef} of the
$\Lambda$-regularized correlator \eqref{corregen}
%, denoted $S^{(1,2)}_\Lambda(q,\omega)$
and then taking the limit $\Lambda\to\infty$, which
turns out to be straightforward.
%%%%%%%%%%%%%%%%%%%%%%%%%%%%%%%%%%%%%%%%%%%%%%%%%%%%%%%%%%%%%%%%%%%%%%%%%%%%%%%%%
\subsubsection{\sfix{One particle-hole contributions to the dynamical
  structure factor}}
%%%%%%%%%%%%%%%%%%%%%%%%%%%%%%%%%%%%%%%%%%%%%%%%%%%%%%%%%%%%%%%%%%%%%%%%%%%%%%%%%
The contribution of the one particle-hole excitations to the DSF
$S^{(1)}(q,\omega)$ is obtained from the relation
\begin{align}
\label{eqbasefourier}
&\int_{-\infty}^\infty \D{t} e^{i\omega t} \int_{-\infty}^\infty \D{x} e^{-iqx}\int_{-\infty}^\infty \D{\lambda} e^{it\lambda^2-ix\lambda} \int_{-\Lambda}^\Lambda \D{\mu} e^{-it\mu^2+ix\mu} f(\lambda,\mu)=\nn
&\hskip5cm
\frac{2\pi^2}{|q|}f(\tfrac{\omega-q^2}{2q},\tfrac{\omega+q^2}{2q}
)\1_{|\tfrac{\omega+q^2}{2q}|<\Lambda}\ .
%1_{|\tfrac{\omega+q^2}{2q}|<\Lambda}\,, 
\end{align}
%where $\theta_H(x)$ is the Heaviside theta function. 
The
$\Lambda\to\infty$ limit of \fr{eqbasefourier} is straightforward and
yields at order ${\cal O}(c^{-2})$ 

\begin{align}
\label{DSF1ph}
S^{(1)}(q,\omega)=&2\pi^2\big(1+\frac{2D}{c}\big)\rho(\tfrac{\omega'-q^{\prime
    2}}{2q'})\rho_h(\tfrac{\omega'+q^{\prime 2}}{2q'})
\bigg[
  \frac{1}{|q'|}-\frac{4\sign(q')}{c}(\tilde{\rho}(\tfrac{\omega'+q^{\prime
      2}}{2q'})-\tilde{\rho}(\tfrac{\omega'-q^{\prime 2}}{2q'}))\nn
&\hskip1cm+\frac{8|q'|}{c^2}(\tilde{\rho}(\tfrac{\omega'+q^{\prime 2}}{2q'})-\tilde{\rho}(\tfrac{\omega'-q^{\prime 2}}{2q'}))^2+\frac{4\pi^2|q'|}{c^2}\rho(\tfrac{\omega'+q^{\prime 2}}{2q'})\rho_h(\tfrac{\omega'+q^{\prime 2}}{2q'})\bigg]\,,
\end{align}
where we have defined
\begin{equation}\label{opqp}
q'=\frac{q}{1+\frac{2D}{c}}\,,\qquad\qquad \omega'=\omega-\frac{4\delta}{c\gam}q\,.
\end{equation}

%%%%%%%%%%%%%%%%%%%%%%%%%%%%%%%%%%%%%%%%%%%%%%%%%%%%%%%%%%%%%%%%%%%%%%%%%%%%%%%%%
\subsubsection{\sfix{Two particle-hole contributions to the dynamical
  structure factor}}
%%%%%%%%%%%%%%%%%%%%%%%%%%%%%%%%%%%%%%%%%%%%%%%%%%%%%%%%%%%%%%%%%%%%%%%%%%%%%%%%%

The two particle-hole contributions involve the functions
$A_{x,t}$, $B_{x,t}(\lambda)$, $C_{x,t}$ and $D_{x,t}$ given in
\eqref{aintrep}, \eqref{bintrep}, \eqref{cintrep} and
\eqref{dintrep} respectively. Their simple dependence on $x$ and $t$
allows for a straightforward computation of their contribution to the
DSF. For example, by first integrating over $x$, then over $v$, then over $t$
and finally over $u$ we find
\begin{align}
&\int_{-\infty}^\infty \D{t} \int_{-\infty}^\infty \D{x} e^{i\omega t-iqx} \int_{-\infty}^\infty \D{\lambda} \int_{-\Lambda}^\Lambda \D{\mu} \rho(\lambda)\rho_h(\mu)(3\mu-2\lambda)C_{x',t}e^{it(\lambda^2-\mu^2)+ix'(\mu-\lambda)}\nn
&=2\pi^2\int_{-\infty}^\infty \D{\lambda} \int_{-\Lambda}^\Lambda \D{\mu}\ \rho(\lambda)\rho_h(\mu)\rho (\bar{\lambda}) \rho_h(\bar{\mu})\frac{3\mu-2\lambda}{(q'+\lambda-\mu)|q'+\lambda-\mu|}\,.
\label{intid1}
\end{align}
Here we have set
\begin{equation}
\bar{\lambda}=\frac{\omega'+\lambda^2-\mu^2-(q'+\lambda-\mu)^2}{2(q'+\lambda-\mu)}\,,\qquad \bar{\mu}=\frac{\omega'+\lambda^2-\mu^2+(q'+\lambda-\mu)^2}{2(q'+\lambda-\mu)}\,.
\end{equation}
The limit $\Lambda\to\infty$ of this expression is again routine. It
is however not immediately obvious that the double integral over
$\lambda$ and $\mu$ in \fr{intid1} is well-defined, since one of
the factors in the integrand exhibits a non-integrable
singularity. A closer inspection reveals that this
singularity is cancelled by the product of root densities.
We will show below by means of a change of variable that the
double integral is indeed well-defined.

All other terms involving the functions $B_{x,t}(\lambda)$,
$B_{x,t}(\mu)$ and $D_{x,t}$ can be computed analogously. The term
involving $A_{x,t}$ however requires a slightly modified approach,
since following through the same steps as before would split the
$1/v^2$ term into a sum of two quantities that are individually
divergent. In order to circumvent this problem we replace the
$1/v^2$ by $\tfrac{1}{v^2+\epsilon^2}$ and send $\epsilon\to 0$ in the
final result.  We obtain  
\begin{align}
&\int_{-\infty}^\infty \D{t} \int_{-\infty}^\infty \D{x} e^{i\omega t-iqx} \int_{-\infty}^\infty \D{\lambda} \int_{-\Lambda}^\Lambda \D{\mu} \rho(\lambda)\rho_h(\mu)(\lambda-\mu)^2A_{x',t}e^{it(\lambda^2-\mu^2)+ix'(\mu-\lambda)}\\
&\hskip1.5cm=2\pi^2\int_{-\infty}^\infty \D{\lambda} \int_{-\Lambda}^\Lambda \D{\mu}\frac{1}{|q'+\lambda-\mu|^3}\bigg[ \rho(\lambda)\rho_h(\mu)\rho(\bar{\lambda})\rho_h(\bar{\mu})(\lambda-\mu)^2\nn
&\hskip5.5cm-|q'(\lambda-\mu)| \rho(\bar{\lambda})\rho_h(\bar{\mu})\rho(\tfrac{\omega'-q^{\prime 2}}{2q'})\rho_h(\tfrac{\omega+q^{\prime 2}}{2q'})\bigg] \,.
\end{align}
Putting everything together we obtain the following result for
the contribution of two particle-hole excitations to the DSF 
\begin{align}
\label{DSF2ph}
S^{(2)}(q,\omega)=&
\frac{8\pi^2}{c^2}\int_{-\infty}^\infty \int_{-\infty}^\infty \rho(\lambda)\rho_h(\mu)\rho(\bar{\lambda})\rho_h(\bar{\mu})\frac{\frac{2(\lambda-\mu)^2}{\lambda-\bar{\lambda}}-\frac{(\lambda-\mu)^2}{\mu-\bar{\lambda}}+3\mu-2\lambda-\bar{\lambda}}{(q'+\lambda-\mu)|q'+\lambda-\mu|}\D{\lambda} \D{\mu}\nn
&+\frac{8\pi^2}{c^2}\int_{-\infty}^\infty \int_{-\infty}^\infty \frac{1}{|q'+\lambda-\mu|^3}\Big[ \rho(\lambda)\rho_h(\mu)\rho(\bar{\lambda})\rho_h(\bar{\mu})(\lambda-\mu)^2\nn
&\qquad\qquad\qquad\qquad\qquad-|q'(\lambda-\mu)| \rho(\bar{\lambda})\rho_h(\bar{\mu})\rho(\tfrac{\omega'-q^{\prime 2}}{2q'})\rho_h(\tfrac{\omega'+q^{\prime 2}}{2q'})\Big] \D{\lambda} \D{\mu}\nn
%&+\frac{8\pi^2}{c^2}\int_{-\infty}^\infty \int_{-\infty}^\infty \rho(\lambda')(\tfrac{1}{2\pi}-\rho(\mu'))\frac{\rho(\lambda)(\tfrac{1}{2\pi}-\rho(\mu))(\lambda-\mu)^2-|q(\lambda-\mu)|\rho(\tfrac{\omega-q^2}{2q})(\tfrac{1}{2\pi}-\rho(\tfrac{\omega+q^2}{2q}))}{|q+\lambda-\mu|^3} d\lambda d\mu\\
&+{\cal O}(c^{-3})\,.
\end{align}
In order to make the convergence of this integral explicit we
perform a change of variables from $\lambda,\mu$ to
\be
z=1+\tfrac{\lambda-\mu}{q}\ ,\qquad
p=\tfrac{\omega'-q'(\lambda+\mu)}{2q'z}\ ,
\ee
and define
\begin{align}
q_1&=\tfrac{\omega'-2q'zp-q^{\prime 2}(1-z)}{2q'}\ ,\qquad
q_2=\tfrac{\omega'-2q'zp+q^{\prime 2}(1-z)}{2q'}\ ,\nn
q_3&=\tfrac{\omega'+2q'p(1-z)-q^{\prime 2}z}{2q'}\ ,\qquad
q_4=\tfrac{\omega'+2q'p(1-z)+q^{\prime 2}z}{2q'}\ .
\end{align}
In terms of the new variables we have

\begin{align}
\label{DSF2ph2}
S^{(2)}(q,\omega)=&\frac{8\pi^2}{c^2}\int_{-\infty}^\infty
\int_{-\infty}^\infty h(q,\omega,z,p)\D{z}\D{p}+{\cal O}(c^{-3})\ ,\nn
h(q,\omega,z,p)=&
\frac{2}{q'z} \rho(q_1)\rho_h(q_2)\rho(q_3)\rho_h(q_4)\Bigg[\frac{5q'}{4}-q'z-\frac{p}{2}+\frac{2q^{\prime 2}(1-z)^2}{(2z-1)q'-2p}-\frac{(1-z)^2q^{\prime 2}}{q'-2p} \Bigg]\nn
+&\frac{\rho(q_3)\rho_h(q_4)}{z^2} \Big[(1-z)^2 \rho(q_1)\rho_h(q_2)-|1-z|\rho\big(\tfrac{\omega'-q^{\prime 2}}{2q'}\big)\rho_h\big(\tfrac{\omega'+q^{\prime 2}}{2q'}\big)\Big].
\end{align}

The integral over $p$ only has singularities that are integrable in a
principal value sense, and after the integral over $p$ has been
carried out the integral over $z$ only has a singularity that
is integrable in a principal value sense.
%As for the second integral,
%the pole is actually only a pole in $z$ since the numerator vanishes
%at $z=0$, which is again integrable in a principal value sense. 
We conclude that \fr{DSF2ph2} is well defined and can be
straightforwardly evaluated numerically.

%%%%%%%%%%%%%%%%%%%%%%%%%%%%%%%%%%%%%%%%%%%%%%%%%%%%%%%%%%%
\section{\sfix{Numerical evaluation of the dynamical structure factor}}
\label{sec5bis}
%%%%%%%%%%%%%%%%%%%%%%%%%%%%%%%%%%%%%%%%%%%%%%%%%%%%%%%%%%%
In this section we numerically evaluate the integral
representations \fr{DSF1ph}, \fr{DSF2ph2} in order to determine
$S(q,\omega)$ for the two examples of root densities introduced in
Section \ref{exrootg}, namely thermal states and the non-equilibrium
steady state after a quantum quench from the ground state at $c=0$.

%%%%%%%%%%%%%%%%%%%%%%%%%%%%%%%%%%%%%%%%%%%
\subsection{\sfix{Zero temperature}}
\begin{figure}[ht]
\begin{center}
\includegraphics[scale=0.18]{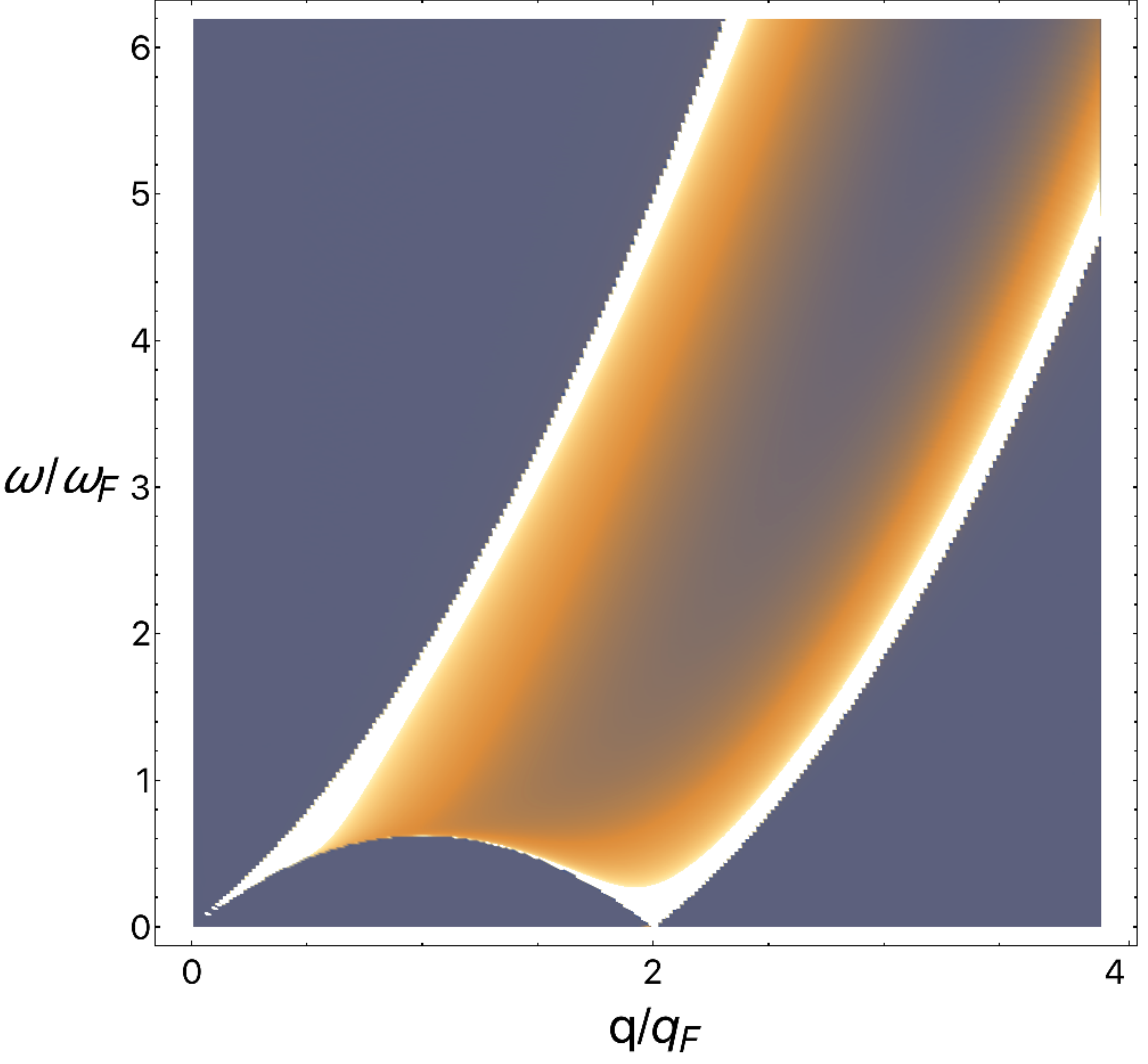}
\qquad
\includegraphics[scale=0.18]{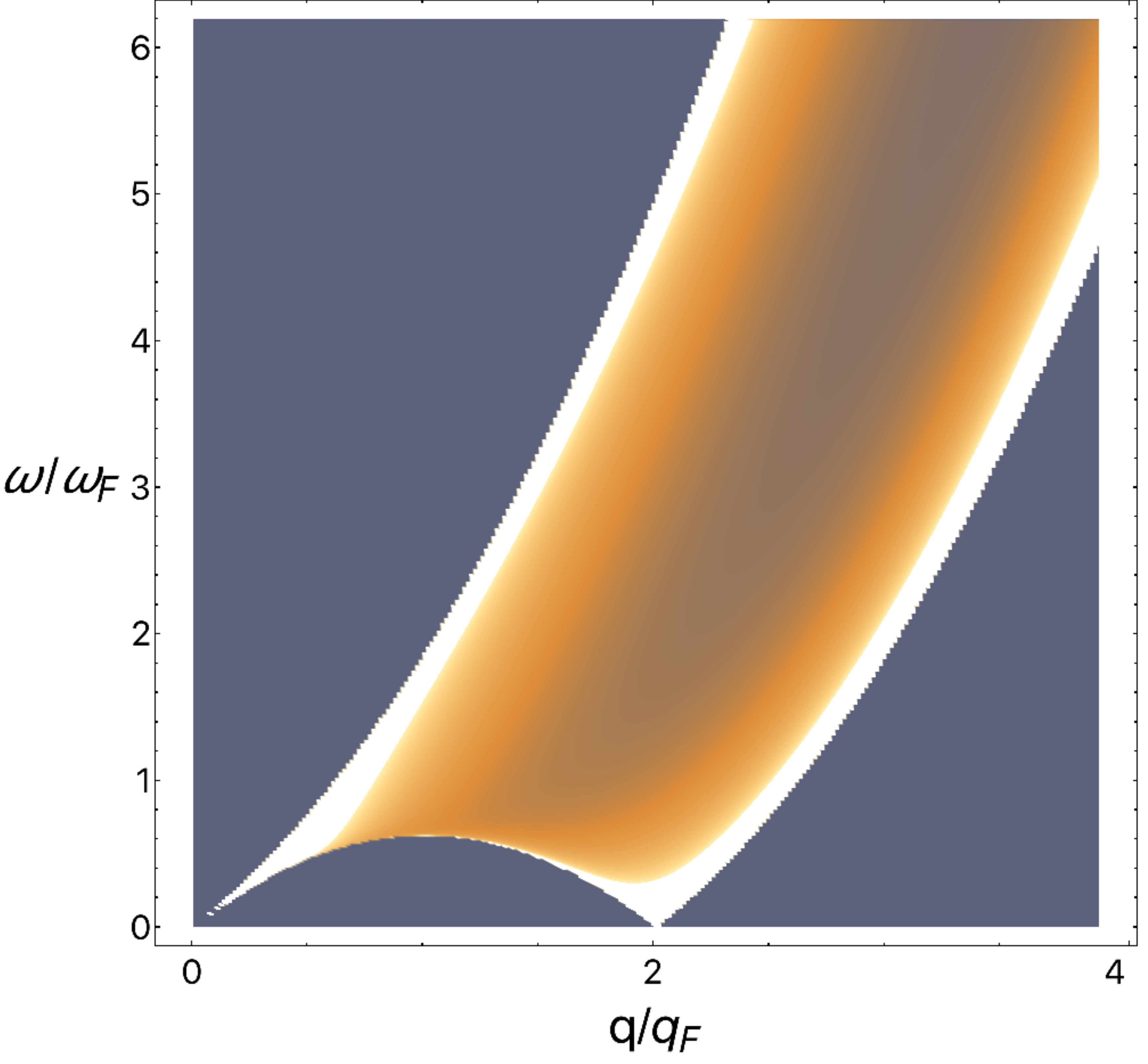}
\end{center}
\caption{$S(q,\omega)$ (left panel) and $S^{(1)}(q,\omega)$ (right
    panel) as functions of $q$ and $\omega$ at zero
    temperature and $D=0.404$, $c=3$ in \fr{rho0t}. The color scale is
  the same for both plots.} 
\label{zerocolor}
\end{figure}

%%%%%%%%%%%%%%%%%%%%%%%%%%%%%%%%%%%%%%%%%%%
We first consider the zero temperature case for density $D=0.404$ and $c=3$ in
\eqref{rho0t}. The value $D/c\approx 0.13$ is well
within the expected range of validity of the $1/c$-expansion. The same
holds true for all other cases considered below.
In Figure~\ref{zerocolor} we present numerical results for the DSF
at order $c^{-2}$ as well as for the one particle-hole
contribution $S^{(1)}(q,\omega)$. It is well known
that at zero temperature the one particle-hole contribution to the
DSF is non-zero only in a certain region of the $q,\omega$ plane
for kinematic reasons, and exhibits (not necessarily divergent) singularities at the
edges of its support \cite{IG08,ISG12,kitanineetcformfactor,kozlowski1}. We note that although the DSF is expected to diverge near the upper threshold, the divergence near the lower thresholds is a consequence of the $1/c$ expansion, that produces logarithms instead of a finite behaviour with a fractional $c$-dependent exponent.
Comparing the full result (left panel) to $S^{(1)}(q,\omega)$ (right panel)
we observe that the contributions due to two particle-hole
excitations significantly modify the numerical values of the DSF
within this region. $S^{(2)}(q,\omega)$ is also non-zero outside the
region, but this effect is barely visible in the plot. 
%%%%%%%%%%%%%%%%%%%%%%%%%%%%%%%
\subsection{\sfix{Finite temperature}}
%%%%%%%%%%%%%%%%%%%%%%%%%%%%%%%
We next turn to the DSF at finite temperatures.
\begin{figure}[ht!]
\begin{center}
\includegraphics[scale=0.18]{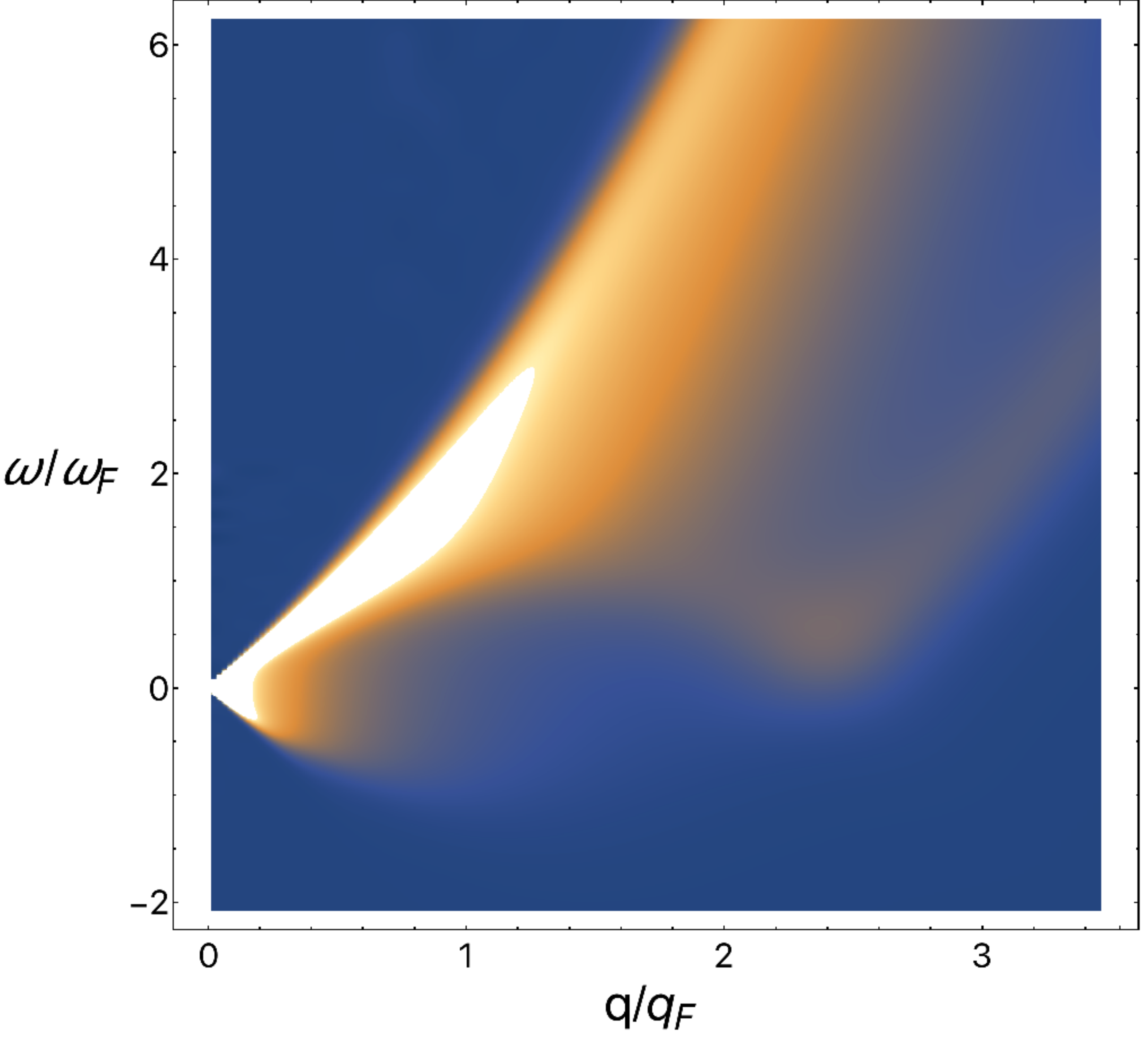}
\qquad
\includegraphics[scale=0.18]{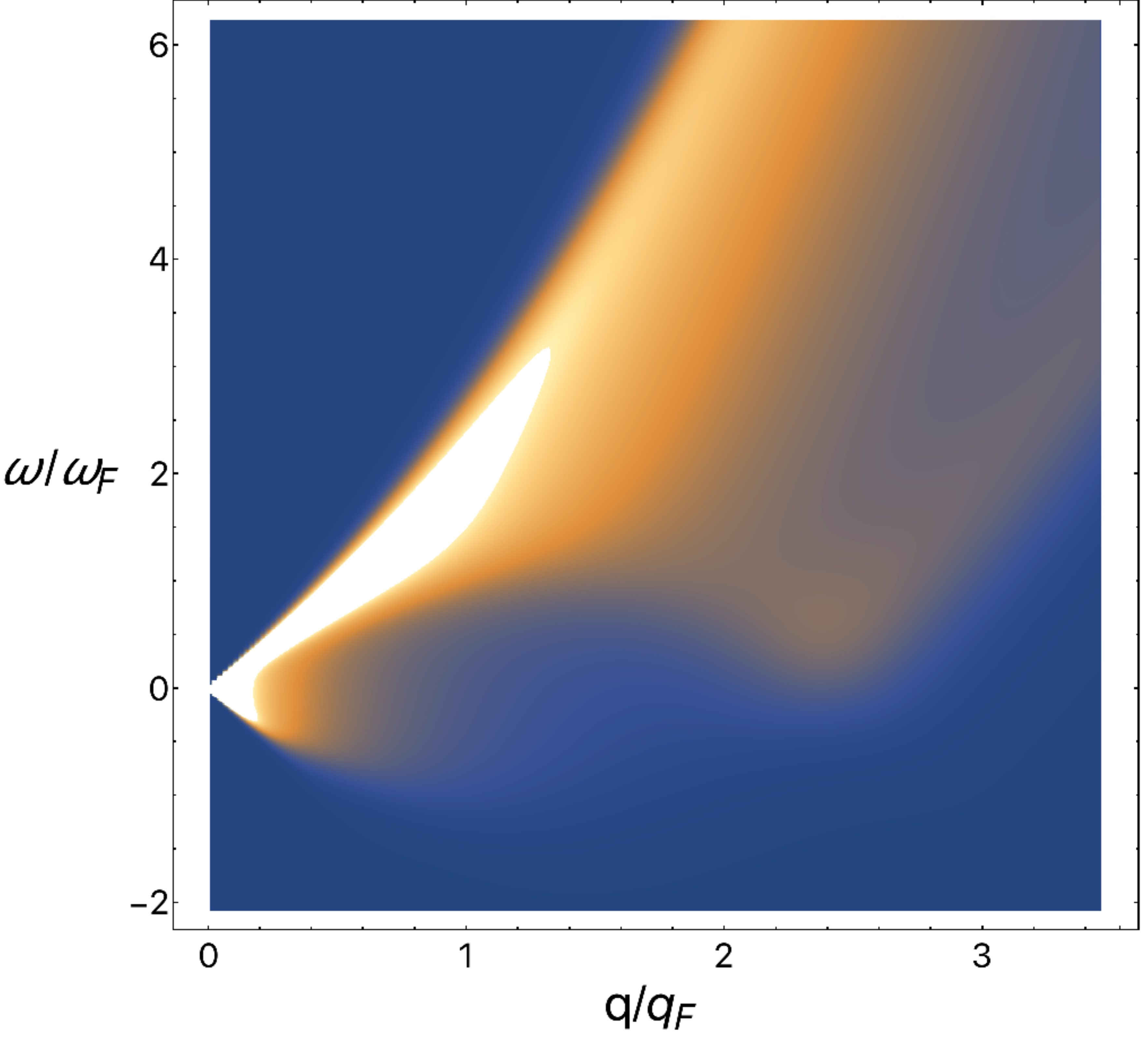}
\end{center}
\caption{$S(q,\omega)$ (left panel) and $S^{(1)}(q,\omega)$ (right panel)
as functions of $q$ and $\omega$ for a thermal state with $\beta=5$,
$D=0.396$ and $c=4$. The color scale is the same for both plots.}
\label{t5color}
\end{figure}
Figure~\ref{t5color} presents numerical results for the full DSF
$S(q,\omega)$ at order $c^{-2}$ for thermal states with
$\beta=5$, $c=4$ and $D=0.396$. For comparison we also plot
the one particle-hole contribution $S^{(1)}(q,\omega)$.
Like in zero temperature case, for these parameter values the one
particle-hole contribution already gives a fairly good account of the
full DSF. The two-particle-hole contribution modifies some details
that become increasingly significant for $q>2q_F$. 
The main difference to the zero temperature case is the emergence
of spectral weight at negative frequencies and the ``washing out'' of the
threshold singularities.

\begin{figure}[ht]
\begin{center}
\includegraphics[scale=0.18]{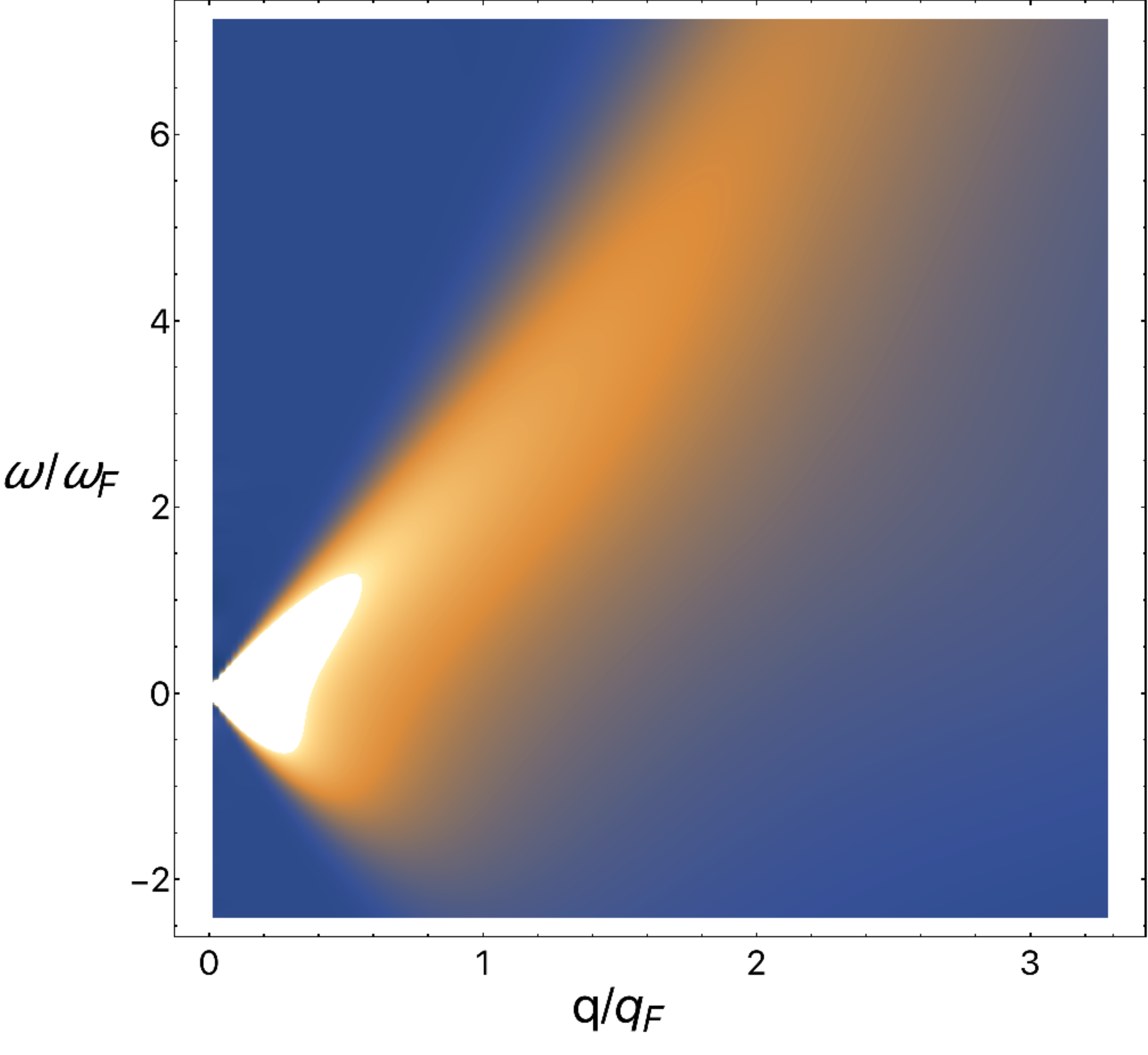}\qquad
\includegraphics[scale=0.18]{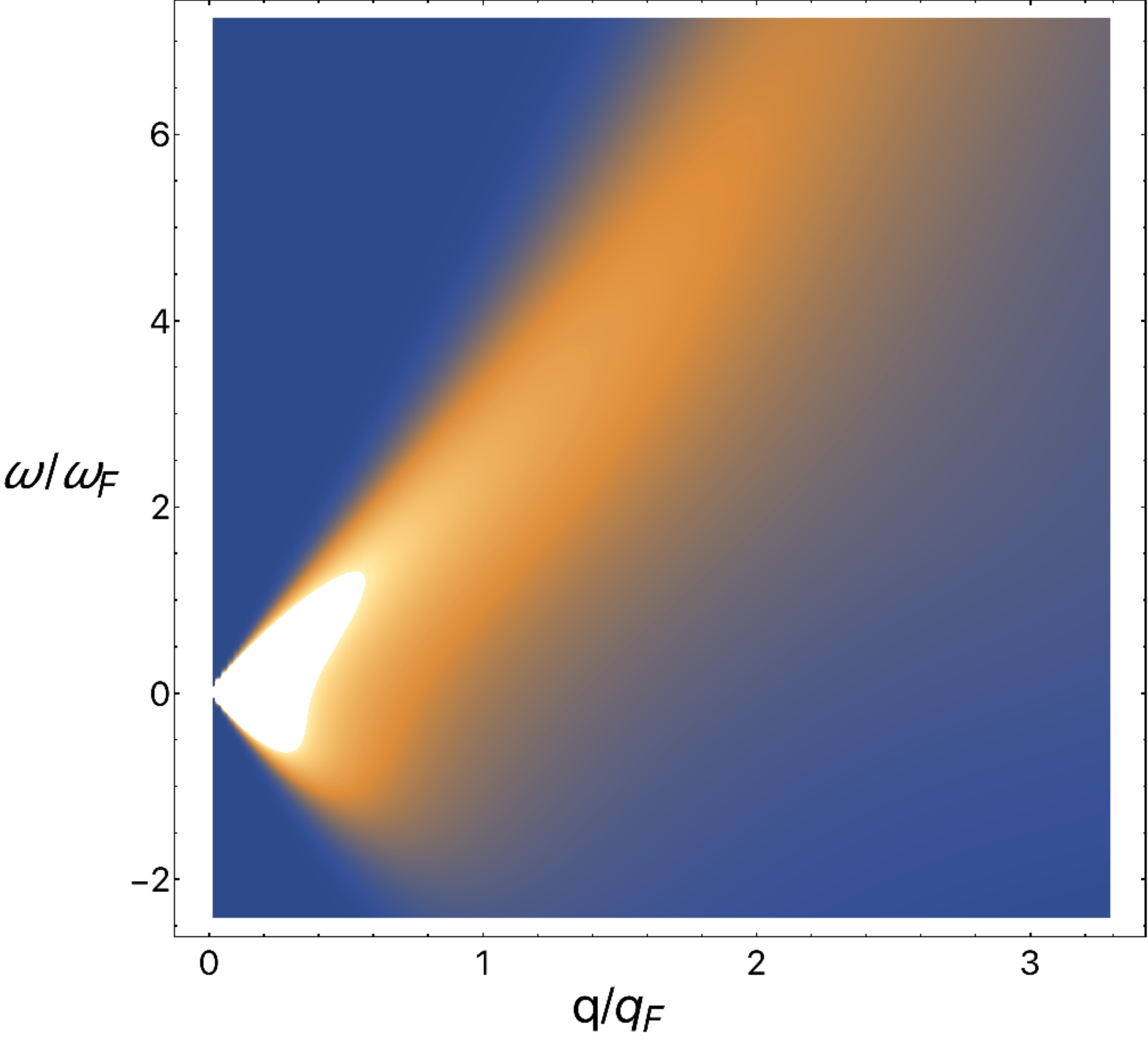}
\end{center}
\caption{Left panel: DSF $S(q,\omega)$ as a function of $q$ and
$\omega$ for a thermal state at inverse temperature $\beta=1$,
$D=0.38$ and $c=4$. Right panel: Same for the one particle-hole contribution
$S^{(1)}(q,\omega)$. The color scale is the same for both plots.}
\label{t1color}
\end{figure}

\begin{figure}[ht!]
\begin{center}
\includegraphics[scale=0.2]{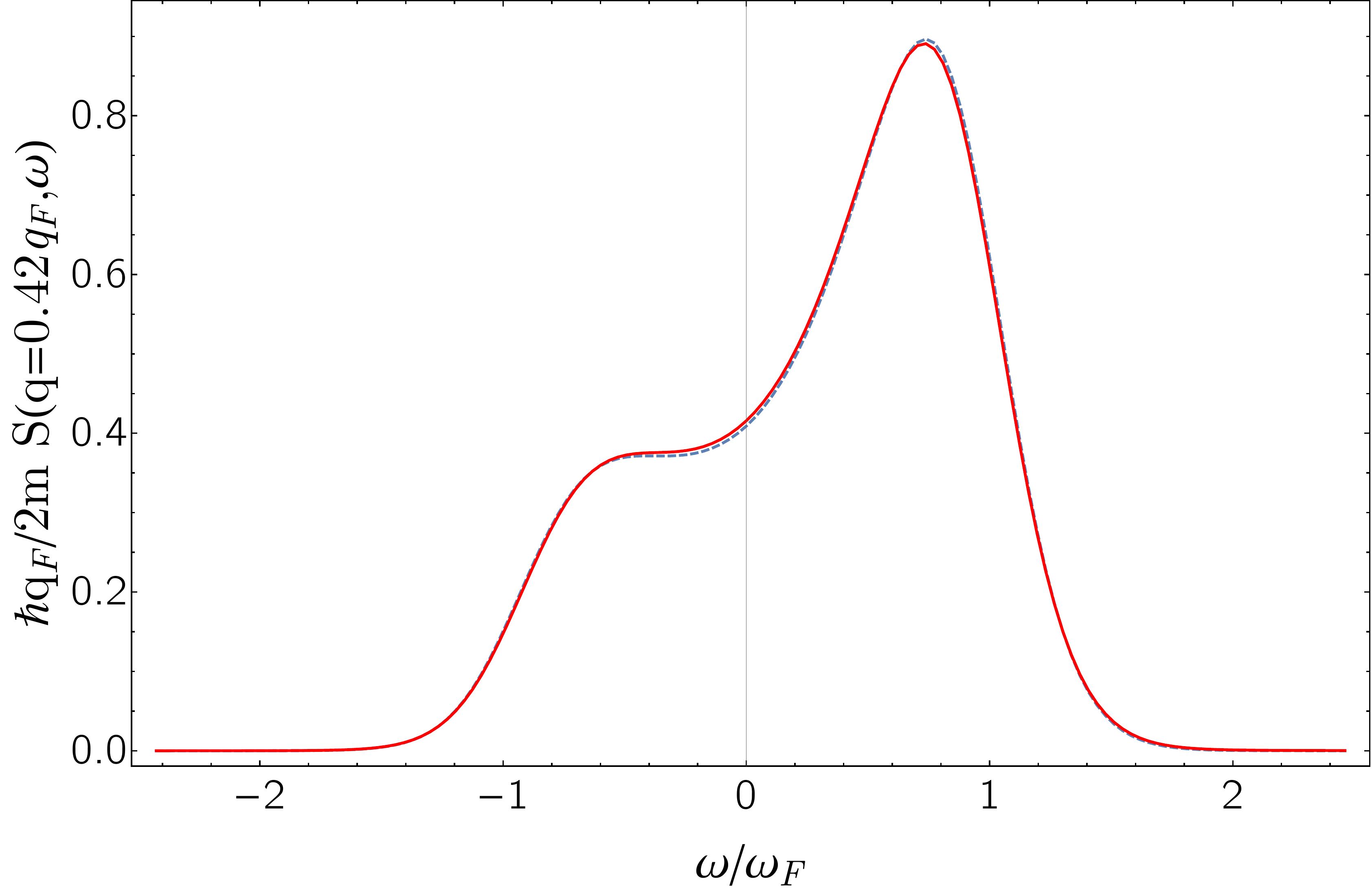}\quad
\includegraphics[scale=0.2]{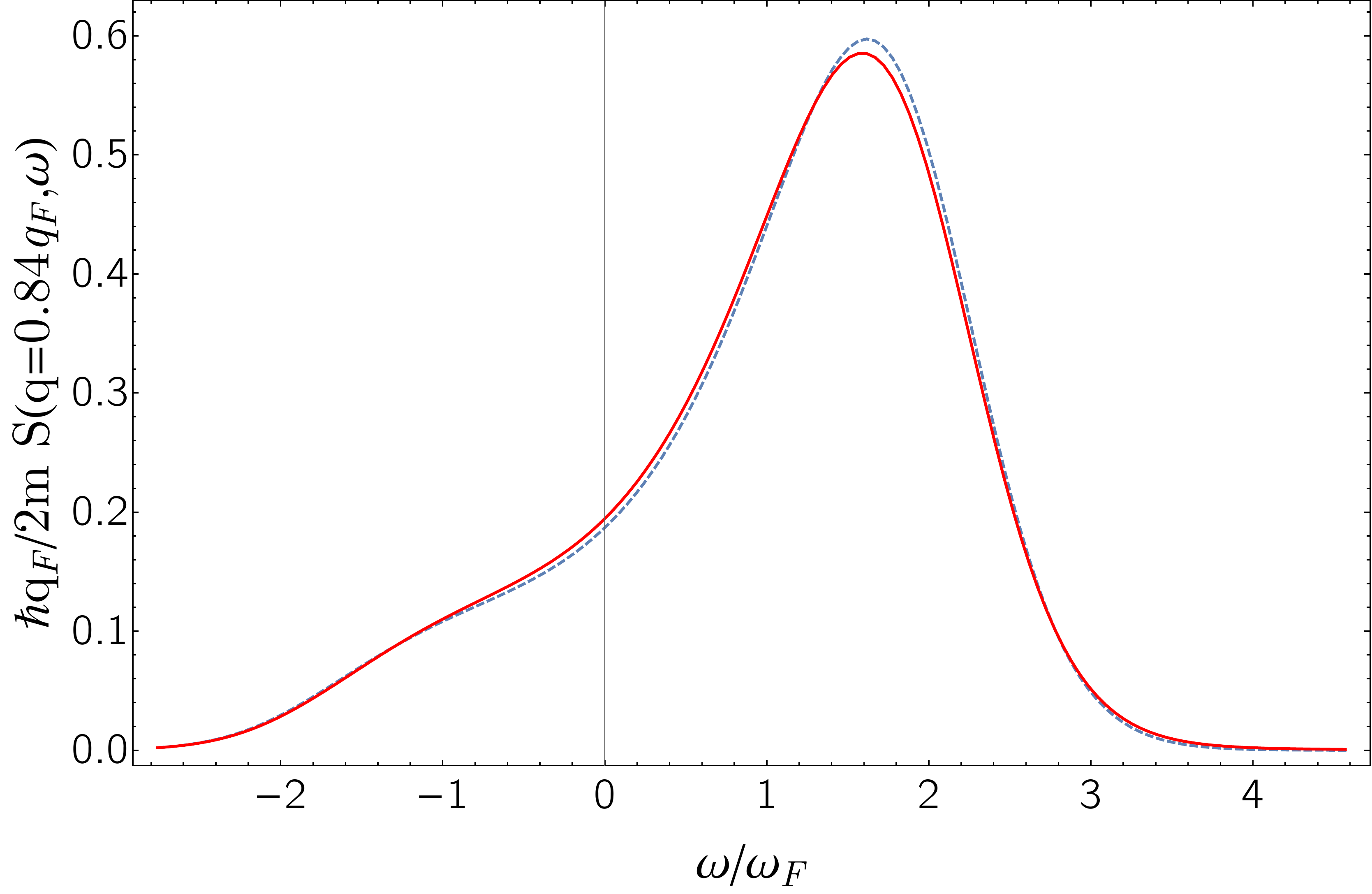}
\end{center}
\caption{$S(q,\omega)$ (red) and $S^{(1)}(q,\omega)$ (blue, dotted) as
functions of $\omega$ for $q=0.42q_F$ (left
panel) and $q=0.84q_F$ (right panel). The parameters are the same
as in Figure~\ref{t1color}.}   
\label{thermalcut}
\end{figure}

\begin{figure}[ht!]
\begin{center}
\includegraphics[scale=0.2]{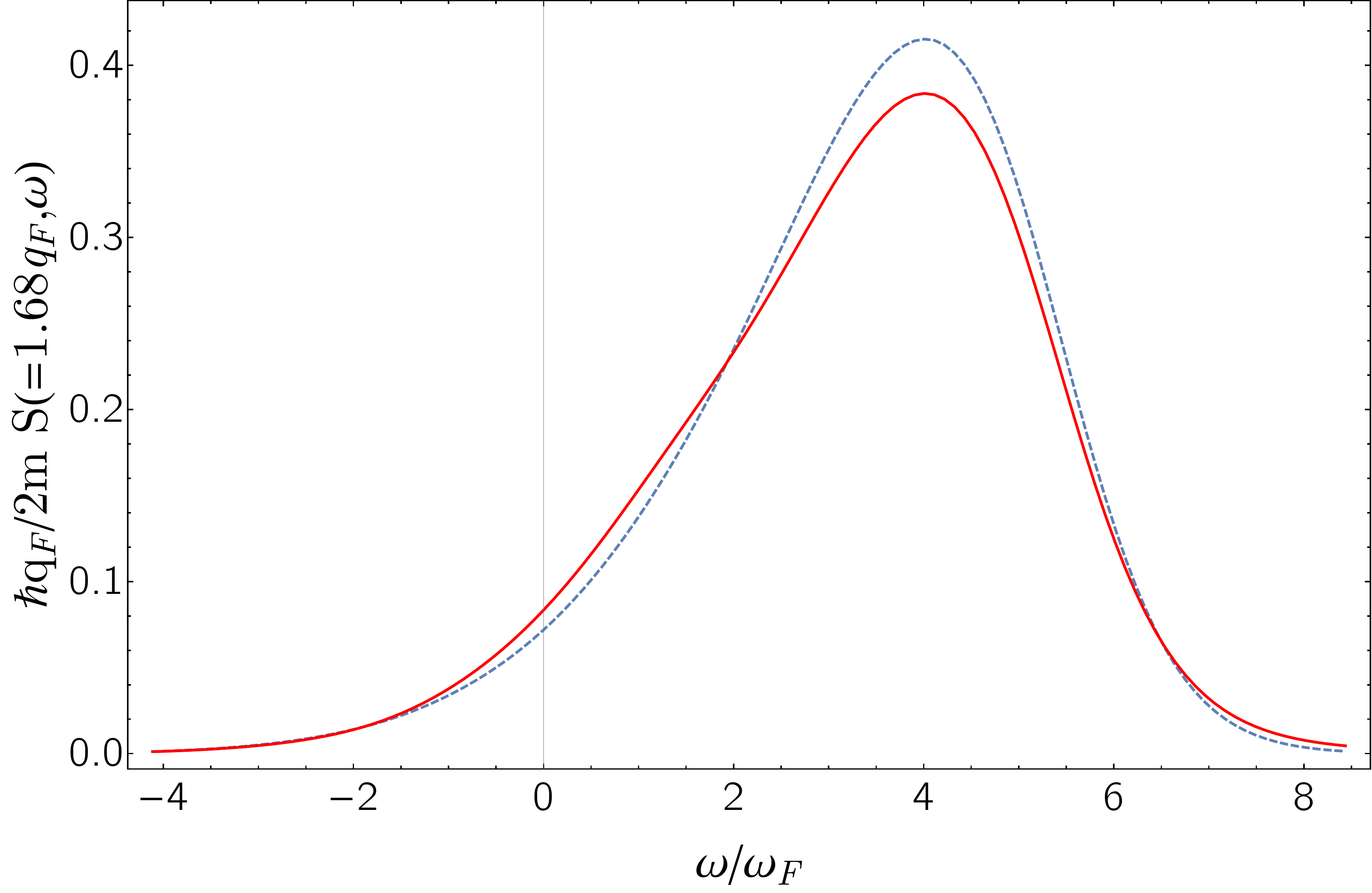}\quad
\includegraphics[scale=0.185]{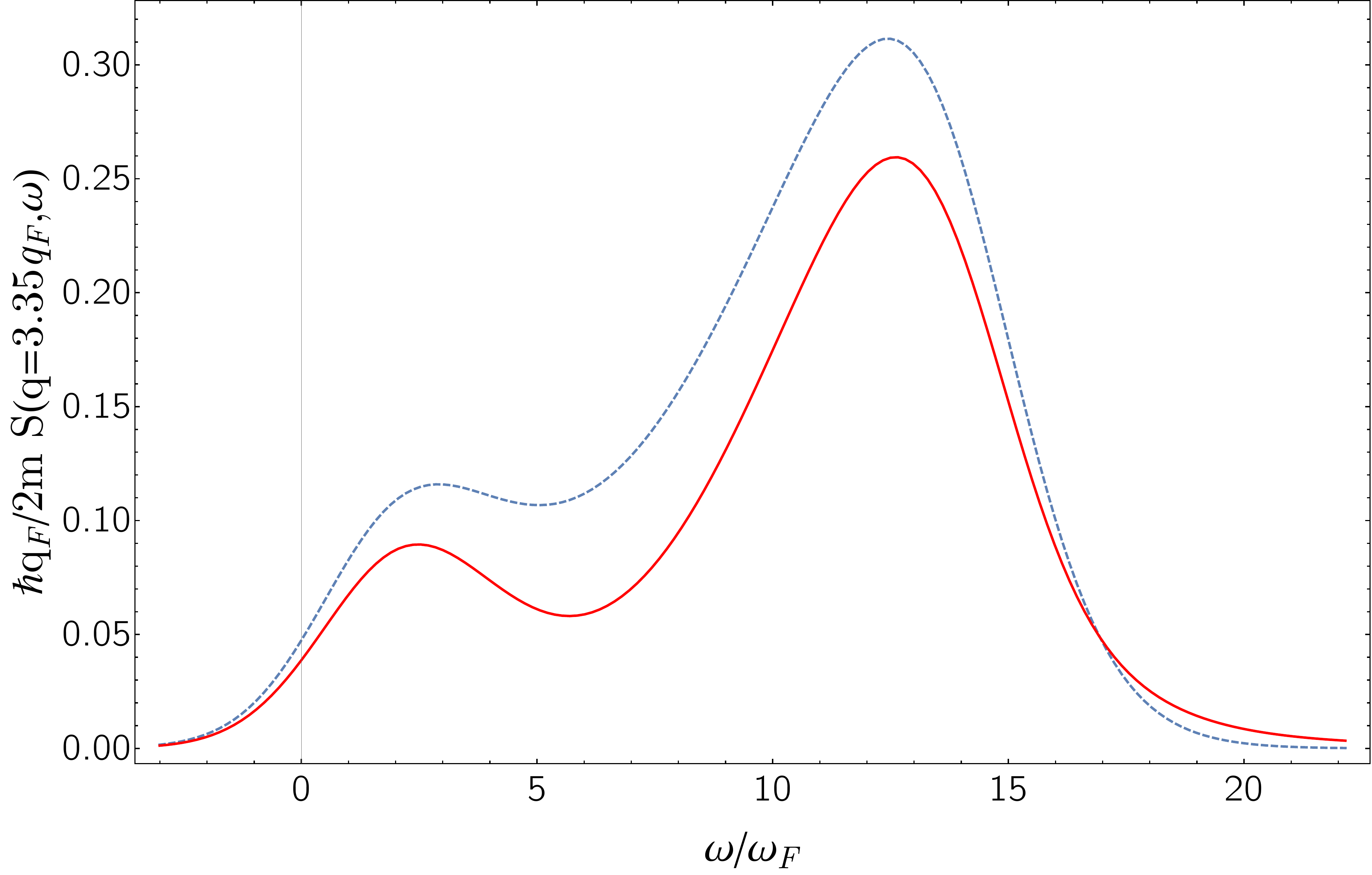}
\end{center}
\caption{Same as Figure~\ref{thermalcut} but for $q=1.68q_F$ (left
  panel) and $q=3.35q_F$ (right panel).}
\label{thermalcut22}
\end{figure}

In Figure~\ref{t1color} we consider the DSF for a different thermal
state characterized by a higher temperature $\beta=1$ and
$D=0.38$, $c=4$. The differences between the $S(q,\omega)$ and
$S^{(1)}(q,\omega)$ are difficult to discern in these plots.
In order to get a more precise notion of the relative contributions of
$S^{(1)}(q,\omega)$ and $S^{(2)}(q,\omega)$ to the DSF for these
values of $D$, $c$ and $\beta$, we show a number of ``constant momentum
cuts", i.e. plots of $S(q,\omega)$ as a function of $\omega$ for
fixed $q$, in Figs~\ref{thermalcut} and \ref{thermalcut22}.
Fig~\ref{thermalcut} gives representative results at ``small''
momenta, defined as $q\lesssim q_F$. We see that the contribution from
two particle-hole excitations is negligibly small. This is in
perfect agreement with observations made in Ref.~\cite{deNP15} based on
comparisons with numerical computations for a finite
number of particles. Our results makes this observation fully quantitative
in the thermodynamic limit in the framework of a $1/c$-expansion.

Figure \ref{thermalcut22} shows how the relative magnitude of
$S^{(2)}(q,\omega)$ evolves at larger values of momentum. We see
that it grows with $q$ and for the values shown is no longer
negligible. 
We further note that while the DSF is expressed as a spectral sum with
only positive terms, the contribution $S^{(2)}(q,\omega)$ can be negative.
The explanation of this behaviour is that the contributions of one and
two particle-hole excitations include terms that arise from
cross-cancellations of divergences occurring in their 'bare' 
spectral sum. Stated differently, each of them incorporates
contributions due to one and two particle-hole excitations so as to
have well-defined thermodynamic limits. If we consider the
spectral sum in a large finite volume, the bare (without
cross-cancelling divergences) one and two particle-hole contributions
are indeed separately positive in the following sense: The leading
$c^0$ term of the bare contribution of one particle-hole excitations is
positive, and we see from \eqref{omega2l} that the same holds true
for the divergent part of the leading $c^{-2}$ term of the
two particle-hole excitations. The interpretation of the resulting contributions as one or two-particle-hole excitations is imposed by whether they are expressed as a double integral (one for the particle and one for the hole) or a quadruple integral (two particles and two holes). Finally, we note that the fact that $S^{(2)}(q,\omega)$ can
be negative is an inherent feature of the $1/c$ expansion as can be
seen by considering the zero temperature limit. Here the successive terms of the $1/c$ expansion of the DSF exhibit a singularity with negative spectral weight,
see Section \ref{zeroTthresholds}, although at finite $c$ all the higher order terms exponentiate into a positive spectral weight.

%%%%%%%%%%%%%%%%%%%%%%%%%%%%%%%%%%%%%%%%%
\subsection{\sfix{DSF in a non-equilibrium steady state}}
\begin{figure}[ht]
\begin{center}
\includegraphics[scale=0.18]{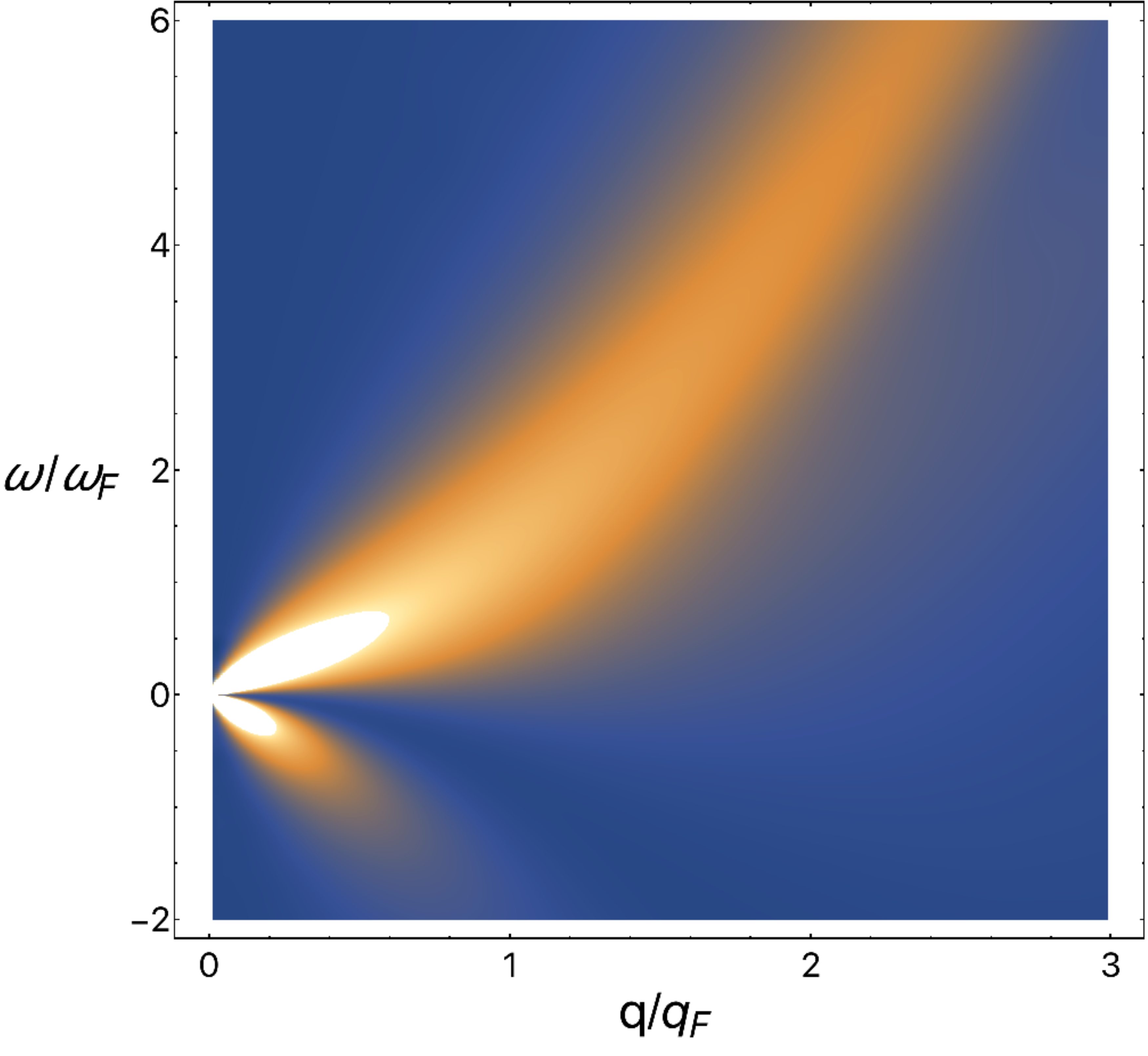}\qquad
\includegraphics[scale=0.18]{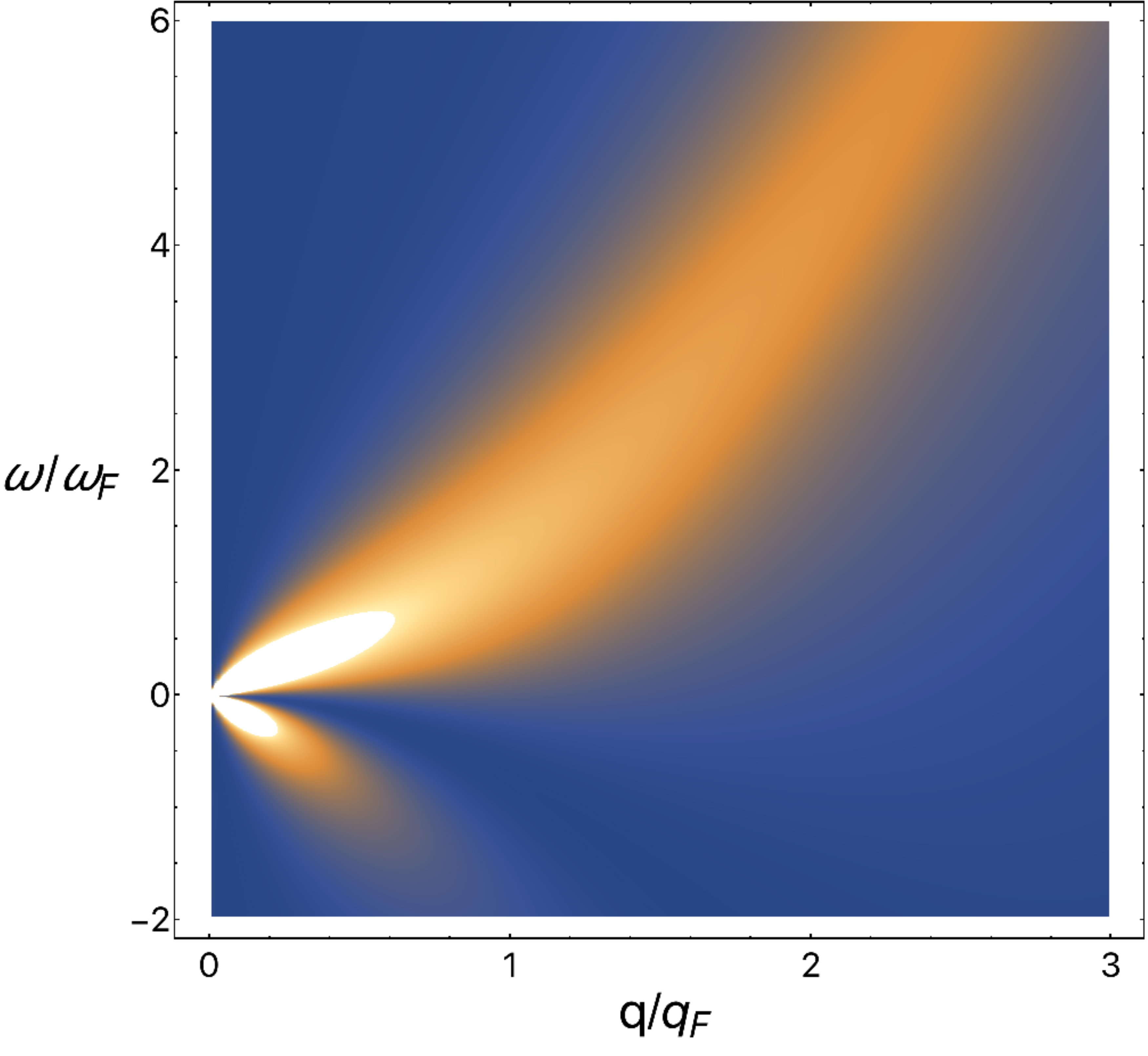}
\end{center}
\caption{$S(q,\omega)$ (left panel) and one particle-hole contribution
$S^{(1)}(q,\omega)$ (right panel) as functions of $q,\omega$ for the
steady state root density \eqref{rhoss} with $c=4$ and $D=1/\pi$. The
color scale is the same for both plots.}  
\label{sscolor}
\end{figure}

%%%%%%%%%%%%%%%%%%%%%%%%%%%%%%%%%%%%%%%%%
In Figure~\ref{sscolor} we show numerical results for $S(q,\omega)$
  and $S^{(1)}(q,\omega)$ for the root density given in Section
\ref{exroot2}. The latter describes the stationary state reached
for the interaction quench of Refs~\cite{Kormos13,denardis14,DNPC15},
where the system is initialized in the ground state at $c=0$ and
density $D=1/\pi$ and time-evolved with the Lieb-Liniger Hamiltonian
at $c=4$. We observe that the two particle-hole contributions lead to a
slight narrowing of the DSF for $q>q_F$. 
%%%%%%%%%%%%%%%%%%%%%%%%%%%%%%%%%%%%%%%%%%%%%%%%
\section{\sfix{Analysis of the result in limiting cases}}
\label{sec6}
%%%%%%%%%%%%%%%%%%%%%%%%%%%%%%%%%%%%%%%%%%%%%%%%
In this section we report a detailed analysis of our results for
the density-density correlation function \eqref{corregenalt}, \eqref{chi2} and the dynamical structure factor \eqref{DSF1ph},
\eqref{DSF2ph2}. The details of the derivations of the results in this
section are reported in Appendix \ref{dev6}. 
%%%%%%%%%%%%%%%%%%%%%%%%%%%%%%%%%%%%%%%%
\subsection{\sfix{ Density-density correlation function}}
%%%%%%%%%%%%%%%%%%%%%%%%%%%%%%%%%%%%%%%%
%%%%%%%%%%%%%%%%%%%%%%%%%%%%%%%%%%%%%%%%%%%%%%%
\subsubsection{\sfix{Asymptotics of equal-time correlations at
    zero temperature}}
\label{cftcompare}
%%%%%%%%%%%%%%%%%%%%%%%%%%%%%%%%%%%%%%%%%%%%%%%
At zero temperature with the root density \eqref{rho0t}, we obtain the following asymptotic behaviour at large $x$ and at order $c^{-2}$
\begin{align}
\langle \sigma(x,0)\sigma(0,0) \rangle =&D^2-\frac{1+\tfrac{4D}{c}+\tfrac{4D^2}{c^2}}{2\pi^2 x^2}\nn
&+A\frac{\cos(2q_F x)}{x^{2}}\left[1-\left(\tfrac{8D}{c}+\tfrac{8D^2}{c^2}\right)\log [2q_F e^{\gamma_{\rm E}}x]+\tfrac{32D^2}{c^2}\log^2[2q_F e^{\gamma_{\rm E}}x]\right]\\
&+o(x^{-2})\,,
\label{cftlarge}
\end{align}
where
\begin{equation}\label{AAA}
A=\frac{1+\frac{4D}{c}}{2\pi^2}+{\cal O}(c^{-2})\,,
\end{equation}
and $\gamma_{\rm E}$ is Euler's constant. This expression is the
large $x$ behaviour of the $1/c$ expansion of the correlation
functions, hence one has $c$  large first and then $x$ large.

Combining CFT/Luttinger liquid theory with exact results
provides the following prediction for the correlations at large $x$ at
fixed $c$ \cite{BIK86,IKR87,haldane,cazalilla,kitanineetcformfactor} 
\begin{equation}\label{cftlargeth}
\langle \sigma(x,0)\sigma(0,0) \rangle =D^2-\frac{K}{2\pi^2 x^2}+A_1\frac{\cos(2q_F x)}{x^{2K}}+\cdots\,,
\end{equation}
with $K$ given in \eqref{K}, and with $A_1$ a known constant
\cite{kitanineetcalgebraic,shashipanfilcaux,shashi}. If one wishes to compare this expression with
\eqref{cftlarge}, one is a priori faced with two problems:   
\begin{enumerate}
\item[(i)] If one expands \eqref{cftlargeth} in powers of $c^{-1}$
one has to take first $x$ large and then $c$ large, which is the
reverse of \eqref{cftlarge}. Hence comparing \eqref{cftlarge} and
\eqref{cftlargeth} entails to commute two limits. This commutation is
possible if our expansion in $c^{-1}$ \eqref{cftlarge} is
\textit{uniform} in space.  
\item[(ii)] There could be corrections to \eqref{cftlargeth} that are
subleading in $x$ at fixed $c$, but become of the same order as
the dominant term once expanded in $c^{-1}$ (i.e. give rise to
$\log(x)$ terms). An example would be a term $\propto 
x^{-4K+2}$. These corrections would be visible in \eqref{cftlarge},
but not in \eqref{cftlargeth}.  
\end{enumerate}
In the case of density correlations in the Lieb-Liniger model it
  follows from the exact large $x$ expansion at fixed $c$
\cite{kitanineetcformfactor} that there are no subleading corrections
with the property described in (ii).
We thus expand \eqref{cftlargeth} in powers of $c^{-1}$.
Since $K\to 1$ when $c\to\infty$ the power-law $x^{-2K}$ becomes 
$x^{-2}$ corrected by logarithms and we find
\begin{align}
\langle \sigma(x,0)\sigma(0,0) \rangle
&=D^2-\frac{1+\tfrac{4D}{c}+\tfrac{4D^2}{c^2}}{2\pi^2 x^2}\nn 
&+A_1\frac{\cos(2q_F
  x)}{x^{2}}\left[1-\left(\tfrac{8D}{c}+\tfrac{8D^2}{c^2}\right)\log
  x+\tfrac{32D^2}{c^2}\log^2x\right]+{\cal O}(c^{-3})\,. 
\end{align}
The coefficient $A_1$ depends on $c$ but as its representation is
rather complicated \cite{shashipanfilcaux,shashi} (with approximations in \cite{BrandCherny09,Lang17}) we left calculating its
$c^{-1}$ expansion for future work. We see that it agrees with
\eqref{cftlarge} if we  identify
\begin{equation}
A_1=A(2q_Fe^{\gamma_{\rm E}})^{2-2K}+{\cal O}(c^{-2})\,.
\end{equation}
In particular
the critical exponents are reproduced at order $c^{-2}$. This both
provides a check of our formula, and shows that our $1/c$ expansion is
\textit{uniform} in space. 
%
%By comparing the two expressions, we identify
%\begin{equation}
%\label{a1pref}
%A_1 D^{2-2K}=\frac{K}{2\pi^2}(2q_F e^{\gamma_{\rm E}})^{2-2K}+O(c^{-2})\,.
%\end{equation}
%This expression,  despite being at order $c^{-1}$, actually gives good agreement with the exact expression at all $c$.

%%%%%%%%%%%%%%%%%%%%%%%%%%%%%%%%%%%%%%%%%%%%%%%
\subsubsection{Dynamical correlations asymptotics at zero temperature}
%%%%%%%%%%%%%%%%%%%%%%%%%%%%%%%%%%%%%%%%%%%%%%%
At zero temperature (with the root density \eqref{rho0t}) we can
evaluate the asymptotic behaviour of the dynamical correlation
function at large $x,t$ at fixed 
\begin{equation}
\alpha=\frac{x}{2t}\,.
\label{alphadef}
\end{equation}
It is convenient to define
\begin{equation}\label{alphap}
\alpha'=\frac{x'}{2t}=\gamb\alpha\,,
\end{equation}
and set
\begin{equation}
s=\begin{cases}
1\,\quad\text{if }|\alpha|>q_F\\
-1\,\quad\text{if }|\alpha|<q_F\\
\end{cases}\,.
\end{equation}
We obtain

\begin{align}
\label{resdynam}
\langle \sigma(x,t)\sigma(0,0) \rangle =D^2&+\sum_{\sigma=\pm}B_\sigma
\frac{e^{ist(Q+\sigma\alpha')^2}}{|t|^{3/2}}
\Bigg[1-\nu_\sigma
%  \left(\frac{2(Q+\sigma\alpha)}{\pi c}
  %  +\frac{2(\alpha^2+4\sigma\alpha Q+3Q^2)}{\pi^2 c^2}  \right)
  \log (i  \varpi_\sigma t)
+\frac{\nu_\sigma^2}{2}\log^2 (i \varpi_\sigma t)\Bigg]\\
&+ o(t^{-3/2}),
%+\frac{2(Q+\sigma\alpha)^2}{\pi^2 c^2}\log^2|t|\Bigg],
%\nn
%&+B_+ \frac{e^{it(Q+\alpha')^2}}{|t|^{3/2}}\bigg[1-\Big(\frac{2(Q+\alpha)}{\pi
%    c}  +\frac{2(\alpha^2+4\alpha  Q+3Q^2)}{\pi^2 c^2}  \Big)\log|t|\nn
%&\hskip4cm  +\frac{2(Q+\alpha)^2}{\pi^2 c^2}\log^2|t|\bigg]\,,
\end{align}
with
\begin{align}
B_\sigma=&\frac{\sign(t)e^{-s\frac{i\pi}{4}\sign(t)}(1+\frac{2D}{c})^4}{8i\pi^\frac{3}{2}(Q+\sigma\alpha')}+{\cal O}(c^{-2}),\nn
%B_\pm=&\frac{\sign(t)e^{-s\frac{i\pi}{4}\sign(t)}(1+\frac{2D}{c})^4}{8i\pi^\frac{3}{2}(Q\pm\alpha')}
%\Bigg[1-\frac{2(Q\pm\alpha)}{\pi c}
%  \left[\log \left|2\frac{(Q\pm\alpha)^2}{Q\mp\alpha}\right|+\gamma_{\rm E}\pm \frac{i\pi}{2} \right]\Bigg]+{\cal O}(c^{-2}),\nn
\nu_\sigma=& \left(1+\frac{2Q}{\pi c}\right)\frac{2(Q+\sigma\alpha')}{\pi c}
  +\frac{2(Q+\sigma\alpha')^2}{\pi^2 c^2}+{\cal O}(c^{-3})\,,\nn
%\nu_\sigma=& (1+\frac{2Q}{\pi c})\frac{2(Q+\sigma\alpha')}{\pi c}
%  +\frac{2(\alpha^{\prime 2}+4\sigma\alpha' Q+3Q^2)}{\pi^2 c^2}+{\cal O}(c^{-3})\,,\nn
  \varpi_\sigma=&s\sigma 4Q \frac{(Q+\sigma\alpha')^2}{|Q-\sigma\alpha'|}e^{\gamma_{\rm E}}\ .
\end{align}

Ref.~\cite{kitanineetcformfactor} derived the full asymptotic
expansion at large $x,t$ for any value of $c$ at zero temperature. The 
$c^{-1}$ expansion of this result at order $c^{-2}$ (without
expanding the prefactors) is in agreement with \eqref{resdynam}. In
particular the critical exponents $\nu_\pm$ are reproduced at order
$c^{-2}$. This both provides a check of our calculation and shows that
our $1/c$ expansion is uniform in time as well, since the
large $x,t$ and large $c$ limits commute. 

\subsubsection{\sfix{Asymptotics of dynamical correlations for a generic root
  density and Generalized Hydrodynamics}}
For a generic continuous root density $\rho$ in the large $x,t$ regime
at fixed $\alpha$ \fr{alphadef} we obtain the following
asymptotic behaviour 

\begin{align}
\label{ghd}
\langle \sigma(x,t)\sigma(0,0) \rangle
=&D^2+\frac{\pi\gam^2\rho(\alpha')\rho_h(\alpha')}{|t|}
+\frac{i\pi\gam^2\left[\rho''(\alpha')\rho_h(\alpha')-\rho(\alpha')\rho_h''(\alpha') \right]}{4t|t|}\nn
&+\frac{\pi^2}{t^2c^2}\bigg[12(\rho\rho_h)^2(\alpha')+8(\rho\rho_h)'(\alpha')\int_{-\infty}^\infty \sign(\alpha'-\zeta)\rho(\zeta)\rho_h(\zeta)\D{\zeta}\nn
&\qquad\qquad+2(\rho\rho_h)''(\alpha')\int_{-\infty}^\infty |\alpha'-\zeta|\rho(\zeta)\rho_h(\zeta)\D{\zeta}
\bigg]+o(t^{-2})\,,
\end{align}
where we recall the definition of $\alpha'$ in \eqref{alphap}. The
first line arises from one particle-hole contributions,
while the second and third lines are two particle-hole
contributions. If the root density is not continuous the leading term
in $1/t$ is still correct, but the higher order corrections may
change.  

GHD \cite{CADY16,BCDF16} makes predictions for the coefficient
of the $1/t$ term in the density-density correlator for any
value of $c$ \cite{DS17,Doyon18}. For the sake of
  completeness we summarize the $1/c$-expansion of the GHD results in
  Appendix \ref{app:GHD}. The leading term proportional to
$1/|t|$ of \eqref{ghd} is in perfect agreement with the order
  $c^{-2}$ expansion of the GHD results. To the best of our knowledge
this constitutes the most non-trivial check to date of GHD
  predictions in an interacting integrable model.

\begin{figure}[ht]
\begin{center}
\includegraphics[scale=0.19]{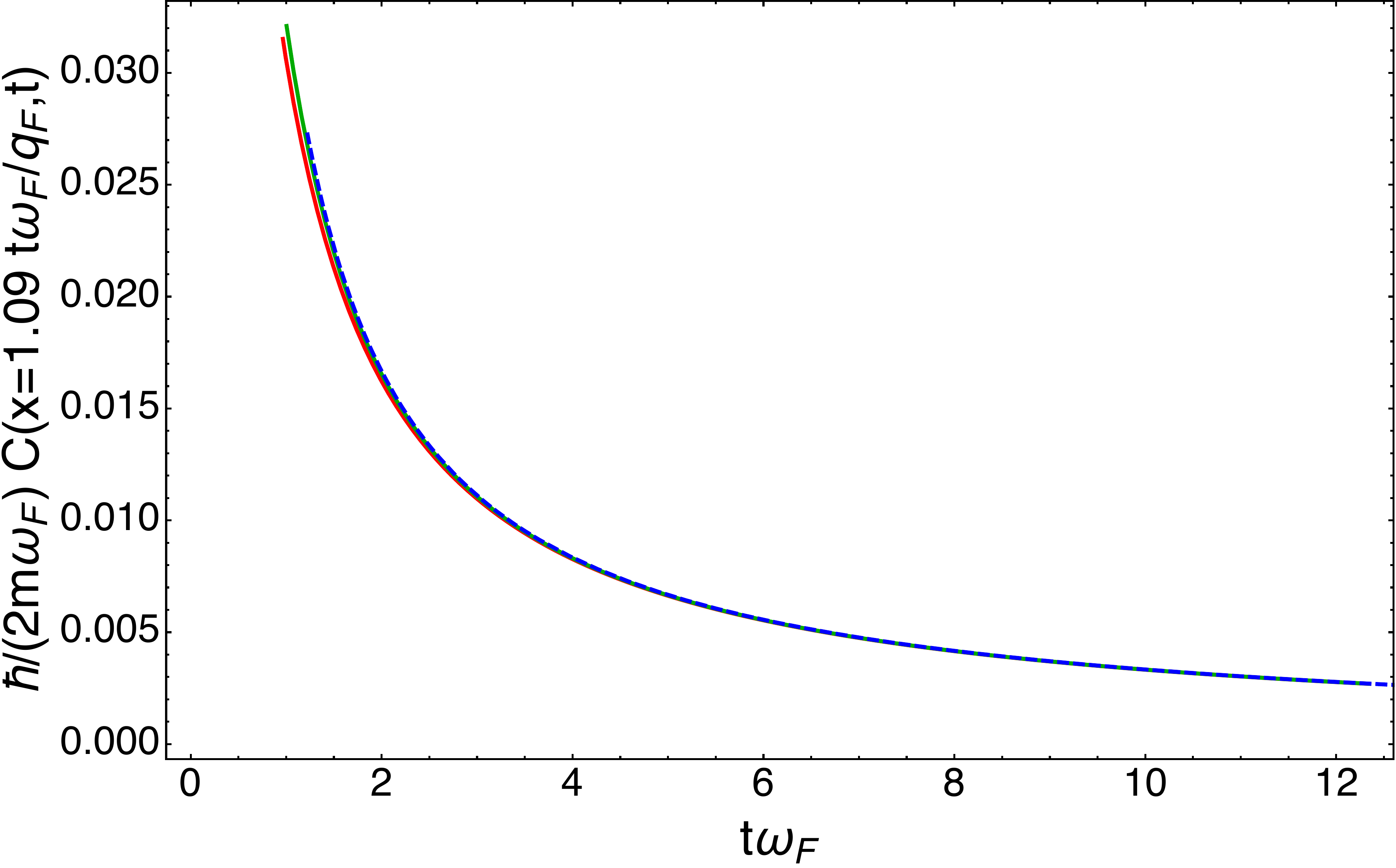}\qquad
\includegraphics[scale=0.19]{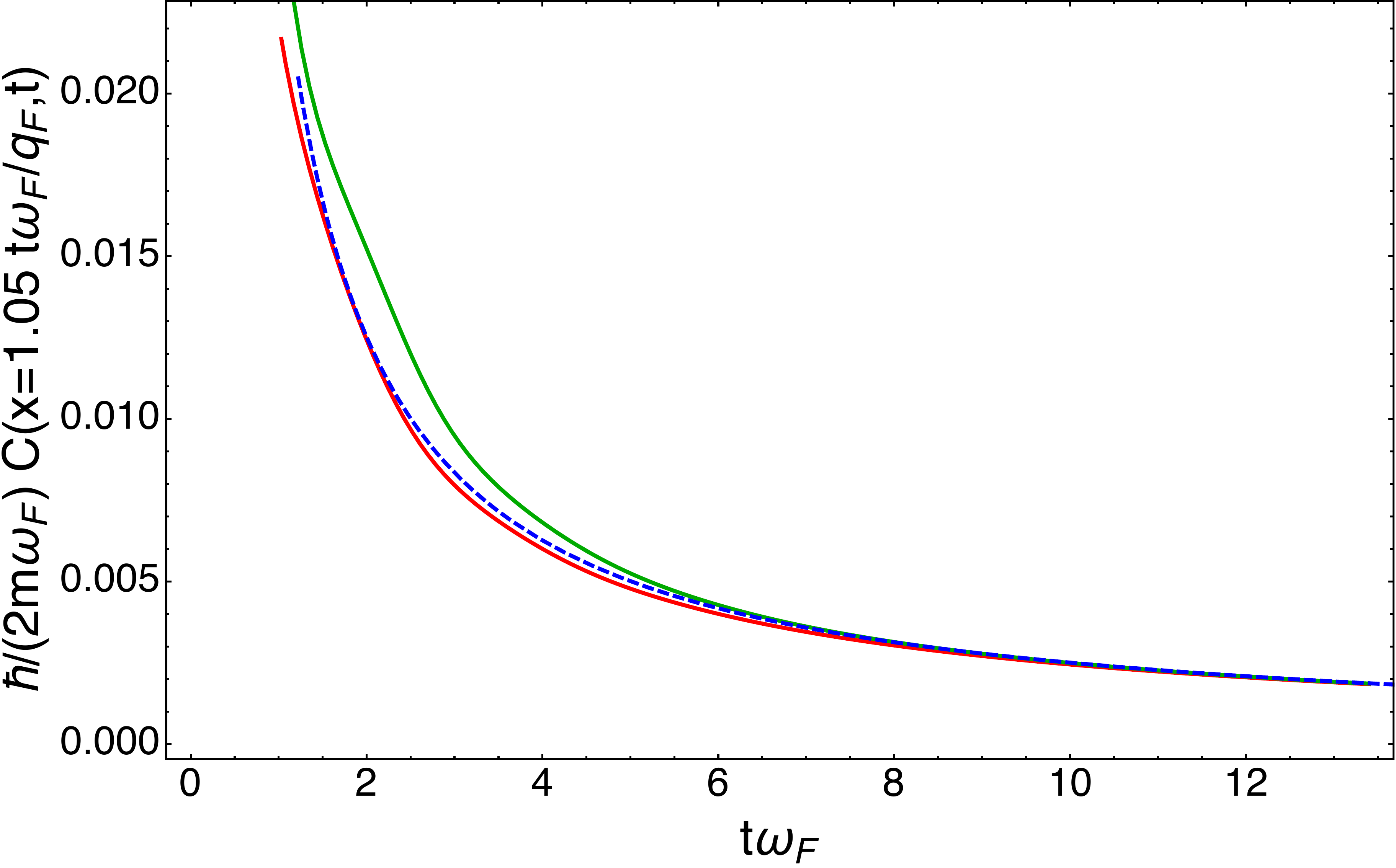}
\end{center}
\caption{Correlation function $C(x,t)\equiv\langle
\sigma(x,t)\sigma(0,0) \rangle$ in a thermal state for $x=2\alpha t$
as a function of $t$ at $c=4$, for $\beta=1$ and $D=0.38$
(left) and $\beta=3$ and $D=0.386$ (right). The three curves
  depict the real part (red), the modulus (green) and the GHD
  approximation (dashed blue).} 
\label{corrthermfig}
\end{figure}

Importantly, we can assess the accuracy of the GHD approximation
outside the asymptotic large space and time regime by comparing it
  to the full correlations at order $c^{-2}$. In
Figure~\ref{corrthermfig} we show our results for the real part and
the modulus of  $\langle \sigma(x,t)\sigma(0,0) \rangle$ for two thermal
states at $c=4$ together with the GHD approximation. We see that at high
temperature $\beta=1$ the GHD approximation is surprisingly good even
at short times. At lower temperatures $\beta=3$ the correlation is
still very well approximated by GHD, but is seen to display damped
oscillations in the absolute value that arise from the imaginary
part of the correlations that decay as $t^{-2}$ and is not
accounted for by GHD.   

\begin{figure}[ht]
\begin{center}
\includegraphics[scale=0.19]{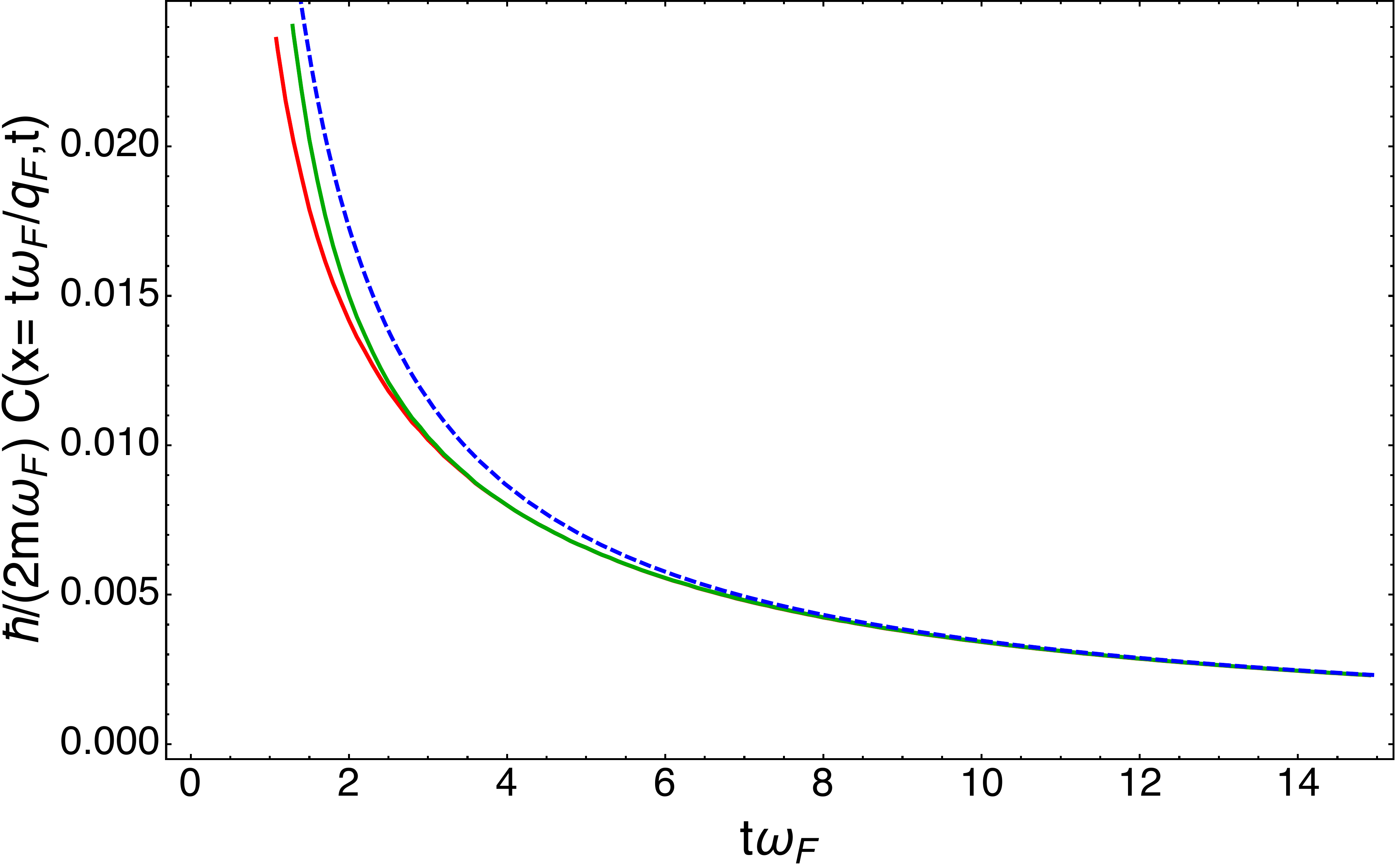}\qquad
\includegraphics[scale=0.19]{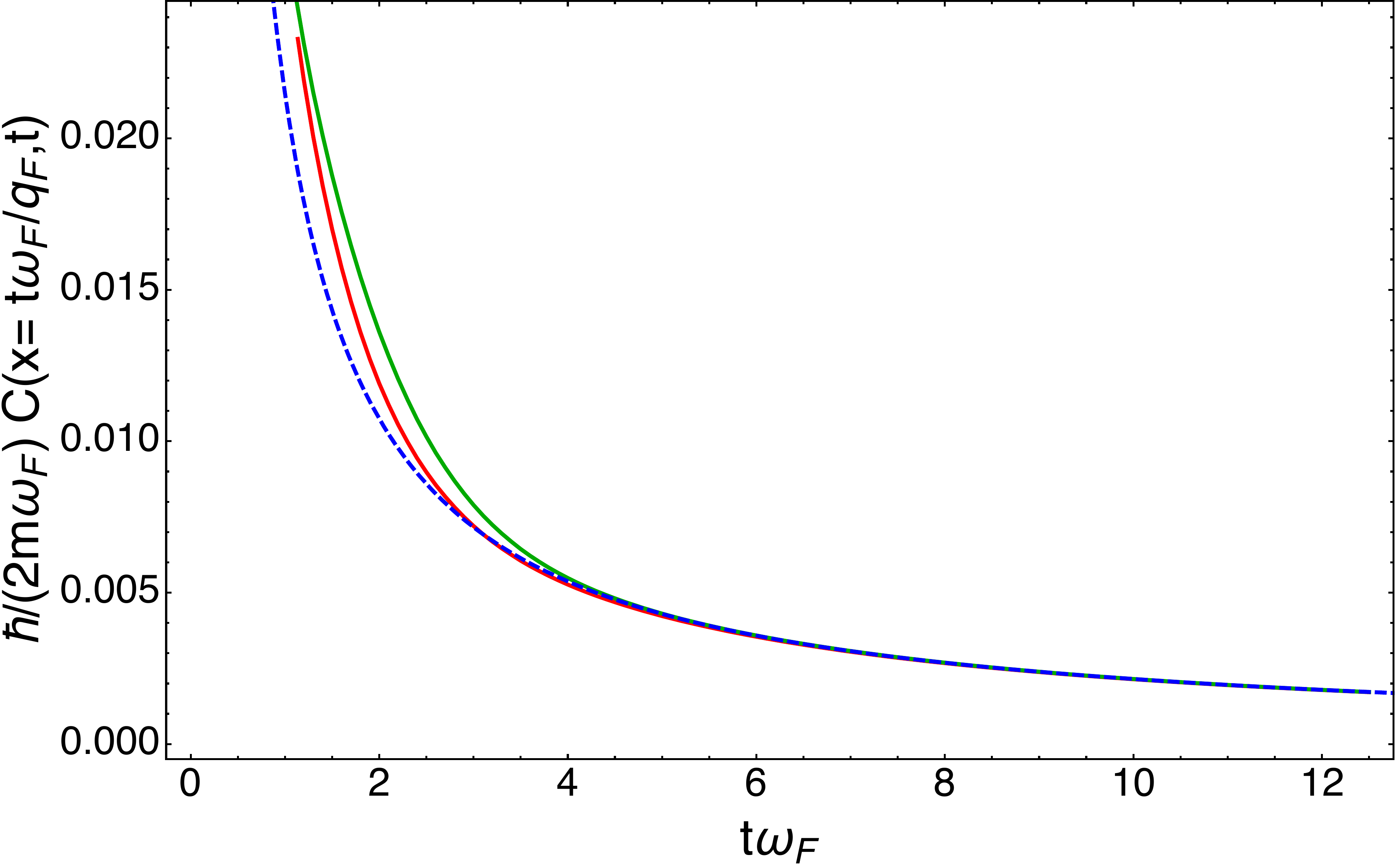}
\end{center}
\caption{Correlation function $C(x,t)\equiv\langle
\sigma(x,t)\sigma(0,0) \rangle$ in the non-equilibrium steady state
\eqref{rhoss} for $x=2\alpha t$ as a function of $t$ at $c=4$,
for $D=1/\pi$ (left) and $D=1/(2\pi)$ (right).
The three curves 
  depict the real part (red), the modulus (green) and the GHD
  approximation (dashed blue).} 
\label{corrthermfig2}
\end{figure}

In Figure~\ref{corrthermfig2} we present the analogous
comparison for a non-equilibrium steady state with root
density \eqref{rhoss}. This root density is ``less regular" than
thermal densities in the sense that it has a narrower peak
at zero. As a consequence, we expect higher Fourier-like corrections
to the oscillatory integral, whereas GHD describes saddle-point-like
corrections only. We indeed observe a more pronounced discrepancy for
short or intermediate times, but the agreement at later times is still
excellent, and globally remains very good. 

%%%%%%%%%%%%%%%%%%%%%%%%%%%%%%%%%%%%%%%%%%
\subsection{Dynamical structure factor}
%%%%%%%%%%%%%%%%%%%%%%%%%%%%%%%%%%%%%%%%%%
\subsubsection{\sfix{A simplified expression at zero temperature}}
\label{simplify}
%%%%%%%%%%%%%%%%%%%%%%%%%%%%%%%%%%%%%%%%%%
Equations \eqref{DSF1ph}, \eqref{DSF2ph2} for the DSF can be
simplified at zero temperature. The one particle-hole
contribution can be written as 
\begin{align}
\label{dsf1pht0}
S^{(1)}(q,\omega)&=\frac{\gamb^4}{2|q|}\left[1-\frac{2q}{\pi c}\log \left|\frac{\omega^2-\omega_+^2}{\omega^2-\omega_-^2} \right|+\frac{2q^2}{\pi^2 c^2}\log^2 \left|\frac{\omega^2-\omega_+^2}{\omega^2-\omega_-^2} \right| \right]\1_{\omega_-<\omega<\omega_+}\nn
&+{\cal O}(c^{-3})\,,
\end{align}
where we have defined
\begin{equation}\label{opm}
\omega_\pm(q)=\left|\qp^2\pm 2|\qp|Q\right|\,.
\end{equation}
The contribution from two particle-hole excitations
can be simplified by carrying out the integral over $p$ in \eqref{DSF2ph2} 
\begin{align}
\label{dsf2pht0}
&S^{(2)}(q,\omega)=\frac{1}{4\pi^2 q' c^2}\int_{-\infty}^\infty \Bigg[\frac{1}{z}\Big[(5-4z)q'(Z_+-Z_-)+Z_-^2-Z_+^2+2q'(1-z)^2\frac{Z_+-Z_-}{z}\nn
&\qquad+2q^{\prime 2}(1-z)^2\log \left|\frac{q'-2Z_+}{q'-2Z_-}\right|-4q^{\prime 2}(1-z)^2\log \left|\frac{(2z-1)q'-2Z_+}{(2z-1)q'-2Z_-}\right|\Big]\1_{Z_-<Z_+}\nn
&\qquad-\frac{2q' \min(|q'z|,2Q)}{z^2}\1_{\omega_-(q)<\omega<\omega_+(q)}\Bigg]\D{z}\,.
\end{align}
Here we have defined
\begin{equation}
Z_+(z)=\begin{cases}
\min\left[\frac{\omega+2q'Q+q^{\prime 2}(z-1)}{2q'z},\frac{\omega-|2q'Q-q^{\prime 2}z|}{2q'(z-1)}\right]\quad\text{if }z\geq 1\\
\min\left[\frac{\omega-|2q'Q-q^{\prime 2}(1-z)|}{2q'z},\frac{-\omega+2q'Q+q^{\prime 2}z}{2q'(1-z)}\right]\quad\text{if }0\leq z\leq 1\\
\min\left[\frac{\omega-2q'Q-q^{\prime 2}(1-z)}{2q'z},\frac{-\omega-|2q'Q+q^{\prime 2}z|}{2q'(1-z)}\right]\quad\text{if }z\leq 0\,,
\end{cases}
\end{equation}
and
\begin{equation}
Z_-(z)=\begin{cases}
\max\left[\frac{\omega+|2q'Q-q^{\prime 2}(z-1)|}{2q'z},\frac{\omega-2q'Q-q^{\prime 2}z}{2q'(z-1)}\right]\quad\text{if }z\geq 1\\
\max\left[\frac{\omega-2q'Q-q^{\prime 2}(1-z)}{2q'z},\frac{-\omega+|-2q'Q+q^{\prime 2}z|}{2q'(1-z)}\right]\quad\text{if }0\leq z\leq 1\\
\max\left[\frac{\omega-|2q'Q-q^{\prime 2}(1-z)|}{2q'z},\frac{-\omega-2q'Q+q^{\prime 2}z}{2q'(1-z)}\right]\quad\text{if }z\leq 0\,.
\end{cases}
\end{equation}

%%%%%%%%%%%%%%%%%%%%%%%%%%%%%%%%%%%%%%%%%%%%%%%%%%%%%%%%%%%%%%%%%
\subsubsection{Behaviour near the thresholds at zero temperature}
\label{zeroTthresholds}
%%%%%%%%%%%%%%%%%%%%%%%%%%%%%%%%%%%%%%%%%%%%%%%%%%%%%%%%%%%%%%%%%
At zero temperature, the DSF exhibits divergences at certain
threshold energies $\omega_{\rm th}(q)$. In our case, at order
${\cal O}(c^{-2})$, the two thresholds occur at $\omega_\pm(q)$ defined in
\eqref{opm}. For $q<2Q$ we find the following singular behaviour
of the one particle-hole DSF near them  
\begin{align}
\label{dsflog1}
S^{(1)}(q,\omega_{+}+\delta\omega)&=\frac{\gamb^2}{2|q|}\left(1-\frac{2q}{\pi c}\log \left|\delta\omega \tfrac{2\omega_+}{\omega_+^2-\omega_-^2}\right|+\frac{2q^2}{\pi^2c^2}\log^2 \left|\delta\omega \tfrac{2\omega_+}{\omega_+^2-\omega_-^2}\right|\right)\1_{\delta\omega<0}\nn
S^{(1)}(q,\omega_{-}+\delta\omega)&=\frac{\gamb^2}{2|q|}\left(1+\frac{2q}{\pi c}\log \left|\delta\omega \tfrac{2\omega_-}{\omega_+^2-\omega_-^2}\right|+\frac{2q^2}{\pi^2c^2}\log^2 \left|\delta\omega \tfrac{2\omega_-}{\omega_+^2-\omega_-^2}\right|\right)\1_{\delta\omega>0}\,.
\end{align}
The analogous results for the two particle-hole contribution are
\begin{align}
\label{dsflog2}
S^{(2)}(q,\omega_{+}+\delta\omega)&=\left(\frac{q}{2\pi^2 c^2}\log
|\delta\omega|\right)\1_{\delta\omega<0}-\left(\frac{q}{2\pi^2
  c^2}\log |\delta\omega|\right)\1_{\delta\omega>0}\ , \nn
S^{(2)}(q,\omega_{-}+\delta\omega)&=\left(\frac{q}{\pi^2 c^2}\log |\delta\omega|\right)\1_{\delta\omega>0}\,.
\end{align}
These limiting behaviours are obtained at large $c$ first, and then $\omega$ close to $\omega_\pm$. 

At fixed $c$ non-linear Luttinger liquid theory predicts the exponent
of the power-law divergence near these thresholds
\cite{IG08,PAW09,ISG12,P12,shashipanfilcaux,kitanineetcformfactor}
\begin{align}
\label{NLL}
S(q,\omega_{+}+\delta\omega)&=C_0|\delta\omega|^{\mu_+}\1_{\delta\omega>0}+C_1|\delta\omega|^{\mu_+}\1_{\delta\omega<0}+C_2+\ldots\nn
S(q,\omega_{-}+\delta\omega)&=C_3(\delta\omega)^{\mu_-}\1_{\delta\omega>0}+C_4\1_{\delta\omega>0}+\ldots\,.
\end{align}
Here $C_{0,1,2,3,4}$ are $c$-dependent constants and the exponents
$\mu_\pm$ have simple expressions that depend on $c$. 
Results for the non-universal prefactors $C_{0,1,2,3,4}$ at finite $c$ are available in the literature
\cite{shashipanfilcaux}.
%, but as their respective expressions are
%rather involved we have not attempted to work out their $1/c$ expansions
The dots encompass less singular pieces
$|\delta\omega|^\mu$ with a $c$-dependent exponent $\mu>\mu_\pm$,
and regular pieces $C_5\delta\omega+O((\delta\omega)^2)$. In the
framework of the $1/c$ expansion these power-laws give rise to
logarithms 
\begin{align}
|\delta\omega|^{\mu_+}&=1-\frac{2q}{\pi c}\log|\delta\omega|+\frac{2q^2}{\pi^2 c^2}\log^2|\delta\omega|+\frac{2q^2}{\pi^2 c^2}\log|\delta\omega|+{\cal O}(c^{-3})\nn
|\delta\omega|^{\mu_-}&=1+\frac{2q}{\pi c}\log|\delta\omega|+\frac{2q^2}{\pi^2 c^2}\log^2|\delta\omega|+\frac{2q^2}{\pi^2 c^2}\log|\delta\omega|+{\cal O}(c^{-3})\,.
\end{align}
These expansions are valid if we take $\omega$ close to
$\omega_\pm$ first and then consider the large-$c$ limit, and
in order to compare with our result we have to commute these two
limits. Importantly, the less singular pieces that are subleading in
$\delta \omega$ can also produce logarithms if their exponent goes to
$0$ when $c\to\infty$. However, it follows from our asymptotic
analysis in real space that there are no such terms. Comparing
\eqref{NLL} with \eqref{dsflog1} and \eqref{dsflog2} we find that our
result is in agreement with the non-linear Luttinger liquid
predictions if we identify  
\begin{align}
C_0&=\frac{1}{4\pi c}+{\cal O}(c^{-2})\nn
C_1&=\frac{(1+\frac{2D}{c})^2}{2|q|}\left( \frac{2\omega_+}{\omega_+^2-\omega_-^2}\right)^{\mu_+}+\frac{1}{4\pi c}+{\cal O}(c^{-2})\nn
C_2&=-\frac{1}{4\pi c}+{\cal O}(c^{-2})\nn
C_3&=\frac{(1+\frac{2D}{c})^2}{2|q|}\left( \frac{2\omega_-}{\omega_+^2-\omega_-^2}\right)^{\mu_-}+{\cal O}(c^{-2})\nn
C_4&=0+{\cal O}(c^{-2})\,.
\end{align}
In particular we obtain the correct exponents at order $c^{-2}$. This
provides a check of our result for the DSF and shows that it
is uniform in $q$ and $\omega$.

% In order to make contact with our $1/c$ expansion one
%still requires the two properties mentioned in Section
%\ref{cftcompare}, namely that our $1/c$ expansion for the DSF 
%is uniform is $q,\omega$, and that there are no subleading corrections
%to \eqref{NLL} with vanishing exponent for $c\to\infty$.
%{\color{red}
%We note that
%contrarily to real space, where the analogous version of the second
%property was previously shown to be satisfied
%\cite{kitanineetcformfactor}, this has not been investigated to our
%knowledge in Fourier space. 
%
%We observe that \eqref{dsflog1} and \eqref{dsflog2} agree with the
%$1/c$ expansion of \eqref{NLL} at the lower threshold and above the
%upper threshold. However it does not agree below the upper threshold,
%which means that one of the two properties is not satisfied. Since we
%have strong evidence for the uniformity of the $1/c$ expansion of the
%DSF -- detailed balance, sum rules, high frequency tail are all
%perfectly recovered at order $c^{-2}$ -- we conclude that there has to
%be subleading corrections to \eqref{NLL} that are not subleading once
%expanded in $1/c$. We note that even a non-divergent and vanishing
%piece of the form $\propto |\omega-\omega_+|^{1/c}$ would yield
%logarithms in a $1/c$ expansion, on the same footing as those in
%\eqref{dsflog1} and \eqref{dsflog2}, whereas it would be subleading
%even in front of a constant in \eqref{NLL}. To confirm this
%conclusion, one would have to compute the higher orders in $c^{-1}$ in
%\eqref{dsflog1} and \eqref{dsflog2}.
%}

%%%%%%%%%%%%%%%%%%%%%%%%%%%%%%%%%%%%%%%%%%%%%
\subsubsection{Sum rule at zero temperature}
%%%%%%%%%%%%%%%%%%%%%%%%%%%%%%%%%%%%%%%%%%%%%
The f-sum rule for the dynamical structure factor in equilibrium
states reads\cite{PSbook}
\begin{equation}
\int_{-\infty}^\infty S(q,\omega)\omega \D{\omega}=2\pi Dq^2\ .
\end{equation}
In our calculation, this sum rule has to be perfectly satisfied at
order $c^{-2}$. It is a stringent test of validity of our formula
since it has to be satisfied for all $q$ and encompasses every single
piece of the DSF. At zero temperature we obtain from Equation
\eqref{dsf1pht0} that
\begin{equation}
\int_{-\infty}^\infty S^{(1)}(q,\omega)\omega \D{\omega}=2\pi
Dq^2+\frac{4}{3c^2}q^4Q+{\cal O}(c^{-3})\, .
\end{equation}
This means that the two-particle-hole DSF \eqref{dsf2pht0}
$\bar{S}^{(2)}(q,\omega)\equiv c^2 S^{(2)}(q,\omega)$ evaluated at
$c=\infty$, must satisfy 
\begin{equation}\label{tobesat}
\int_{-\infty}^\infty \bar{S}^{(2)}(q,\omega)\omega \D{\omega}=-\frac{4}{3}q^4Q\,.
\end{equation}
We computed this integral numerically from \eqref{dsf2pht0} for
several values of $q$. We find that \fr{tobesat} is indeed
satisfied within the numerical accuracy of our calculation.
The relative deviations of our results from \fr{tobesat} are around
$10^{-4}$, which is quite satisfactory.

%%%%%%%%%%%%%%%%%%%%%%%%%%%%%%%%%%%%%%%%%%%%%%%%%%%%
\subsubsection{Detailed balance for thermal states}
%%%%%%%%%%%%%%%%%%%%%%%%%%%%%%%%%%%%%%%%%%%%%%%%%%%%
The dynamical structure factor of a thermal state at inverse
temperature $\beta$ should satisfy the detailed balance relation for
all values of $q,\omega$
\begin{equation}\label{DBf}
S(q,-\omega)=e^{-\beta\omega}S(q,\omega)\,.
\end{equation}
In our calculation the detailed balance relation for $S(q,\omega)$
should be perfectly satisfied at order $c^{-2}$. We note it is a very
stringent test of validity of our formulas for $S(q,\omega)$, given
that a thermal state at finite temperature corresponds to a
generic root density with a complicated $c$ dependence, while
for an arbitrary root density there is no particular relation between
$S(q,-\omega)$ and $S(q,\omega)$.

In order to check that our formulas for the DSF satisfy detailed
balance at order $c^{-2}$, we need to evaluate \eqref{DBf} with
$\rho(\lambda)$ given at order $c^{-2}$ in \eqref{rhotemperature}. We
found convenient to define 
\begin{equation}
\tilde{S}^{(1)}(q,\omega)=S^{(1)}(q,\omega)-\frac{8\pi^4|q'|}{c^2}\rho(\tfrac{\omega'-q^{\prime
    2}}{2q'})\rho(\tfrac{\omega'+q^{\prime
    2}}{2q'})\rho_h(\tfrac{\omega'+q^{\prime 2}}{2q'})^2\,, 
\end{equation}
i.e. the one-particle-hole DSF without the dressed piece coming from
two particle-hole excitations. We recall that $\omega'$ and $q'$ were
previously defined in \fr{opqp}. It is straightforward to
check numerically that $\tilde{S}^{(1)}(q,\omega)$ satisfies
detailed balance at order $c^{-2}$   
\begin{equation}
\tilde{S}^{(1)}(q,-\omega)=e^{-\beta\omega}\tilde{S}^{(1)}(q,\omega)+{\cal O}(c^{-3})\,.
\end{equation}
Hence the following quantity
\begin{equation}
\tilde{S}^{(2)}(q,\omega)=c^2S^{(2)}(q,\omega)+8\pi^4|q'|\rho(\tfrac{\omega'-q^{\prime
    2}}{2q'})\rho(\tfrac{\omega'+q^{\prime
    2}}{2q'})\rho_h(\tfrac{\omega'+q^{\prime 2}}{2q'})^2\,, 
\end{equation}
evaluated at $c=\infty$ should also independently satisfy detailed balance
\begin{equation}
\tilde{S}^{(2)}(q,-\omega)=e^{-\beta\omega}\tilde{S}^{(2)}(q,\omega)\,.
\end{equation}
We find that this indeed holds within the accuracy of our
numerical computation, i.e. within a relative error of $10^{-5}$. This
is quite satisfactory.

%%%%%%%%%%%%%%%%%%%%%%%%%%%%%%%%%%%%%%%%%%%%%%%%%%%%%
\subsubsection{\sfix{Behaviour at small $q,\omega$}}
%%%%%%%%%%%%%%%%%%%%%%%%%%%%%%%%%%%%%%%%%%%%%%%%%%%%%
At small $q,\omega$ with fixed
\begin{equation}
\gamma=\frac{\omega}{2q}\,,
\end{equation}
we find the following behaviour of the DSF

\begin{align}
S(q,\omega)&=\frac{2\pi^2\gamb}{|q'|}\rho\left(\gamma'-\tfrac{q'}{2} \right)\rho_h\left(\gamma'+\tfrac{q'}{2} \right)\nn
&+\frac{8\pi^2}{c^2}\int_{-\infty}^\infty\int_{-\infty}^\infty\frac{\rho(u)\rho_h(u)}{(\gamma'-\lambda)^2}\Big[\sign(u-\lambda)(2\gamma'+u-3\lambda)\rho(\lambda)\rho_h(\lambda)\\
&\qquad\qquad\qquad\qquad\qquad\qquad\qquad\qquad-|\gamma'-u|\rho(\gamma')\rho_h(\gamma')\Big]\D{\lambda} \D{u}\nn
&+o(q^0)\,.
%&+\frac{8\pi^2}{c^2}\int_{-\infty}^\infty \rho(\lambda)\rho_h(\lambda)\rho(u)\rho_h(u)\frac{\sign(u-\lambda)}{\gamma'-\lambda}\left[2+\frac{u-\lambda}{\gamma'-\lambda}\left(1-\left|\frac{\gamma'-u}{\lambda-u}\right| \right)\right]\D{\lambda} \D{u}\nn
%&+\frac{8\pi^2}{c^2}\left(\int_{-\infty}^\infty \rho(u)\rho_h(u)|u-\gamma'|\D{u}\right)\left(\dashint_{-\infty}^\infty \frac{\rho(u)\rho_h(u)-\rho(\gamma')\rho_h(\gamma')}{(u-\gamma')^2}\D{u}\right)+o(q^0)\,.
\label{smallq}
\end{align}
We have set
\begin{equation}
\gamma'=\frac{\omega'}{2q'}\,.
\end{equation}
Here the term proportional to $1/|q|$ arises only from the
one particle-hole contribution, while the constant term is due to two particle-hole
excitations. This result can again be compared to GHD predictions,
which at order $c^{-2}$ give  \cite{CADY16,BCDF16,Doyon18}
\begin{equation}
S_{\rm GHD}(q,\omega)=\frac{2\pi^2\gamb^2}{|q|}
\rho\Big(\gamma(1+\frac{2D}{c})\Big)\rho_h\Big(\gamma(1+\frac{2D}{c})\Big)\, .
\end{equation}
This is indeed in agreement with the leading term in \fr{smallq}
at small $q$. It would be interesting to see whether the subleading
terms in \fr{smallq} can be obtained by considering corrections to GHD
following Ref.~\cite{deNBD19}.

%%%%%%%%%%%%%%%%%%%%%%%%%%%%%%%%%%%%
\subsubsection{High frequency tail}
%%%%%%%%%%%%%%%%%%%%%%%%%%%%%%%%%%%%
Finally we consider the large-$\omega$ behaviour of the DSF
$S(q,\omega)$ (at fixed $q$) in an arbitrary eigenstate
$|\pmb{\lambda}\rangle$ with a root density $\rho(\lambda)$ that 
decays faster than any power law $|\lambda|^{-n}$ at infinity. For
such states we find

\be
S(q,\omega)=\frac{32\sqrt{2}\qp^4}{c^2{\omegap}^{7/2}}(\varepsilon D-\delta^2)+{\cal O}({\omegap}^{-9/2})\,,
\label{tail}
\ee
where $\varepsilon=\int u^2\rho(u)\D{u}$ is the energy of the state and
$\delta$  its momentum defined in \eqref{delta}. The result \fr{tail} arises
entirely from the two particle-hole contribution since the
one particle-hole contribution decays faster than any power in
$\omega$ for $\omega\to\infty$. The corrections to this leading
behaviour can all be computed and expressed as a series in
${\omegap}^{-1/2}$. For example the next term is 

\begin{align}
&\frac{4\sqrt{2}\qp^4}{c^2{\omegap}^{9/2}}\int_{-\infty}^\infty\int_{-\infty}^\infty(u-v)^2[(u-v)^2+14\qp(u-v)+15\qp^2+28\qp
    v]\rho(u)\rho(v)\D{u}\D{v}\nn
&+{\cal O}({\omegap}^{-11/2})\,.
\end{align}

For eigenstates $|\pmb{\lambda}\rangle$ corresponding to root
densities that instead decay like a power law at infinity it is
straightforward to see that the one particle-hole contribution to the
DSF decays at large $\omega$ with the same power-law. For such root
densities the large-$\omega$ expansion of the two particle-hole
contribution breaks down at some order because the coefficients would
diverge.  

It has been shown some time ago that the large $\omega$ behaviour of
the DSF $S(q,\omega)$ in equilibrium states is universal, with a
decay $\propto q^4\omega^{-7/2}$ for quantum fluids with short-range
interactions \cite{Wong74,Wong77,Kirkpatrick84}. This behaviour was
also observed to be in good agreement with scattering
experiments\cite{Wong77}. Our result \fr{tail} is in perfect agreement
with these findings, which again confirms that our $1/c$ expansion is
indeed uniform in $q,\omega$.  

%%%%%%%%%%%%%%%%%%%%%%%%
\section{Conclusions}
%%%%%%%%%%%%%%%%%%%%%%%%
In this work we have introduced and developed an \emph{ab initio}
expansion of dynamical density-density correlation functions in the
Lieb-Liniger model that can be performed within any energy
eigenstate. It is a combined expansion in
$1/c$ and in the number of particle-hole excitations taken into
account in the spectral representation of the dynamical
correlation function. The expansion has a well-defined
thermodynamic limit and is uniform in all $x$ and $t$, or 
equivalently all $\omega$ and $q$. We have obtained fully
explicit and readily usable expressions for both the correlator
and the dynamical structure factor at order ${\cal O}(c^{-2})$ which
take into account all one- and two particle-hole excitations, Equations \eqref{corregenalt}, \eqref{chi2}, \eqref{DSF1ph} and \eqref{DSF2ph2}. 

The main obstacle we faced in deriving these results occurs at
order ${\cal O}(c^{-2})$. Indeed, the leading ${\cal O}(c^0)$ term of
the expansion is simply the result for impenetrable bosons, which can
be straightforwardly obtained using the mapping to free fermions
\cite{Creamer80,korepin}. In terms of the form factor expansion
the only non-zero form factors are those involving a single
particle-hole excitation, and they are all equal.
The ${\cal O}(c^{-1})$ term is almost as simple since its form
factor expansion is identical to the impenetrable limit case albeit with a
root density dependent numerical modification of the form factors. In contrast the
${\cal O}(c^{-2})$ contribution comes with a number of complications.
%which is supposedly why such an expansion has not been carried out
%before.  

As is well-known the form factor expansion generally exhibits
non-integrable singularities whenever two rapidities coincide. In the
framework of the $1/c$ expansion these first arise at order ${\cal  O}(c^{-2})$
for contributions involving both one- and two particle-hole
excitations. The presence of such singularities precludes
directly taking the thermodynamic limit and expressing the spectral
sum as integrals over root densities in a simple way. Indeed, we
find that the contributions from both one- and two particle-hole
excitations are individually divergent in the thermodynamic limit, but
their sum is not. Even after compensating the divergent parts
they individually depend on the particular choice of representative
state and cannot be expressed in terms of the root densities. But
remarkably, and reassuringly, their sum -- and thus the
correlation function -- is representative-state-independent, i.e. depends only on the root density. These
cancellations eventually leave a piece that can be interpreted as a  
dressing of the contribution due to one particle-hole excitations by
two particle-hole excitations. Although this vanishes for the
zero-temperature ground state as well as for any zero-entropy states
it is non-zero in general and is crucial for detailed balance to
be satisfied in thermal states. Such a fine-tuned ``regularisation" of
the divergences could only be achieved with a careful treatment of the
thermodynamic limit of the exact spectral sum in a finite
volume. Anticipating that for other quantities the
representative-state-dependent parts may not always compensate
one another we derived a formula for their average over all
representative states for a given root density.  

We have verified that our results are in full accord with known results
including CFT and (non-linear) Luttinger liquid theory predictions
for zero-temperature critical exponents, thresholds singularities, sum rules, detailed balance
  relations and high frequency behaviour.
We have also recovered the order ${\cal O}(c^{-2})$ GHD predictions for
Euler scale density correlations in finite entropy states. This
constitutes the most non-trivial verification of GHD in an interacting
integrable model. We have also determined corrections to GHD
and compared the GHD result to the full correlator at order ${\cal
  O}(c^{-2})$ outside the asymptotic regime. We found that GHD
provides a rather good description of the correlator even at
  short times and distances.

The framework developed in this work is not restricted to density
correlations in the Lieb-Liniger model but is expected to apply to any
\emph{local} operator in any integrable model that has a
well-behaved expansion around a strong coupling limit. One example
is the large anisotropy regime of the spin-1/2 Heisenberg XXZ chain
\cite{CE20,JGE09}. A significant complication that occurs in that case is
the presence of string solutions to the Bethe Ansatz equations.
The restriction to local operators is crucial as the
spectral representation of two-point functions of semi-local
operators such as the field $\psi(x)$ are dominated by a completely
different set of excited states \cite{GFE20} and does not allow
for an expansion in the number of particle-hole-excitations.  

Our work opens up several interesting lines of further enquiry.
First, our analysis should be extended to higher orders
in the expansion. The ${\cal O}(c^{-3})$ term still involves at most
two particle-hole excitations, but the expansions of the Bethe
equations and the determinant in the expression for the form factors
become more involved. Second, the repulsive Lieb-Liniger model is
particularly simple in that the Bethe equations have only real
  roots. It would be very interesting to extend our analysis to a
model with complex roots, e.g. the spin-1/2 Heisenberg XXZ
chain. Third, our framework is readily generalized to quench 
dynamics \cite{TBP} by combining it with the quench action approach
\cite{CE13,caux16}. Here the novel feature is that the spectral sum
involves ``overlaps" that multiply the form factors. Finally, it
would be interesting to recover results obtained from the
$1/c$-expansion considering corrections to GHD as well as the thermodynamic
bootstrap program \cite{CCP20}.

%%%%%%%%%%%%%%%%%%%%%%%%%%%%%%%%%%%
\paragraph{Acknowledgements}
%%%%%%%%%%%%%%%%%%%%%%%%%%%%%%%%%%%
We are grateful to Jean-S\'ebastien Caux, Jacopo de Nardis and Karol
Kozlowski for very helpful discussions and comments. This work was
supported by the EPSRC under grant EP/S020527/1.

\begin{appendix}
\section{Double principal values}

\subsection{\sfix{Proof of Equation \eqref{PB} \label{proofPB}}}

We start by recalling that a single principal value can be expressed
as a regular integral 
\begin{equation}\label{eqprincieq}
\dashint \frac{F(\lambda,\mu)}{\lambda-\mu}\D{\lambda}=\frac{1}{2}\int
\frac{F(\lambda,\mu)-F(-\lambda+2\mu,\mu)}{\lambda-\mu}\D{\lambda}\, .
\end{equation}
Hence successive principal value triple integrals can be written as
\begin{align}
\label{eqintermed}
&\dashint \frac{F(\lambda,\mu,\nu)}{(\lambda-\mu)(\mu-\nu)}\D{\lambda} \D{\mu} \D{\nu}=\nn
&\frac{1}{4}\iiint \left[\frac{F(\lambda,\mu,\nu)-F(2\mu-\lambda,\mu,\nu)}{(\lambda-\mu)(\mu-\nu)}-\frac{F(\lambda,\mu,2\mu-\nu)-F(2\mu-\lambda,\mu,2\mu-\nu)}{(\lambda-\mu)(\mu-\nu)}\right]\D{\lambda} \D{\mu} \D{\nu}\,.
\end{align}
If $G(\lambda,\mu,\nu)$ is a function without singularities, then we have
\begin{equation}\label{GG}
\iiint G(\lambda,\mu,\nu)\D{\lambda} \D{\mu} \D{\nu}=\underset{L\to\infty}{\lim}\, \frac{1}{L^3}\sum_{i,j,k} G(x_i,x_j,x_k)\,,
\end{equation}
where
\begin{equation}
x_i=\frac{i}{L}\,,
\end{equation}
and $i,j,k$ range e.g. between $-L^2$ and $L^2$. In \fr{GG} one
has the freedom to exclude some values, e.g. consider $i\neq j$,
since this only amounts to subleading corrections in $L$ that
vanish when taking the limit. The integrand of \eqref{eqintermed}
is of this type. Hence one can write 
\begin{align}
\dashint\frac{F(\lambda,\mu,\nu)}{(\lambda-\mu)(\mu-\nu)}\D{\lambda} \D{\mu} \D{\nu}&=\frac{1}{4}\underset{L\to\infty}{\lim}\, \frac{1}{L^3}\sum_{\substack{i,j,k\\ i\neq j\\ j\neq k}}\Big[\frac{F(x_i,x_j,x_k)-F(-x_i+2x_j,x_j,x_k)}{(x_i-x_j)(x_j-x_k)}\nn
&-\frac{F(x_i,x_j,2x_j-x_k)-F(2x_j-x_i,x_j,2x_j-x_k)}{(x_i-x_j)(x_j-x_k)} \Big]\,.
\end{align}
Separating the four sums and changing variables so that the
argument of $f$ is always $x_i,x_j,x_k$ leads to
\begin{align}
\dashint\frac{F(\lambda,\mu,\nu)}{(\lambda-\mu)(\mu-\nu)}\D{\lambda} \D{\mu} \D{\nu}&=\underset{L\to\infty}{\lim}\, \frac{1}{L^3}\sum_{\substack{i,j,k\\ i\neq j\\ j\neq k}}\frac{F(x_i,x_j,x_k)}{(x_i-x_j)(x_j-x_k)}\,.
\end{align}
Finally we turn this into a simultaneous principal value
integral by adding the condition $i\neq k$
\be
\dashint\frac{F(\lambda,\mu,\nu)}{(\lambda-\mu)(\mu-\nu)}\D{\lambda} \D{\mu} \D{\nu}=\ddashint\frac{F(\lambda,\mu,\nu)}{(\lambda-\mu)(\mu-\nu)}\D{\lambda} \D{\mu} \D{\nu}-\underset{L\to\infty}{\lim}\, \frac{1}{L^3}\sum_{\substack{i,j\\ i\neq j}}\frac{F(x_i,x_j,x_i)}{(x_i-x_j)^2}\,.
\ee
Using $\sum_{i\neq 0}\frac{1}{i^2}=\tfrac{\pi^2}{3}$, we have
\begin{equation}
\underset{L\to\infty}{\lim}\, \frac{1}{L^3}\sum_{\substack{i,j\\ i\neq j}}\frac{F(x_i,x_j,x_i)}{(x_i-x_j)^2}=\frac{\pi^2}{3}\int F(x,x,x)\D{x}\,,
\end{equation}
and obtain Equation \eqref{PB}.

%%%%%%%%%%%%%%%%%%%%%%%%%%%%%%%%%%%%%%%%%%%%%
\subsection{\sfix{Proof of Equation \eqref{dashint33}}}
\label{dashint22p}
%%%%%%%%%%%%%%%%%%%%%%%%%%%%%%%%%%%%%%%%%%%%%
We note that formulae \eqref{dashint22} are direct consequences of
Equation \eqref{dashint33}, as can be seen by interchanging the
dummy variables. 

We start with representation \eqref{eqintermed}. As the integrand is
regular one can impose that $|\lambda-\mu|>\epsilon$ and
$|\mu-\nu|>\epsilon'$ with an error ${\cal O}(\epsilon)+{\cal
  O}(\epsilon')$. This allows one to separate the integral into
four pieces and make appropriate changes of variables so that 
the argument of $F$ is always $\lambda',\mu',\nu'$. One sees that
in the four cases one has  $|\lambda'-\mu'|>\epsilon$ and
$|\mu'-\nu'|>\epsilon'$. Hence  
\begin{equation}
\begin{aligned}
&\dashint \frac{F(\lambda,\mu,\nu)}{(\lambda-\mu)(\mu-\nu)}\D{\lambda} \D{\mu} \D{\nu}=\int_{\substack{|\lambda'-\mu'|>\epsilon\\|\nu'-\mu'|>\epsilon'}} \frac{F(\lambda',\mu',\nu')}{(\lambda'-\mu')(\mu'-\nu')}\D{\lambda'} \D{\mu'} \D{\nu'}+{\cal O}(\epsilon)+{\cal O}(\epsilon')\,,
\end{aligned}
\end{equation}
which is precisely \eqref{dashint33}.

\section{\sfix{Proof of Equation \eqref{thmregula} \label{proofthm}}}

\subsection{Reduction to a combinatorial problem}
For a given solution to the Bethe equations
$\{\lambda_i\}_i\in\mathfrak{S}_L$ we define the set of pairs of
rapidities that belong to the same bin
\begin{equation}
B=\Big\{(\lambda_i,\lambda_j)\Big|\,\, i\neq j\,,\,\exists k\in\{1,...,n_L\},\, \lambda_i,\lambda_j\in [x_{L,k},x_{L,k+1}]\Big\}\,,
\end{equation}
We have
\begin{equation}
\frac{1}{L^3}\sum_{i\neq j} \frac{f(\lambda_i,\lambda_j)}{(\lambda_i-\lambda_j)^2}=\frac{1}{L^3}\sum_{(\lambda_i,\lambda_j)\in B} \frac{f(\lambda_i,\lambda_j)}{(\lambda_i-\lambda_j)^2}+\frac{1}{L^3}\sum_{(\lambda_i,\lambda_j)\notin B} \frac{f(\lambda_i,\lambda_j)}{(\lambda_i-\lambda_j)^2}\,.
\end{equation}
Let us show that when the pairs of rapidities are not in $B$, the sum
is negligible. We observe that
\begin{equation}
\frac{1}{L^3}\sum_{\substack{\lambda_i\in
    [x_{L,k},x_{L,k+1}]\\ \lambda_j\in [x_{L,p},x_{L,p+1}]}}
\frac{|f(\lambda_i,\lambda_j)|}{(\lambda_i-\lambda_j)^2}\leq
\frac{(L\epsilon_L)^2}{(|k-p|-1)^2L^3}
\frac{{\rm max_{\lambda,\lambda'}}|f(\lambda,\lambda')|}{C_0^2\epsilon_L^2}\,,
\end{equation}
provided that $|k-p|>1$, i.e. if the bins to which $\lambda_i$ and
$\lambda_j$ belong are not adjacent. Here $C_0={\rm
  min}_y[\rho(y)]^{-1}$ is a constant independent of the
representative state and of the bins. Indeed, in this case we have
$|\lambda_i-\lambda_j|>(|k-p|-1)C_0\epsilon_L$ and there are
$(L\epsilon_L)^2$ pairs of roots. Since there are $D/\epsilon_L$ bins,
by summing over $p$ and $k$ these contributions are ${\cal
  O}(\tfrac{1}{L\epsilon_L})$, and since $L\epsilon_L\to\infty$, they
are negligible in the thermodynamic limit.  

If the bins are adjacent we have
\begin{equation}
\begin{aligned}
\frac{1}{L^3}\sum_{\substack{\lambda_i\in [x_{L,k},x_{L,k+1}]\\ \lambda_j\in [x_{L,k+1},x_{L,k+2}]}} \frac{|f(\lambda_i,\lambda_j)|}{(\lambda_i-\lambda_j)^2}&\leq \frac{C_1}{L^3}\sum_{\substack{0\leq n,m\leq L\epsilon_L\\ n+m\neq 0}}\frac{1}{(n+m)^2/L^2}\
&={\cal O}(\tfrac{\log (L\epsilon_L)}{L})\,,
\end{aligned}
\end{equation}
with $C_1$ another constant independent of the representative state and of the bins. Since there are $D/\epsilon_L$ bins and $L\epsilon_L\to\infty$, these contributions are also negligible in the thermodynamic limit. Hence we have
\begin{equation}
\frac{1}{L^3}\sum_{i\neq j} \frac{f(\lambda_i,\lambda_j)}{(\lambda_i-\lambda_j)^2}=\frac{1}{L^3}\sum_{(\lambda_i,\lambda_j)\in B} \frac{f(\lambda_i,\lambda_j)}{(\lambda_i-\lambda_j)^2}+o(L^0)\,,
\end{equation}
with the $o(L^0)$ being independent of the representative state. Hence we also have 
\begin{equation}
\underset{L\to\infty}{\lim}\,\frac{1}{|\mathfrak{S}_L|}\sum_{\{\lambda_i\}_i\in \mathfrak{S}_L}\frac{1}{L^3}\sum_{i\neq j} \frac{f(\lambda_i,\lambda_j)}{(\lambda_i-\lambda_j)^2}=\underset{L\to\infty}{\lim}\,\frac{1}{|\mathfrak{S}_L|}\sum_{\{\lambda_i\}_i\in \mathfrak{S}_L}\frac{1}{L^3}\sum_{(\lambda_i,\lambda_j)\in B} \frac{f(\lambda_i,\lambda_j)}{(\lambda_i-\lambda_j)^2}\,.
\end{equation}
Writing
\begin{equation}
\sum_{(\lambda_i,\lambda_j)\in B} \frac{f(\lambda_i,\lambda_j)}{(\lambda_i-\lambda_j)^2}=\sum_{k=1}^{n_L}\sum_{\substack{i\neq j\\\lambda_i,\lambda_j\in [x_{L,k},x_{L,k+1}]}} \frac{f(\lambda_i,\lambda_j)}{(\lambda_i-\lambda_j)^2}\,,
\end{equation}
we have
\begin{equation}
\frac{1}{|\mathfrak{S}_L|}\sum_{\{\lambda_i\}_i\in \mathfrak{S}_L}\frac{1}{L^3}\sum_{(\lambda_i,\lambda_j)\in B} \frac{f(\lambda_i,\lambda_j)}{(\lambda_i-\lambda_j)^2}=\sum_{k=1}^{n_L}\frac{1}{|\mathfrak{S}_L|}\sum_{\{\lambda_i\}_i\in \mathfrak{S}_L}\frac{1}{L^3}\sum_{\substack{i\neq j\\\lambda_i,\lambda_j\in [x_{L,k},x_{L,k+1}]}} \frac{f(\lambda_i,\lambda_j)}{(\lambda_i-\lambda_j)^2}\,.
\end{equation}
To go further and decouple the average over the representative states
one needs to ensure that the modification of rapidities in one bin
does not notably affect the distance between rapidities in another
bin. From the Bethe equations, a modification of order $\epsilon_L$ of
$DL$ rapidities modifies the distance between two rapidities $i,j$ in
the same bin by an order
$\frac{1}{L}DL\epsilon_L(\lambda_i-\lambda_j)$, which is indeed
subleading compared to $\lambda_i-\lambda_j$. Hence one can assume at
leading order in $L$ that the rapidities decouple, and given a bin
$k$, sum over the rapidities of the other bins without modifying the
values of the rapidities inside bin $k$. Thus one can write 
\begin{align}
&\frac{1}{|\mathfrak{S}_L|}\sum_{\{\lambda_i\}_i\in \mathfrak{S}_L}\frac{1}{L^3}\sum_{\substack{i\neq j\\\lambda_i,\lambda_j\in [x_{L,k},x_{L,k+1}]}} \frac{f(\lambda_i,\lambda_j)}{(\lambda_i-\lambda_j)^2}\nn
  &\qquad\qquad=\frac{1}{\binom{K_{L,k}}{\lfloor L\epsilon_L \rfloor}}
\sum_{\{\lambda_i\}_i\in \mathfrak{S}_L^k}\frac{1}{L^3}\sum_{\substack{i\neq j\\\lambda_i,\lambda_j\in [x_{L,k},x_{L,k+1}]}} \frac{f(\lambda_i,\lambda_j)}{(\lambda_i-\lambda_j)^2}+o(L^0)\,,
\end{align}
with $\mathfrak{S}_L^k\subset \mathfrak{S}_L$ the subset of $\mathfrak{S}_L$ containing states whose rapidities outside the bin $[x_{L,k},x_{L,k+1}]$ are fixed to those of an arbitrary representative state, and
\begin{equation}
K_{L,i}=\lfloor L (x_{L,i+1}-x_{L,i})(\rho(x_{L,i})+\rho_h(x_{L,i}))\rfloor\,.
\end{equation}
is the number of vacancies in $[x_{L,i},x_{L,i+1}]$.
Since around $\lambda$ two consecutive vacancies are separated by $\tfrac{1}{L(\rho(\lambda)+\rho_h(\lambda))}$ at leading order in $L$, we can write $\lambda_i-\lambda_j$ as an integer times $\tfrac{1}{L(\rho(x_{L,k})+\rho_h(x_{L,k}))}$, for $\lambda_i$ and $\lambda_j$ in the same bin $[x_{L,k},x_{L,k+1}]$. This yields
\begin{align}
\label{intermed}
&\frac{1}{|\mathfrak{S}_L|}\sum_{\{\lambda_i\}_i\in \mathfrak{S}_L}\frac{1}{L^3}\sum_{(\lambda_i,\lambda_j)\in B} \frac{f(\lambda_i,\lambda_j)}{(\lambda_i-\lambda_j)^2}=\nn
  &\qquad\qquad\frac{1}{L}\sum_{k=1}^{n_L}\frac{f(x_{L,k},x_{L,k})}{
    \binom{K_{L,k}}{\lfloor L\epsilon_L \rfloor}}(\rho(x_{L,k})+\rho_h(x_{L,k}))^2\sum_{\substack{I\subset \{1,...,K_{L,k}\}\\ |I|=\lfloor L\epsilon_L\rfloor}} \sum_{\substack{i,j\in I\\ i\neq j}}\frac{1}{(i-j)^2}\nn
&\qquad\qquad+o(L^0)\,.
\end{align}
This reduces the problem to evaluating the large $K,M$ limit at fixed $K/M$ of the following combinatorial quantity
\begin{equation}
\mathcal{C}_{M,K}=\sum_{\substack{I\subset \{1,...,K\}\\ |I|=M}} \sum_{\substack{i,j\in I\\ i\neq j}}\frac{1}{(i-j)^2}\,.
\end{equation}

\subsection{The generating functions}
To simplify the expression of $\mathcal{C}_{M,K}$, we would like to recast the sum over pairs of integers into a sum over (next-nearest-)neighbouring integers.  We exactly rewrite $\mathcal{C}_{M,K}$ in the form
\begin{equation}
\mathcal{C}_{M,K}=2\sum_{m=1}^M \sum_{j=1}^m  \sum_{\substack{a_1<...<a_M \\\in \{1,...,K\}}} \sum_{\substack{i\geq 0\\ j+(i+1)m\leq M}}\frac{1}{(a_{j+im}-a_{j+(i+1)m})^2}\,.
\end{equation}
Introducing
\begin{equation}
C^{[m]}_{M,K}=\sum_{\substack{a_1<...<a_M \\\in \{1,...,K\}}} \sum_{\substack{i\geq 0\\ (i+1)m\leq M}}\frac{1}{(a_{im}-a_{(i+1)m})^2}
\end{equation}
with $a_0=0$, we have
\begin{equation}\label{intermede}
\mathcal{C}_{M,K}=2\sum_{m=1}^M \sum_{j=1}^m \sum_{a=j}^K \binom{a-1}{j-1} C^{[m]}_{M-j,K-a}\,.
\end{equation}
Let us now determine the asymptotic behaviour of $C^{[m]}_{M,K}$ for large $K,M$ at fixed $K/M$.
Summing separately over $a_1,...,a_m$, one obtains the following recurrence relation
\begin{equation}
C_{M,K}^{[m]}=\sum_{a_m=m}^{K-M+m}\binom{a_m-1}{m-1}\left[\frac{\binom{K-a_m}{M-m}}{a_m^2}+C_{M-m,K-a_m}^{[m]} \right]\,,
\end{equation}
where we use conventions such that $C_{M,K}^{[m]}=0$ if $K<M$ or $M<m$. Indeed, the
factor $\binom{a_m-1}{m-1}$ counts the number of possibilities for the
first $m-1$ particles between $1$ and $a_m-1$, while the factor
$\binom{K-a_m}{M-m}$ counts the number of times this $1/a_m^2$ term
will appear in all the subsequent configurations for
$a_{m+1},...,a_M$. Introducing the generating functions 
\begin{equation}
\begin{aligned}
C^{[m]}(x,y)=\sum_{M,K\geq 0}C_{M,K}^{[m]}x^M y^K\,,\qquad S^{[m]}(x,y)=\sum_{M,K\geq 0}\sum_{a=m}^{K-M+m}\binom{a-1}{m-1}\frac{\binom{K-a}{M-m}}{a^2}x^M y^K\,,
\end{aligned}
\end{equation}
this recurrence relation implies that
\begin{equation}
C^{[m]}(x,y)=S^{[m]}(x,y)+\frac{x^my^m}{(1-y)^m}C^{[m]}(x,y)\,.
\end{equation}
Expressing $S^{[m]}(x,y)$ as
\begin{equation}
S^{[m]}(x,y)=\frac{x^m}{1-y(1+x)}\sum_{a\geq 1}\frac{\binom{a-1}{m-1}}{a^2}y^a\,,
\end{equation}
we obtain the following generating function
\begin{equation}
C^{[m]}(x,y)=\frac{x^m(1-y)^m}{(1-y(1+x))^2\sum_{k=0}^{m-1}(xy)^k(1-y)^{m-1-k}} \sum_{a\geq 1}\frac{\binom{a-1}{m-1}}{a^2}y^a\,.
\end{equation}

\subsection{Asymptotics of the coefficients}
We now use Ref. \cite{pemantle} which shows how to determine the
asymptotic behaviour of combinatorial coefficients from the analytic
behaviour of their generating function\footnote{Specifically, in order
  to have only a simple pole in the generating function as in 
  \cite{pemantle}, we define 
\begin{equation}
\bar{C}^{[m]}(x,y)=\int_0^x C^{[m]}(u,y)\D{u}\,.
\end{equation}
We then integrate by parts
\begin{equation}
\begin{aligned}
\bar{C}^{[m]}(x,y)&=\frac{x^m(1-y)^m}{(1-y(1+x))\sum_{k=0}^{m-1}(xy)^k(1-y)^{m-1-k}}\sum_{a\geq 1}\frac{\binom{a-1}{m-1}}{a^2}y^{a-1}\\
&-\int_0^x\D{u}\frac{1}{1-y(1+u)}\frac{d}{du}\frac{u^m(1-y)^m}{\sum_{k=0}^{m-1}(uy)^k(1-y)^{m-1-k}}  \sum_{a\geq 1}\frac{\binom{a-1}{m-1}}{a^2}y^{a-1}\,.
\end{aligned}
\end{equation}
and use Theorem 1.3 and Corrolary 3.21 of \cite{pemantle} on the first
term, where in their notations $x=\tfrac{M/K}{1-M/K}$ and
$y=1-M/K$. The second term gives negligible contributions because
the $x$ integral will give rise to a multiplicative factor $M^{-1}$.}. One
obtains 
%One now has to determine the asymptotic behaviour of $C_{K,N}^{(n)}$ when $K,N\to\infty$ at fixed $K/N$. To that end we define
%\begin{equation}
%\bar{C}^{(n)}(x,y)=\int_0^x C^{(n)}(u,y)du\,,
%\end{equation}
%that we write as, using an integration by part
%\begin{equation}
%\begin{aligned}
%\bar{C}^{(n)}(x,y)&=\frac{x^n(1-y)^n}{(1-y(1+x))\sum_{k=0}^{n-1}(xy)^k(1-y)^{n-1-k}}\sum_{a\geq 1}\frac{{a-1\choose n-1}}{a^2}y^{a-1}\\
%&-\int_0^x\frac{1}{1-y(1+u)}\frac{d}{du}\frac{u^n(1-y)^n}{\sum_{k=0}^{n-1}(uy)^k(1-y)^{n-1-k}}  \sum_{a\geq 1}\frac{{a-1\choose n-1}}{a^2}y^{a-1}\,.
%\end{aligned}
%\end{equation}
%The pole $1-y(1+x)$ being of order $1$, we can use \cite{pemantle} to obtain that the coefficient in $x^Ky^N$ of the first line is asymptotically when $K,N\to\infty$ at fixed $K/N$
%\begin{equation}
%{N\choose K}\left( \frac{K/N}{1-K/N}\right)^n \frac{1}{n}\sum_{a\geq 1}\frac{(1-K/N)^a}{a^2}{a-1\choose n-1}\,,
%\end{equation}
%while the coefficients of the second line are, because of the integration, ${N\choose K} O(K^{-1})$. We obtain that
\begin{equation}\label{asympcc}
C_{M,K}^{[m]}= M\binom{K}{M}\left( \frac{M/K}{1-M/K}\right)^m \frac{1}{m}\sum_{a\geq 1}\frac{(1-M/K)^a}{a^2}\binom{a-1}{m-1}+{\cal O}(\binom{K}{M})\,.
\end{equation}
%We now have to plug this expression in \eqref{intermede}.
This implies that
\begin{equation}
C_{M-j,K-a}^{[m]}=(\tfrac{M}{K})^j(1-\tfrac{M}{K})^{a-j}C_{M,K}^{[m]}+{\cal O}(\binom{K}{M})\,,
\end{equation}
and substituting this into \eqref{intermede} we obtain at
leading order 
\begin{equation}
\mathcal{C}_{M,K}=\left[2\sum_{m=1}^\infty mC^{[m]}_{M,K}\right](1+{\cal O}(M^{-1}))\,.
\end{equation}
Using the asymptotics \eqref{asympcc} one then finds in the limit
$M,K\to\infty$ at fixed $K/M$ 
\begin{equation}
\mathcal{C}_{M,K}=\frac{\pi^2}{3}M\frac{M}{K}\binom{K}{M}+{\cal O}(\binom{K}{M})\,.
\end{equation}
\subsection{Conclusion}
Coming back to \eqref{intermed}, we have when $L\to\infty$
\begin{equation}
M\sim L\epsilon_L\,,\qquad K\sim L\epsilon_L \frac{\rho(x_{L,k})+\rho_h(x_{L,k})}{\rho(x_{L,k})}\,,
\end{equation}
which yields
\begin{equation}
\begin{aligned}
&\frac{1}{|\mathfrak{S}_L|}\sum_{\{\lambda_i\}_i\in \mathfrak{S}_L}\frac{1}{L^3}\sum_{(\lambda_i,\lambda_j)\in B} \frac{f(\lambda_i,\lambda_j)}{(\lambda_i-\lambda_j)^2}\\
&\qquad\qquad\qquad=\epsilon_L \frac{\pi^2}{3}\sum_{k=1}^{n_L}f(x_{L,k},x_{L,k})(\rho(x_{L,k})+\rho_h(x_{L,k}))\rho(x_{k,L})+o(L^0)\,.
\end{aligned}
\end{equation}
In the limit $L\to\infty$ we then arrive at the result \fr{thmregula}
\begin{equation}
\underset{L\to\infty}{\lim}\,\frac{1}{|\mathfrak{S}_L|}\sum_{\{\lambda_i\}_i\in \mathfrak{S}_L}\frac{1}{L^3}\sum_{i\neq j} \frac{f(\lambda_i,\lambda_j)}{(\lambda_i-\lambda_j)^2}=\frac{\pi^2}{3}\int_{-\infty}^\infty f(\lambda,\lambda)(\rho(\lambda)+\rho_h(\lambda))\rho(\lambda)^2\D{\lambda}\,.
\end{equation}

\section{Derivations of the results presented in Section \ref{sec6}\label{dev6}}
\subsection{Correlation functions}
\subsubsection{Asymptotics of static correlators at
  zero-temperature\label{dervstat}} 
The study of the asymptotic behaviour of \eqref{corregenx} at large $x$ at zero temperature reduces to the asymptotics of the Fourier transform
\begin{equation}
\label{oscill}
\hat{f}(x)=\int_{-\infty}^\infty f(u) e^{-ixu}\D{u}\,,
\end{equation}
of a given function $f(u)$. These asymptotics depend on the regularity of the integrand, hence at leading order on points of non-analyticity of $f$ on the real axis. We have the following behaviours.
\begin{itemize}
\item If $f(u)$ has a discontinuity $\Delta=f(u_0^+)-f(u_0^-)$ at $u_0$ and is otherwise regular, then for $x\to\infty$
\begin{equation}
\label{decay}
\hat{f}(x)=\Delta  \frac{e^{-ixu_0}}{ix}+{\cal O}(x^{-2})\,.
\end{equation}
This is straightforwardly obtained with an integration by part.
\item If $f(u)\sim \Delta \log^n |u-u_0|$ for $u> u_0$ and is regular and bounded for $u< u_0$, then
\begin{equation}
\label{logdecay}
\hat{f}(x)=\begin{cases}
- \Delta  \frac{e^{-ixu_0}}{ix}\left(\log |x|+p_1\right)+o(\tfrac{1}{x})\qquad\text{if }n=1\\
 \Delta  \frac{e^{-ixu_0}}{ix}\left(\log^2 |x|+2p_1\log |x|+p_2\right)+o(\tfrac{1}{x})\qquad\text{if }n=2\,.
\end{cases}
\end{equation}
Here the constants $p_{1,2}$ are given by
\begin{equation}
\begin{aligned}
p_1&=\gamma_E+\frac{i\pi}{2}\sign(x)\\
p_2&=\gamma_E^2+i\pi\gamma_E\sign(x)-\frac{\pi^2}{12}\,,
\end{aligned}
\end{equation}
where $\gamma_{\rm E}$ is Euler's gamma constant. If $f(u)\sim \Delta \log^n |u-u_0|$ for $u< u_0$ and is regular and bounded for $u> u_0$, then the result is multiplied by $-1$ and $p_1,p_2$ are changed to their complex conjugates  $p_1^*,p_2^*$.

These relations are obtained from the relation
\begin{equation}
\int_0^\infty e^{-ixu}u^{\alpha}\D{u}=\Gamma(1+\alpha)[i(x-i0)]^{-1-\alpha}\,,
\end{equation}
 expanded around $\alpha=0$.

\item If $f(u)\sim \Delta \log^n |u-u_0|$ for both $u<u_0$ and $u>u_0$, then we have
\begin{equation}
\label{logdecay2}
\hat{f}(x)=\begin{cases}
- \pi\Delta  \frac{e^{-ixu_0}}{x}+o(\tfrac{1}{x})\qquad\text{if }n=1\\
 2\pi\Delta \frac{e^{-ixu_0}}{x}\left(\log |x|+\gamma_{\rm E}\right)+o(\tfrac{1}{x})\qquad\text{if }n=2\,.
 \end{cases}
\end{equation}
These equations directly follow from the previous results.
\end{itemize}

In order to determine the large $x$ behaviour of the correlation functions, we also need zero-temperature result for $\tilde\rho(x)$ defined in \fr{rhotilde}
\begin{equation}
\tilde{\rho}(\lambda)=\frac{1+2D/c}{2\pi}\log \left|\frac{\lambda+Q}{\lambda-Q} \right|\,,
\end{equation}
and the large $x$ behaviour of the functions $A_{x,0}$, $C_{x,0}$ and
$D_{x,0}$ defined in \fr{aintrep}, \fr{cintrep} and \fr{dintrep} respectively 
\begin{equation}
\begin{aligned}
A_{x,0}&=-\frac{\log |x|}{2\pi^2}+{\cal O}(x^0)\\
C_{x,0}&=o(x^{-1})\\
D_{x,0}&=o(x^{-1})\,.
\end{aligned}
\end{equation}
The asymptotics of $C_{x,0}$ and $D_{x,0}$ follow from
\eqref{largeC} and \eqref{largeD} for a generic root density. As for
$A_{x,0}$, integrating \fr{aintrep} by parts we obtain for a generic root
density 
\begin{equation}
A_{x,0}=\dashint \frac{\rho(u)\rho_h'(v)}{v-u}(e^{ix(v-u)}-1)\D{u}\D{v}+ixC_{x,0}\,.
\end{equation}
Specializing to zero temperature at leading order in $c^{-1}$, it yields
\begin{equation}
A_{x,0}=\frac{1}{4\pi^2}\int_{-Q}^Q \frac{e^{ix(Q-u)}-1}{Q-u}\D{u}-\frac{1}{4\pi^2}\int_{-Q}^Q \frac{e^{ix(-Q-u)}-1}{-Q-u}\D{u}+ixC_{x,0}\,,
\end{equation}
that is
\begin{equation}
\begin{aligned}
A_{x,0}&=\frac{1}{4\pi^2}\int_{0}^{2Qx}\frac{e^{iu}+e^{-iu}-2}{u}\D{u}+ixC_{x,0}\\
&=-\frac{\log |x|}{2\pi^2}+{\cal O}(x^0)\,.
\end{aligned}
\end{equation}
As for the $B_{x,0}(\lambda)$ and $B_{x,0}(\mu)$ terms, they require a special treatment since they cannot be decoupled from the $\lambda,\mu$ integrals. The $B_{x,0}(\lambda)$ term  involves the following functions
\begin{equation}\label{fn}
f_n(x)=\int_{-\infty}^\infty \rho(\lambda) \lambda^n B_{x,0}(\lambda) e^{-i\lambda x}\D{\lambda}\,,
\end{equation}
for $n=0,1,2$, whose we wish to determine the asymptotic behaviour at large $x$, by computing its Fourier transform $\hat{f}_n(q)=\int_{-\infty}^\infty e^{-iqx}f(x)\D{x}$.  We have (at leading order in $c^{-1}$)
\begin{equation}
\hat{f}_n(q)=2\pi\int_{-\infty}^\infty\int_{-\infty}^\infty\rho(\lambda)\rho(u) \lambda^n \frac{\tfrac{1}{2\pi}-\rho(u+\lambda+q)}{(\lambda+q)(\lambda-u)}\D{u}\D{\lambda}\,.
\end{equation}
Specializing this relation to the ground state root density we obtain
\begin{equation}
\hat{f}_n(q)=\frac{1}{4\pi^2}\int_{-Q}^Q\frac{\lambda^n}{|\lambda+q|}\log \left| \frac{\lambda+\min(Q,-Q+|\lambda+q|)\sign(\lambda+q)}{\lambda-Q\sign(\lambda+q)}\right| \D{\lambda}\,.
\end{equation}
We note that the non-integrable divergence near $\lambda=q$ is compensated by the argument of the $\log$ going to $1$ in this limit.
In the vicinity of $q=Q$ we have for $\eta>0$
\begin{equation}
\begin{aligned}
\hat{f}_n(Q+\eta)&=\frac{1}{4\pi^2}\int_{-Q}^Q\frac{\lambda^n}{\lambda+Q}\log \left|\frac{2\lambda}{\lambda-Q} \right|\D{\lambda}+o(\eta^0)\\
\hat{f}_n(Q-\eta)&=\frac{1}{4\pi^2}\int_{-Q}^Q\frac{\lambda^n}{\lambda+Q}\log \left|\frac{2\lambda}{\lambda-Q} \right|\D{\lambda}+\frac{(-Q)^n}{4\pi^2} \int_0^1 \frac{1}{v-1}\log \left|\frac{v}{2v-1} \right|\D{v}+o(\eta^0)\,,
\end{aligned}
\end{equation}
where the last integral is $\int_0^1 \frac{1}{v-1}\log \left|\frac{v}{2v-1} \right|\D{v}=-\frac{\pi^2}{12}$, so that $\hat{f}_n$ has a discontinuity at $Q$ of

\begin{equation}
\lim_{\eta\to 0}\left[\hat{f}_n(Q+\eta)-\hat{f}_n(Q-\eta)\right]=\frac{(-Q)^n}{48}\,.
\end{equation}
Similarly we find
\begin{equation}
\lim_{\eta\to 0}\left[\hat{f}_n(-Q+\eta)-\hat{f}_n(-Q-\eta)\right]=-\frac{Q^n}{48}\,,
\end{equation}
and $\hat{f}_n(q)$ does not have discontinuities elsewhere. This implies that
\begin{equation}
\begin{aligned}
f_0(x)&=-\frac{1}{48\pi}\frac{\sin(Q x)}{x}+o(x^{-1})\\
f_1(x)&=-\frac{iQ}{48\pi}\frac{\cos(Q x)}{x}+o(x^{-1})\\
f_2(x)&=-\frac{Q^2}{48\pi}\frac{\sin(Q x)}{x}+o(x^{-1})\,.
\end{aligned}
\end{equation}
Then one builds the full $B_{x,0}(\lambda)$ term from the functions \eqref{fn}, in particular by taking into account the remaining oscillatory $\mu$ integral. One obtains that the $B_{x,0}(\lambda)$ term gives contributions that decay as $\cos(2q_Fx)/x^2$ and of order $c^{-2}$, which are encapsulated in the result in the ${\cal O}(c^{-2})$ term of the $A$ in \eqref{AAA}. A similar analysis shows that the $B_{x,0}(\mu)$ term also gives contributions that decay as $\cos(2q_Fx)/x^2$.

From these various relations, it is straightforward albeit tedious to determine the asymptotics of the static correlation functions. Putting everything together we find that
$\chi^{(1,2)}_{x,0}(\lambda,\mu)$ given in \fr{chi2} contributes to the large-$x$
  behaviour of the density-density correlator as follows:

\begin{itemize}
\item{${\cal O}(c^0)$ contribution of $\chi^{(1)}_{x,0}(\lambda,\mu)$}
\begin{equation}
-\frac{1+\tfrac{4D}{c}+\tfrac{4D^2}{c^2}}{2\pi^2x^2}(1-\cos(2q_Fx))\,,
\end{equation}
\item{${\cal O}(c^{-1})$ contribution of $\chi^{(1)}_{x,0}(\lambda,\mu)$}
\begin{equation}
-\frac{4D(1+\tfrac{4D}{c})}{c\pi^2}\frac{\cos(2q_F x)}{x^2}\log|2q_Fe^{\gamma_{\rm E}}x|+o(x^{-2})\,,
\end{equation}
\item{${\cal O}(c^{-2})$ contribution of $\chi^{(1)}_{x,0}(\lambda,\mu)$}
\begin{equation}
\frac{16D^2}{\pi^2 c^{2}} \frac{\cos(2q_Fx)}{x^2}\log^2|2q_Fe^{\gamma_{\rm E}}x|+{\cal O}(\tfrac{\cos 2q_Fx}{x^2})\,,
\end{equation}
\item{Contribution of $\chi^{(2)}_{x,0}(\lambda,\mu)$}
\begin{equation}
-\frac{4D^2}{\pi^2 c^2} \frac{\cos(2q_Fx)}{x^2}\log|2q_Fe^{\gamma_{\rm E}}x|+{\cal O}(\tfrac{\cos 2q_Fx}{x^2})\,.
\end{equation}
\end{itemize}
This establishes \eqref{cftlarge}.

\subsubsection{\sfix{Asymptotics of dynamical correlations zero
    temperature}}

The study of the asymptotic behaviour of \eqref{corregen} at large
$x,t$ at fixed $\alpha=\tfrac{x}{2t}$ at zero temperature reduces to
the study of an oscillatory integral of the type 
\begin{equation}
I(x,t)=\int_{-\infty}^\infty f(u) e^{itu^2-ix'u}\D{u}\,.
\end{equation}
In this regime, the integral is dominated by the point where the phase
has an extremum as a function of $u$, which is $\alpha'$ defined in
\eqref{alphap}. If $f$ is regular and $\alpha'$ in the support of $f$,
then we have 
\begin{equation}
\label{oscalsallde}
I(x,t)=\frac{\sqrt{\pi}e^{i\sign(t)\pi/4}e^{-i\alpha^{\prime 2}
    t}}{\sqrt{|t|}}\left(f(\alpha')+\frac{i}{4t}f''(\alpha')-\frac{1}{32t^2}f''''(\alpha')\right)+{\cal
  O}(|t|^{-7/2})\,. 
\end{equation}
If $f$ has singular points one has to combine \fr{oscalsallde} with
the results of Section \ref{dervstat}. 

The correlation function  \eqref{corregen} is expressed as a double
integral, one over $\lambda$ with a factor $\rho(\lambda)$ and one
over $\mu$ with a factor $\rho_h(\mu)$. Because of the very particular
structure of $\rho(\lambda)$ at zero temperature \eqref{rho0t},
the saddle point $\alpha$ necessarily lies within the support of
either $\rho(\lambda)$ or $\rho_h(\mu)$, but not both. Hence if
$|\alpha|>q_F$, the $\lambda$ integral is dominated by boundary
effects as in the static case, while the $\mu$ integral is dominated
by the saddle point. If $|\alpha|<q_F$, the converse holds true.  

Let us detail the case $|\alpha|>q_F$ (the case $|\alpha|<q_F$ is
  very similar). We perform a change of variables
$\lambda\to\lambda+\alpha'$ and $\mu\to\mu+\alpha'$ in \eqref{corregen}
in order to move the saddle point to $0$, which results in
shifting the arguments of the root densities by $\alpha'$. The $\mu$
integral is then simply evaluated at $\mu=0$, while the $\lambda$
integral is dominated by the vicinities of the points $Q-\alpha'$
and $-Q-\alpha'$. Using the results for \eqref{oscill} with
$x=-2t(Q-\alpha')$ we obtain the leading contribution from
  $Q-\alpha'$ to the integral over $\chi^{(1)}_{x,t}(\lambda,\mu)$, with $\pm=\sign(t)$
\begin{align}
&\frac{e^{-i\sign(t)\frac{\pi}{4}}\gamb^4}{4\pi^\frac{3}{2}|t|^\frac{1}{2}}\frac{e^{it(Q-\alpha')^2}}{2it(Q-\alpha')}\bigg[1-\frac{4}{c}\frac{1+\frac{2D}{c}}{2\pi}(Q-\alpha')\Big[\log\big|4Qt\frac{(Q-\alpha')^2}{Q+\alpha'}\big|+\gamma_{\rm E}\mp i\frac{\pi}{2}\Big]\nn
&\qquad\qquad+\frac{8}{c^2}\left(\frac{1+\frac{2D}{c}}{2\pi}\right)^2(Q-\alpha')^2\Big[\log\big|4Qt\frac{(Q-\alpha')^2}{Q+\alpha'}\big|+\gamma_{\rm E}\mp  i\frac{\pi}{2}\Big]^2+{\cal O}(t^0c^{-2})\bigg]\,.
\end{align}

%Using the results for \eqref{oscill} with $x=2t(Q+\alpha)$, one has
%the $-Q-\alpha$ contribution of the one-particle-hole excitations
%$\chi^1_{x,t}$
The leading contribution from $-Q-\alpha'$ to the integral
  over $\chi^{(1)}_{x,t}(\lambda,\mu)$ is obtained analogously
\begin{align}
&\frac{e^{-i\sign(t)\frac{\pi}{4}}\gamb^4}{4\pi^\frac{3}{2}|t|^\frac{1}{2}}\frac{e^{it(Q+\alpha')^2}}{2it(Q+\alpha')}\bigg[1-\frac{4}{c}\frac{1+\frac{2D}{c}}{2\pi}(Q+\alpha')\Big[\log\big|4Qt\frac{(Q+\alpha')^2}{Q-\alpha'}\big|+\gamma_{\rm E}\pm i\frac{\pi}{2}\Big]\nn
&\qquad\qquad+\frac{8}{c^2}\left(\frac{1+\frac{2D}{c}}{2\pi}\right)^2(Q+\alpha')^2\Big[\log\big|4Qt\frac{(Q+\alpha')^2}{Q-\alpha'}\big|+\gamma_{\rm E}\pm  i\frac{\pi}{2}\Big]^2+{\cal O}(t^0c^{-2})\bigg]\,.
\end{align}

%\begin{equation}
%\begin{aligned}
%&\frac{e^{-\sign(t)i\pi/4}(1+2D/c)^4}{4\pi^{3/2}|t|^{1/2}}\frac{e^{it(Q+\alpha')^2}}{2it(Q+%\alpha')}\\
%&\qquad\times\Big[1-\frac{4}{c}\frac{1+2D/c}{2\pi}(Q+\alpha')[\log|4Qt\tfrac{(Q+\alpha')^2}%{Q-\alpha'}|+\gamma_{\rm E}+i\pi/2]\\
%&\qquad\qquad+\frac{8}{c^2}\left(\frac{1+2D/c}{2\pi}\right)^2(Q+\alpha')^2[\log|4Qt\tfrac{(%Q+\alpha')^2}{Q-\alpha'}|+\gamma_{\rm E}+i\pi/2]^2\\
%&\qquad\qquad+{\cal O}(t^0c^{-2})\Big]\,.
%\end{aligned}
%\end{equation}
In order to determine the asymptotic behaviour of
the two particle-hole contribution $\chi^{(2)}_{x,t}(\lambda,\mu)$ we
require the 
asymptotic behaviours of the functions $A_{2\alpha't,t}$,
$C_{2\alpha't,t}$ and $D_{2\alpha't,t}$ defined in \fr{aintrep},
\fr{cintrep} and \fr{dintrep} respectively. We find
\begin{align}
A_{2\alpha' t,t}&=-\frac{\log |t|}{2\pi^2}+{\cal O}(t^0)\ ,\nn
C_{2\alpha' t,t}&=o(t^{-1})\ ,\nn
D_{2\alpha' t,t}&=o(t^{-1})\,.
\end{align}
The results for $C_{2\alpha' t,t}$ and $D_{2\alpha' t,t}$ again follow from
\eqref{largeC} and \eqref{largeD} for a generic root density. As for
$A_{2\alpha' t,t}$, we integrate by parts to express it in the form
\begin{align}
A_{2\alpha' t,t}&=\dashint
\frac{\rho(u+\alpha')\rho_h'(v+\alpha')}{v-u}(e^{it(u^2-v^2)}-1)\D{u}\D{v}\nn
&-2it\iint\rho(u+\alpha')\rho_h(v+\alpha')e^{it(u^2-v^2)}\D{u}\D{v}-2itD_{2\alpha' t,t}\,.
\end{align}
A saddle point approximation on the second double integral shows that
the second line is ${\cal O}(t^0)$. Specializing to zero temperature
we then have at leading order in $c^{-1}$
\begin{align}
A_{2\alpha' t,t}&=\frac{1}{4\pi^2}\int_{0}^{2Q}\frac{\D{u}}{u}[e^{it(u^2-2u(Q-\alpha))}-1]\nn
&+\frac{1}{4\pi^2}\int_{0}^{2Q}\frac{\D{u}}{u}[e^{it(u^2-2u(Q+\alpha))}-1]+{\cal O}(t^0)\,,
\end{align}
which can be further simplified
\begin{align}
A_{2\alpha' t,t}&=\frac{1}{4\pi^2}\int_{0}^{2Qt}\frac{\D{u}}{u}[e^{i(u^2/t-2u(Q-\alpha))}-1]
+\frac{1}{4\pi^2}\int_{0}^{2Qt}\frac{\D{u}}{u}[e^{i(u^2/t-2u(Q+\alpha))}-1]+{\cal O}(t^0)\nn
&=-\frac{\log|t|}{2\pi^2}+{\cal O}(t^0)\,.
\end{align}

Finally there are the contributions of the $B_{x,t}(\lambda)$ and
$B_{x,t}(\mu)$ terms. One can perform an analysis similar to the
static case and obtain that they are both ${\cal O}(t^{-3/2})$.  Their
contributions are encapsulated in the result in the ${\cal O}(c^{-2})$
term of the $B_\pm$ in \eqref{resdynam}. 
% which as in the static case require special treatment because they cannot be decoupled from the integrals over $\lambda,\mu$. Regarding the $B_{x,t}(\lambda)$ term, we refer the reader to Appendix \ref{below2}, below Equation \eqref{below}, where the asymptotics of this term is determined for a generic continuous root density. In this case the dominant term is of order $t^{-2}$, and the presence of discontinuities in $\rho$ can only make this power more negligible, since saddle point effects are predominant. Hence we obtain that the  $B_{x,t}(\lambda)$ term is $o(t^{-3/2})$. 
%
%However, the $B_{x,t}(\mu)$ term is not as clear since its dominant contribution for a generic continuous root density is $t^{-1}$. They give contributions that decay as $1/|t|^{3/2}$ and of order $c^{-2}$, which  
Putting everything together we obtain the leading
contribution from the vicinity of $Q-\alpha'$ to the double integral over
$\chi^{(2)}_{x,t}(\lambda,\mu)$
\begin{align}
\frac{4}{c^2}\frac{e^{-\sign(t)i\pi/4}\gamb^4}{4\pi^{3/2}|t|^{1/2}}\frac{e^{it(Q-\alpha')^2}}{2it(Q-\alpha')}(Q-\alpha)^2\left[-\frac{\log|t|}{2\pi^2}
  +{\cal O}(t^0)\right]\, .
\end{align}
The analogous result for the contribution from the vicinity of $-Q-\alpha'$ is
\begin{equation}
\begin{aligned}
\frac{4}{c^2}\frac{e^{-\sign(t)i\pi/4}\gamb^4}{4\pi^{3/2}|t|^{1/2}}\frac{e^{it(Q+\alpha')^2}}{2it(Q+\alpha')}(Q+\alpha)^2\left[-\frac{\log|t|}{2\pi^2} +{\cal O}(t^0)\right]\,.
\end{aligned}
\end{equation}

\subsubsection{Euler scale asymptotic behaviour \label{below2}}
In this section we will assume $\rho$ to be continuous. If it is not continuous the leading $1/t$ behaviour is unchanged, but the $1/t^2$ corrections might differ.

For a generic continuous root density at large $x,t$ and fixed
$\alpha=\tfrac{x}{2t}$, the two integrals over $\lambda$ and $\mu$ in 
the correlation function \eqref{corregen} are both dominated by the
saddle point at $\alpha'$. Applying \eqref{oscalsallde} to the
one particle-hole contribution gives
\begin{equation}
\frac{\pi\gamb^2\rho(\alpha')\rho_h(\alpha')}{|t|}
+\frac{i\pi\gamb^2\left[\rho''(\alpha')\rho_h(\alpha')-\rho(\alpha')\rho_h''(\alpha')
  \right]}{4t|t|}
+{\cal O}(t^{-3})\,.
\end{equation}

The contribution due to two particle-hole excitations is more
subtle and requires determining the asymptotic behaviour of
oscillatory integrals with principal values, whose saddle point falls
on the singularity. The general strategy is to write each singularity
as 
\begin{equation}
\frac{1}{\lambda-\mu}=\frac{t}{2i}\int_{-\infty}^\infty \sign(\xi)e^{it\xi(\lambda-\mu)}\D{\xi}\,,
\end{equation}
and then to carry out a regular asymptotic analysis of the multiple
oscillatory integrals successively. The $\sign(\xi)$ factors
introduce discontinuities which result in contributions on top of
those from the saddle points.

Let us treat the case of $C_{x,t}$ in detail. We write
\begin{equation}
C_{2\alpha' t,t}=\frac{t}{2i}\iiint \rho(\alpha'+u)\rho_h(\alpha'+v)\sign(\xi)e^{it[(u-\xi/2)^2-(v-\xi/2)^2]}\D{u}\D{v}\D{\xi}\,,
\end{equation}
and then apply a saddle point approximation to the $u$ and $v$
integrals using \eqref{oscalsallde} to obtain 
\begin{align}
C_{2\alpha' t,t}&=\frac{t}{2i}\int_{-\infty}^\infty \D{\xi}\sign(\xi)
\bigg[\frac{\pi}{|t|}\rho\big(\alpha'+\tfrac{\xi}{2}\big)\rho_h\big(\alpha'+\tfrac{\xi}{2}\big)\nn
&\qquad\qquad+\frac{\pi
    i}{4t|t|}\Big(\rho''\big(\alpha'+\tfrac{\xi}{2}\big)\rho_h\big(\alpha'+\tfrac{\xi}{2}\big)-\rho\big(\alpha'+\tfrac{\xi}{2}\big)  \rho_h''\big(\alpha'+\tfrac{\xi}{2}\big)\Big)
  \bigg]
+{\cal O}(t^{-2})\,.
\end{align}
This can be simplified by performing an integration by parts on
the $\xi$ integral of the subleading term
\begin{align}
\label{largeC}
C_{2\alpha' t,t}&=i\pi\sign(t) \int_{-\infty}^\infty
\rho(\xi)\rho_h(\xi)\sign(\alpha'-\xi)\D{\xi}
\nn
&+\frac{\pi}{2|t|}\big(\rho(\alpha')\rho_h'(\alpha')-\rho'(\alpha')\rho_h(\alpha')\big)
+{\cal O}(t^{-2})\,.
\end{align}
Similarly one finds for the $D_{x,t}$ term
\begin{align}
\label{largeD}    
D_{2\alpha' t,t}&=i\pi\sign(t) \int_{-\infty}^\infty \rho(\xi)\rho_h(\xi)\xi\sign(\alpha'-\xi)\D{\xi}\nn
&+\frac{\pi}{2|t|}\Big[\alpha'(\rho(\alpha')\rho_h'(\alpha')-\rho'(\alpha')\rho_h(\alpha'))-\rho(\alpha')\rho_h(\alpha')\Big]+{\cal O}(t^{-2})\,.
\end{align}
To deal with the $A_{x,t}$ term we use that in a distributional sense
\begin{equation}
\frac{1}{(\lambda-\mu)^2}=-\frac{t^2}{2}\int_{-\infty}^\infty |\xi|e^{it\xi(\lambda-\mu)}\D{\xi}\,,
\end{equation}
and then carry out a similar analysis to obtain
\begin{equation}\label{largeA}
\begin{aligned}
A_{2\alpha' t,t}&=-2\pi|t| \int_{-\infty}^\infty \rho(\xi)\rho_h(\xi)|\alpha'-\xi|\D{\xi}+o(t)\,.
\end{aligned}
\end{equation}
This leaves us with the $B_{x,t}(\lambda)$ term. It is not possible
to determine the asymptotics of $B_{x,t}(\lambda)$ at fixed $\lambda$
and then carry out a saddle point approximation of the resulting
integral as the asymptotic expression for $B_{x,t}(\lambda)$
becomes singular at the saddle point $\lambda=\alpha'$.
The full contribution involving $B_{x,t}(\lambda)$ to the correlation
function is
\footnote{Here and in what follows we assume that $\rho_h(\mu)$
is a continuous function of $\mu$ that decays to zero at infinity so
that the integral exists. The case where $\rho_h(\mu)$ is the
actual hole density is then obtained as a limit of the resulting
expression.}
\begin{equation}\label{below}
X_{x,t}\equiv\int \D{\mu} \dashint \D{\lambda} \D{u}\D{v} \frac{e^{it(\lambda^2-\mu^2+u^2-v^2)+ix(\mu-\lambda+v-u)}}{(\lambda-u)(v-u)}(\lambda-\mu)^2\rho(\lambda)\rho_h(\mu)\rho(u)\rho_h(v)\,.
\end{equation}
We rewrite this as a six-fold integral
\begin{align}
X_{2\alpha' t,t}=-\frac{t^2}{4}\idotsint &\rho(\alpha'+\lambda)\rho_h(\alpha'+\mu)\rho(\alpha'+u)\rho_h(\alpha'+v)\sign(\xi)\sign(\zeta)\nn
&\times (\lambda-\mu)^2e^{it(\lambda^2-\mu^2+u^2-v^2)}e^{i\xi t (v-u)+i\zeta t(\lambda-u)}\D{u}\D{v}\D{\lambda} \D{\mu} \D{\xi} \D{\zeta}\,,
\end{align}
and then perform saddle point approximations on the $u,v,\lambda,\mu$
integrals. This gives
\begin{align}
X_{2\alpha' t,t}=-\frac{\pi^2}{4}\iint
&\rho\big(\alpha'-\tfrac{\zeta}{2}\big)\rho_h(\alpha')\rho\big(\alpha'+\tfrac{\zeta+\xi}{2}\big)\rho_h\big(\alpha'+\tfrac{\xi}{2}\big)\frac{\zeta^2}{4}\sign(\xi)\sign(\zeta)\nn
&\times e^{-it\frac{\zeta^2}{2}-it\zeta\frac{\xi}{2}}\ \D{\xi} \D{\zeta}[1+o(t^0)]\,.
\end{align}
We now carry out the integral over $\xi$, which does not have saddle
point and is dominated by the discontinuity of the integrand at
$\xi=0$ using
\begin{equation}\label{help0}
\int f(\xi)e^{is\xi}\D{\xi}=-\frac{f(0+)-f(0-)}{is}-\frac{f'(0+)-f'(0-)}{s^2}+{\cal O}(s^{-3})\,,
\end{equation}
where $f$ is a function with discontinuities only at zero. This gives
\begin{equation}
X_{2\alpha'
  t,t}=-\frac{\pi^2\rho_h(\alpha')^2}{4it}\int_{-\infty}^\infty
|\zeta|\rho\big(\alpha'-\tfrac{\zeta}{2}\big)\rho\big(\alpha'+\tfrac{\zeta}{2}\big)
e^{-it\frac{\zeta^2}{2}}\D{\zeta}[1+o(t^{0})]\,. 
\end{equation}
This last integral also has a saddle point at zero, but with a coefficient that is not differentiable, so that one cannot apply \eqref{oscalsallde}. Approximating $\zeta=0$ in the $\rho$'s at leading order in $t$, one can integrate the remaining terms to obtain
\begin{equation}
X_{2\alpha' t,t}=\frac{\pi^2}{2t^2}\rho(\alpha')^2\rho_h(\alpha')^2+o(t^{-2})\,.
\end{equation}
The contribution involving $B_{x,t}(\mu)$ is given by
\begin{equation}
Y_{x,t}\equiv\int \D{\lambda} \dashint \D{\mu} \D{u}\D{v} \frac{e^{it(\lambda^2-\mu^2+u^2-v^2)+ix(\mu-\lambda+v-u)}}{(\mu-u)(v-u)}(\lambda-\mu)^2\rho(\lambda)\rho_h(\mu)\rho(u)\rho_h(v)\,,
\end{equation}
and can be analyzed in a similar way. We start by rewriting it
as a six-fold integral
\begin{align}
Y_{2\alpha' t,t}=-\frac{t^2}{4}\idotsint &\rho(\alpha'+\lambda)\rho_h(\alpha'+\mu)\rho(\alpha'+u)\rho_h(\alpha'+v)\sign(\xi)\sign(\zeta)\nn
&\times (\lambda-\mu)^2e^{it(\lambda^2-\mu^2+u^2-v^2)}e^{i\xi t (v-u)+i\zeta t(\mu-u)}\D{u}\D{v}\D{\lambda} \D{\mu} \D{\xi} \D{\zeta}\,,
\end{align}
and then perform saddle-point approximations on the $\lambda,\mu,u,v$ integrals
\begin{align}
&Y_{2\alpha' t,t}\approx-\frac{\pi^2}{16}\iint
\sign(\xi)\zeta|\zeta|\rho(\alpha')\rho_h\big(\alpha'+\tfrac{\zeta}{2}\big)
\rho\big(\alpha'+\tfrac{\zeta+\xi}{2}\big)\rho_h\big(\alpha'+\tfrac{\xi}{2}\big)
e^{-it\xi\frac{\zeta}{2}}\D{\xi}\D{\zeta}\nn 
&-\frac{\pi^2 i}{64t}\iint \sign(\xi)|\zeta|
\bigg[\zeta\rho''(\alpha')\rho_h\big(\alpha'+\tfrac{\zeta}{2}\big)
\rho\big(\alpha'+\tfrac{\zeta+\xi}{2}\big)\rho_h\big(\alpha'+\tfrac{\xi}{2}\big)\nn
&-8\rho'(\alpha')\rho_h\big(\alpha'+\tfrac{\zeta}{2}\big)
\rho\big(\alpha'+\tfrac{\zeta+\xi}{2}\big)\rho_h\big(\alpha'+\tfrac{\xi}{2}\big)
-\zeta\rho(\alpha')\rho_h''\big(\alpha'+\tfrac{\zeta}{2}\big)
\rho\big(\alpha'+\tfrac{\zeta+\xi}{2}\big)\rho_h\big(\alpha'+\tfrac{\xi}{2}\big)\nn
&-8\rho(\alpha')\rho_h'\big(\alpha'+\tfrac{\zeta}{2}\big)
\rho\big(\alpha'+\tfrac{\zeta+\xi}{2}\big)\rho_h\big(\alpha'+\tfrac{\xi}{2}\big)
+\zeta\rho(\alpha')\rho_h\big(\alpha'+\tfrac{\zeta}{2}\big)
\rho''\big(\alpha'+\tfrac{\zeta+\xi}{2}\big)\rho_h\big(\alpha'+\tfrac{\xi}{2}\big)\nn
&-\zeta\rho(\alpha')\rho_h\big(\alpha'+\tfrac{\zeta}{2}\big)\rho\big(\alpha'+\tfrac{\zeta+\xi}{2}\big)\rho_h''\big(\alpha'+\tfrac{\xi}{2}\big)\bigg]e^{-it\xi\frac{\zeta}{2}}\D{\xi} \D{\zeta}\,.
\end{align}
We next perform the integral over $\xi$ in the large $t$ limit
using \eqref{help0}. After some rearrangements we obtain 
\begin{align}
Y_{2\alpha' t,t}&=\frac{i\pi^2}{t}\rho(\alpha')\rho_h(\alpha')\int_{-\infty}^\infty |\alpha'-\zeta|\rho(\zeta)\rho_h(\zeta)\D{\zeta}\nn
&-\frac{\pi^2}{4t^2}\int_{-\infty}^\infty|\alpha'-\zeta|\bigg[\rho''(\alpha')\rho_h(\alpha')\rho(\zeta)\rho_h(\zeta)-\rho(\alpha')\rho_h''(\alpha')\rho(\zeta)\rho_h(\zeta)\nn
&\qquad\qquad+\rho(\alpha')\rho_h(\alpha')\rho''(\zeta)\rho_h(\zeta)-\rho(\alpha')\rho_h(\alpha')\rho(\zeta)\rho_h''(\zeta)\bigg]\D{\zeta}\nn
&-\frac{\pi^2}{2t^2}[\rho(\alpha')\rho_h'(\alpha')+2\rho'(\alpha')\rho_h(\alpha')]\int_{-\infty}^\infty\sign(\alpha'-\zeta)\rho(\zeta)\rho_h(\zeta)\D{\zeta}\nn
&-\frac{\pi^2}{2t^2}\rho(\alpha')\rho_h(\alpha')\int_{-\infty}^\infty\sign(\alpha'-\zeta)[\rho'(\zeta)\rho_h(\zeta)+2\rho(\zeta)\rho_h'(\zeta)]\D{\zeta}
+o(t^{-2})\,.
\end{align}
Putting everything together we arrive at \eqref{ghd}.
%%%%%%%%%%%%%%%%%%%%%%%%%%%%%%%%%%%5
\subsubsection{GHD predictions}
\label{app:GHD}
%%%%%%%%%%%%%%%%%%%%%%%%
The GHD result for the asymptotics of the density-density correlator
is \cite{DS17,Doyon18}
\be
\langle\sigma(x,t)\sigma(0,0)\rangle=\int_{-\infty}^\infty
\delta\big(x-v(\lambda)t\big)\ \rho(\lambda)\big(1-\vartheta(\lambda)\big)\left[q^{\rm dr}(\lambda)\right]^2\D{\lambda}\ , 
\label{GHDpred}
\ee
with $\rho,\vartheta$ defined in \eqref{vartheta} and \eqref{rhoh}, and where the other functions are defined as follows:
\begin{align}
%\rho(\lambda)+\rho_h(\lambda)&=\frac{1}{2\pi}+
%\int_{-\infty}^\infty
%\frac{2c}{c^2+(\lambda-\nu)^2}\ \rho(\nu)\frac{\D{\nu}}{2\pi}\ \ ,\nn
%\vartheta(\lambda)&=\frac{\rho(\lambda)}{\rho(\lambda)+\rho_h(\lambda)}\ ,\nn
F(\lambda,\nu)&=\frac{1}{\pi}{\rm   arctan}\bigg(\frac{\lambda-\nu}{c}\bigg)
+\int_{-\infty}^\infty
\frac{2c}{c^2+(\lambda-\lambda')^2}\ \vartheta(\lambda')\ F(\lambda',\nu)\frac{\D{\lambda'}}{2\pi}\ ,\nn
q^{\rm dr}(\lambda)&=1 -\int_{-\infty}^\infty q(\lambda')
\ \vartheta(\lambda')\ \partial_\lambda F(\lambda',\lambda)\D{\lambda'}\ ,\nn
v(\lambda)&=\frac{e'(\lambda)}{2\pi\big(\rho(\lambda)+\rho_h(\lambda)\big)}\ ,\nn
e'(\lambda)&=2\lambda-\int_{-\infty}^\infty
2\nu\ \vartheta(\nu)\partial_\lambda F(\nu,\lambda)\D{\nu}\ .
\end{align}
These equations can be
straightforwardly solved in a $1/c$-expansion up to order ${\cal
  O}(c^{-2})$
\begin{align}
\rho(\lambda)+\rho_h(\lambda)&=\frac{1+\frac{2D}{c}}{2\pi}\ ,\quad
  \vartheta(\lambda)=\frac{2\pi\rho(\lambda)}{1+2D/c}\ ,\quad
F(\lambda,\alpha)=\frac{\lambda-\alpha}{\pi c}+\frac{2}{\pi
  c^2}(\delta-D\alpha)\ ,\nn
v(\lambda)&=\frac{2\lambda-4\delta/c}{1+2D/c}\ ,\quad
q^{\rm dr}(\lambda)=1+\frac{2D}{c}\ ,
\label{GHD1overc}
\end{align}
where $D$ and $\delta$ are defined in \eqref{D} and \eqref{delta}. Substituting \fr{GHD1overc} back into \fr{GHDpred}
precisely recovers the leading contribution in \fr{ghd}.

\subsection{Dynamical structure factor}
\subsubsection{Behaviour near the thresholds at zero temperature}
We start from the simplified expression of the DSF \eqref{dsf2pht0} at
zero temperature and will assume $q>0$ for simplicity. We
note that when $z\to\pm\infty$ we necessarily have $Z_+<Z_-$, so
that the only possible region that can lead to a divergence of the
integral is the region $z$ close to $0$. In this region we first set 
\begin{equation}
\omega=\qp^2+2\qp Q-\eta\,,
\end{equation}
with $\eta>0$ small, and investigate the values taken by $Z_\pm$. We
find for $z$ close to zero
\begin{align}
Z_+(z)&=\begin{cases}
\frac{-\qp^2+\qp^2z+\eta}{2\qp(1-z)} & \text{if }z>0\ ,\\
\frac{-\qp^2-4\qp Q-\qp^2z+\eta}{2\qp(1-z)}& \text{if }z<0\ ,
\end{cases}\nn
Z_-(z)&=\begin{cases}
\frac{-\eta+\qp^2z}{2\qp z} & \text{if }z>\frac{\eta}{2\qp^2}\ ,\\
\frac{-\qp^2-\qp^2z+\eta}{2\qp(1-z)}&\text{if }0<z<\frac{\eta}{2\qp^2}\ ,\\
\frac{-\qp^2-4\qp Q+\qp^2z+\eta}{2\qp(1-z)}& \text{if }z<0\ .
\end{cases}
\end{align}
We observe that for small $z$ we have $Z_-<Z_+$ if and only if
$z<\frac{\eta}{2\qp^2}$, in which case 
$Z_+-Z_-=\frac{\qp|z|}{1-z}$. Substituting this expression
back into \eqref{dsf2pht0}, we find that among the contribution
proportional to $\1_{Z_-<Z_+}$ only the term $\frac{Z_+-Z_-}{z}$ is non-integrable
when $z\to 0$. However, its divergent part is exactly cancelled
by the term in the third line of \fr{dsf2pht0} proportional to
$\1_{\omega_-<\omega<\omega_+}$. All other terms in the first
two lines of \fr{dsf2pht0} give a finite ${\cal O}(\eta^0)$
contribution because they are integrable for $z\to 0$. This leaves
the contribution proportional to $\1_{\omega_-<\omega<\omega_+}$ for
$z>\frac{\eta}{2\qp^2}$, which leads to a logarithmic singularity in
  $\eta$. Setting an arbitrary upper limit in the integral since its
modification amounts to a ${\cal O}(\eta^0)$ correction we have
\begin{align}
S^{(2)}(q,\omega)&=\frac{1}{4\pi^2\qp c^2}\int_{\tfrac{\eta}{2\qp^2}}^1 \left[ -\frac{2\qp \min(|\qp z|,2Q)}{z^2}\right]\D{z}+{\cal O}(\eta^0)\nn
&=\frac{\qp}{2\pi^2c^2}\log |\eta|+{\cal O}(\eta^0)\,.
\end{align}
We now turn to singularities above the upper threshold. Taking
  $\eta>0$ to be small and setting
\begin{equation}
\omega=\qp^2+2\qp Q+\eta\,,
\end{equation}
we find for $z\approx 0$
\begin{align}
Z_+(z)&=\begin{cases}
\frac{-\qp^2+\qp^2z-\eta}{2\qp(1-z)} &\text{if }z>0\ ,\\
\frac{\eta+\qp^2z}{2\qp z}&\text{if }-\frac{\eta}{2\qp(\qp+2Q)}<z<0\ ,\\
\frac{-\qp^2-4\qp Q-\qp^2z-\eta}{2\qp(1-z)}&\text{if
}z<-\frac{\eta}{2\qp(\qp+2Q)}\ ,
\end{cases}\nn
Z_-(z)&=\begin{cases}
\frac{\eta+\qp^2z}{2\qp z}&\text{if }z>0\ ,\\
\frac{-\qp^2-4\qp Q+\qp^2z-\eta}{2\qp(1-z)}& \text{if }z<0\ .
\end{cases}
\end{align}
We observe that for $z$ close to zero we have $Z_-<Z_+$ if and
only if $z<-\frac{\eta}{2\qp(\qp+2Q)}$, in which case
$Z_+-Z_-=\frac{\qp|z|}{1-z}$. Above the threshold we have
$\1_{\omega_-<\omega<\omega_+}=0$ so that the last term in
\fr{dsf2pht0} vanishes. Of the remaining terms only the one
proportional to $\frac{Z_+-Z_-}{z^2}$ diverges near $z=0$, so that
\begin{align}
S^{(2)}(q,\omega)&=\frac{1}{4\pi^2\qp c^2}\int_{-1}^{-\tfrac{\eta}{2\qp(\qp+2Q)}} 2\qp(1-z)^2 \frac{Z_+-Z_-}{z^2}\D{z}+{\cal O}(\eta^0)\nn
&=-\frac{\qp}{2\pi^2c^2}\log |\eta|+{\cal O}(\eta^0)\,.
\end{align}
The behaviour near the lower threshold is obtained through a similar analysis.

\subsubsection{\sfix{Behaviour at small $q,\omega$}}
We start by writing the two particle-hole contribution as
\begin{align}
\label{eqreft}
S^{(2)}(q,\omega)&=
\frac{8\pi^2}{c^2}\int_{-\infty}^\infty
\int_{-\infty}^\infty\frac{\rho(q_3)\rho_h(q_4)}{z^2} |1-z|\Big[
  \rho(q_1)\rho_h(q_2)-\rho\big(\tfrac{\omegap-\qp^{
      2}}{2\qp}\big)\rho_h\big(\tfrac{\omegap+\qp^{ 2}}{2\qp}\big)\Big]\D{z}\D{p}\nn
&+\frac{8\pi^2}{c^2}\int_{-\infty}^\infty \int_{-\infty}^\infty
\frac{\frac{2(\lambda-\mu)^2}{\lambda-\bar{\lambda}}-\frac{(\lambda-\mu)^2}{\mu-\bar{\lambda}}+3\mu-2\lambda-\bar{\lambda}+\tfrac{(\lambda-\mu)^2-|\qp(\lambda-\mu)|}{\qp+\lambda-\mu}}{(\qp+\lambda-\mu)|\qp+\lambda-\mu|}\nn
&\hskip2.5cm\times\
\rho(\lambda)\rho_h(\mu)\rho(\bar{\lambda})\rho_h(\bar{\mu})\
\D{\lambda} \D{\mu}\
\equiv\frac{8\pi^2}{c^2}(\Psi_1+\Psi_2),
\end{align}
where $\Psi_{1,2}$ denote the first and
second terms respectively.
%
%{\color{red} I suggest to keep $q'$ and $\omega'$ in the various
%  eqns as I think this is free of charge. I have restored them (using
%  macros) in (319) and below.
%  
%We removed the primes on $q,\omega$ for simplicity as we work at
%leading order in $c$.}

The integral for $\Psi_1$ with $q,\omega\to 0$ at fixed
$\gamma=\tfrac{\omegap}{2\qp}$ is well-defined and finite. In
this limit we have $q_3=q_4=\gamma+p(1-z)$ and
$q_1=q_2=\gamma-pz$. Changing variables to $v=\gamma-pz$ and
  $u=\gamma+p(1-z)$ we have
\begin{align}
%&\int_{-\infty}^\infty
%\int_{-\infty}^\infty\frac{\rho(q_3)\rho_h(q_4)}{z^2} |1-z|\Big[
%\rh%o(q_1)\rho_h(q_2)-\rho(\tfrac{\omegap-\qp^{
%2}}{2\qp})\rho_h(\tfrac{\omegap+\qp^{ 2}}{2\qp}%)\Big] \D{z}\D{p}=\nn
%&\qquad\qquad
\Psi_1=  \left(\int_{-\infty}^\infty \rho(u)\rho_h(u)|u-\gamma|\D{u}\right)\left(\dashint_{-\infty}^\infty \frac{\rho(v)\rho_h(v)-\rho(\gamma)\rho_h(\gamma)}{(v-\gamma)^2}\D{v}\right)+o(\qp^0)\,.
\end{align}
As for $\Psi_2$
%the \bl{second} term in \fr{eqreft}
, we first perform a change of
variables from $\mu$ to $v=q+\lambda-\mu$
%{\color{brown}
%\begin{align}
%&\int_{-\infty}^\infty \int_{-\infty}^\infty \rho(\lambda)\rho_h(\mu)\rho(\bar{\lambda})\rho_h(\ba%r{\mu})\frac{\frac{2(\lambda-\mu)^2}{\lambda-\bar{\lambda}}-\frac{(\lambda-\mu)^2}{\mu-\bar{%\lambda}}+3\mu-2\lambda-\bar{\lambda}+\tfrac{(\lambda-\mu)^2-|\qp(\lambda-\mu)|}{\qp+\lambda%-\mu}}{(\qp+\lambda-\mu)|\qp+\lambda-\mu|}\D{\lambda} \D{\mu}\nn
%&=\int_{-\infty}^\infty \int_{-\infty}^\infty \rho(\lambda)\rho_h(q+\lambda-v)\rho(\qp+\lambda-v+\%qp\tfrac{2\gamma-2\lambda-\qp}{2v})\rho_h(\qp+\lambda+\qp\tfrac{2\gamma-2\lambda-\qp}{2v})\f%rac{1}{v|v|}\nn
%  &\qquad\qquad\times\biggl[\frac{2(v-\qp)^2}{v-\qp-\qp\tfrac{2\gamma-2\lambda-\qp}{2v}}+\fr%ac{(v-\qp)^2}{\qp\tfrac{2\gamma-2\lambda-\qp}{2v}}+2\qp-2v\nn
%    &\qquad\qquad\qquad-\qp\tfrac{2\gamma-2\lambda-\qp}{2v}+\frac{(v-\qp)^2-|\qp(v-\qp)|}{v}% \biggr]\D{\lambda} \D{v}\,.
%\end{align}
%}
\begin{align}
\Psi_2
&=\int_{-\infty}^\infty \int_{-\infty}^\infty
\rho(\lambda)\rho_h(q+\lambda-v)\rho(\qp+\lambda-v+\qp\tfrac{2\gamma-2\lambda-\qp}{2v})\rho_h(\qp+\lambda+\qp\tfrac{2\gamma-2\lambda-\qp}{2v})\nn
  &\qquad\qquad\times\frac{1}{v|v|}
\biggl[\frac{2(v-\qp)^2}{v-\qp-\qp\tfrac{2\gamma-2\lambda-\qp}{2v}}+\frac{(v-\qp)^2}{\qp\tfrac{2\gamma-2\lambda-\qp}{2v}}+2\qp-2v\nn
    &\qquad\qquad\qquad\qquad-\qp\tfrac{2\gamma-2\lambda-\qp}{2v}+\frac{(v-\qp)^2-|\qp(v-\qp)|}{v} \biggr]\D{\lambda} \D{v}\,.
\end{align}
We now observe that the four $\rho$ factors are invariant under
the change of variable 
\begin{equation}
v'=-\qp\tfrac{2\gamma-2\lambda-\qp}{2v}\,.
\end{equation}
We apply this change of variable to all the terms except
\begin{equation}
\frac{(v-\qp)^2}{\qp\tfrac{2\gamma-2\lambda-\qp}{2v}}\,,
\end{equation}
and express the term
\begin{equation}
\frac{2v^2}{v-\qp-\qp\tfrac{2\gamma-2\lambda-\qp}{2v}}
\end{equation}
as one half of itself plus one half of itself after the change of
variables. We obtain for $\qp>0$ 
%\begin{align}
%&\int_{-\infty}^\infty \int_{-\infty}^\infty \rho(\lambda)\rho_h(\mu)\rho(\bar{\lambda})\rh%o_h(\bar{\mu})\frac{\frac{2(\lambda-\mu)^2}{\lambda-\bar{\lambda}}-\frac{(\lambda-\mu)^2}{\%mu-\bar{\lambda}}+3\mu-2\lambda-\bar{\lambda}+\tfrac{(\lambda-\mu)^2-|q(\lambda-\mu)|}{q+\l%ambda-\mu}}{(q+\lambda-\mu)|q+\lambda-\mu|}\D{\lambda} \D{\mu}=\\
%\begin{align}
%\Psi_2&=\qp\int_{-\infty}^\infty \int_{-\infty}^\infty \rho(\lambda)\rho_h(\qp+\lambda-v)\rho\big(\qp+\lambda-v+\qp\tfrac{2\gamma-2\lambda-\qp}{2v}\big)\rho_h\big(\qp+\lambda+\qp\tfrac{2\gamma-2\lambda-\qp}{2v}\big)\nn
%&\hskip1cm\times\frac{1}{v|v|}
%\left[2+\frac{2v}{2\gamma-2\lambda-\qp}+\frac{\qp-v}{v-\qp-\qp\tfrac{2\gamma-2\lambda-\qp}{2v}}+\frac{\qp-|v-\qp|}{v} \right]\D{\lambda} \D{v}\,.
%\end{align}
\begin{align}
\Psi_2&=-2\int_{-\infty}^\infty \int_{-\infty}^\infty \rho(\lambda)\rho_h(\qp+\lambda-v)\rho\big(\qp+\lambda-v+\qp\tfrac{2\gamma-2\lambda-\qp}{2v}\big)\rho_h\big(\qp+\lambda+\qp\tfrac{2\gamma-2\lambda-\qp}{2v}\big)\nn
&\hskip1cm\times\frac{\sign(v)}{2\gamma-2\lambda-\qp}
\Bigg[1-\frac{\qp}{v}+\frac{2v^2}{2v^2-2\qp(v+\gamma-\lambda)+\qp^2}\\
&\qquad\qquad\qquad\qquad\qquad\qquad-\frac{2v}{2\gamma-2\lambda-\qp}\left(1-\left|1+\frac{2\gamma-2\lambda-\qp}{2v} \right| \right) \Bigg]\D{\lambda} \D{v}\,.
\end{align}
Since there are no non-integrable divergences in the integrand at small $v$, in this representation one can set $q'=0$ in the $\rho$ terms as well as in the integrand, at small $q'$. It yields
\be
\Psi_2=-\int_{-\infty}^\infty \int_{-\infty}^\infty \rho(\lambda)\rho_h(\lambda)\rho(u)\rho_h(u)\frac{\sign(\lambda-u)}{\gamma-\lambda}\left[ 2-\frac{\lambda-u}{\gamma-\lambda}\left(1-\left|\frac{\gamma-u}{\lambda-u}\right|\right)\right]\D{\lambda}\D{u}+o(q^0)\,.
\ee
We obtain then the claimed result.
%\be
%\Psi_2=\int_{-\infty}^\infty \int_{-\infty}^\infty \rho(\lambda)\rho_h(\lambda)\rho(\lambda+\tfrac{\gamma-\lambda}{w})\rho_h(\lambda+\tfrac{\gamma-\lambda}{w})\frac{1}{w|w|}\left[2+\frac{1-|w-1|}{w} \right]\D{\lambda} \D{w}+o(q^0)\,.
%\ee
\subsubsection{High frequency tail}
%{\color{red} I suggest to restore the $\qp$ and $\omegap$ here as well.}
We start with the representation \eqref{DSF2ph} for the two
particle-hole contribution to the DSF expressed as a single
double integral. We first decompose the double integral into the
  two regions $|q'+\lambda-\mu|>\epsilon$ and $|q'+\lambda-\mu|<\epsilon$
and focus on the latter part. Since we have assumed that $\rho$
decays faster than any power law at infinity $\lambda$ has to remain
smaller than any power law of $\omega$ for the integral not to vanish
at any order ${\cal O}(\omega^{-n})$. Since
$|q'+\lambda-\mu|<\epsilon$ the same holds true for $\mu$. But
this implies that $\overline{\lambda}$ necessarily grows as
$\omega^{1/2}$, which makes this contribution vanish at any order
${\cal O}(\omega^{-n})$. Hence at any given order ${\cal 
  O}(\omega^{-n})$ we can impose $|q'+\lambda-\mu|>\epsilon$. This removes
all poles in the integrand of \eqref{DSF2ph} and one can consider
  all contributions separately. Moreover, since $\rho$
decays faster than a power law at infinity the term proportional to
$\rho(\tfrac{\omega'-q^{\prime
    2}}{2q'})\rho_h(\tfrac{\omega'+q^{\prime 2}}{2q'})$ is negligible
at order ${\cal O}(\omega^{-n})$.   

We then split the $\mu$ integral into the sum of positive
and negative $\mu$ parts and perform the change of variables
\be
z=\begin{cases}
2\mu-\qp-\lambda-\sqrt{2\omegap} & \text{if } \mu>0\ ,\\
2\mu-\qp-\lambda+\sqrt{2\omegap} & \text{if } \mu<0\ .
\end{cases}
\ee
This way the DSF can be brought to the form
\begin{align}
&S^{(2)}(q,\omega)=\int_{-\infty}^\infty \D{\lambda}
  \int_{-\qp-\lambda-\sqrt{2\omegap}}^\infty \D{z}
  \Big[\rho(\lambda)\rho\big(z+f_+(z,\lambda,\omegap)\big)\ g_+(z,\lambda,\omegap)\nn
    &\qquad\qquad\qquad\qquad\qquad\times\rho_h\big(\tfrac{z+\qp+\lambda+\sqrt{2\omegap}}{2}\big)
\rho_h\big(z+f_+(z,\lambda,\omegap)+\tfrac{\qp+\lambda-z-\sqrt{2\omegap}}{2}\big)\Big]\nn
&+\int_{-\infty}^\infty \D{\lambda} \int_{-\infty}^{-\qp-\lambda+\sqrt{2\omegap}} \D{z} \Big[\rho(\lambda)\rho\big(z+f_-(z,\lambda,\omegap))g_-(z,\lambda,\omegap\big)\nn
&\qquad\qquad\qquad\qquad\qquad\times\rho_h\big(\tfrac{z+\qp+\lambda-\sqrt{2\omegap}}{2}\big)\rho_h\big(z+f_-(z,\lambda,\omegap)+\tfrac{\qp+\lambda-z+\sqrt{2\omegap}}{2}\big)\Big]+\dots\ ,
%&+{\cal O}(\omega^{-n})\,,
\end{align}
where the dots indicate subleading corrections that decay faster than
any inverse power in $\omega$ and
\begin{align}
f_\pm(z,\lambda,\omega)&=\pm\frac{1}{2}
\frac{\qp^2+2\qp z+2\qp\lambda-(\lambda-z)^2}{\sqrt{2\omega}\pm(z-\qp-\lambda)}\ ,\nn
g_\pm(z,\lambda,\omega)=&\frac{16\pi^2}{c^2(\lambda+\qp-z\mp\sqrt{2\omega})|\lambda+\qp-z\mp\sqrt{2\omega}|}\Bigg[-4\lambda+\frac{(\lambda-\qp-z\mp\sqrt{2\omega})^2}{\lambda+\qp-z\mp\sqrt{2\omega}}\nn
&+2\frac{(\lambda+\qp-z\mp\sqrt{2\omega})(\lambda-\qp-z\mp\sqrt{2\omega})^2}{4\omega-8\qp\lambda-4\qp^2}+3(\lambda+\qp+z+\sqrt{2\omega})\nn
&+\frac{-\lambda^2+2\lambda \qp+\qp^2+z^2\pm 2\sqrt{2\omega}z}{\lambda+\qp-z-\sqrt{2\omega}}\nn
&+2\frac{(\lambda+\qp-z\mp\sqrt{2\omega})(\lambda-\qp-z\mp\sqrt{2\omega})^2}{\lambda^2+\qp^2\pm2\sqrt{2\omega}z+z^2+4\lambda \qp-2\lambda z\mp 2\lambda\sqrt{2\omega}}
\Bigg]\,.
\end{align}
We now observe that any part of the integral where the argument of
one of the two $\rho$'s grows as a power-law in $\omega$ will give
contributions that decay faster than any power-law, since $\rho$ is
assumed to decay faster than any power-law at infinity. From the
expression of $f_\pm$ one sees that $z$ cannot grow faster than
$\omega^{1/4}$. Consequently, with an error that goes to zero faster
than any power law in $\omega$
%a precision ${\cal O}(\omega^{-n})$ for any $n$,
one can replace the limits of the integrals and the
arguments of the $\rho_h$'s by $\pm\infty$.
%One obtains for any $n>0$ at order $c^{-2}$
This gives
\begin{align}
S^{(2)}(q,\omega)&=\frac{1}{4\pi^2}\int_{-\infty}^\infty
\int_{-\infty}^\infty
\rho(\lambda)\rho\big(z+f_+(z,\lambda,\omegap)\big)\ g_+(z,\lambda,\omegap)\D{z}\D{\lambda}\nn
&+\frac{1}{4\pi^2}\int_{-\infty}^\infty \int_{-\infty}^\infty
\rho(\lambda)\rho\big(z+f_-(z,\lambda,\omegap)\big)\ g_-(z,\lambda,\omegap)\D{z}\D{\lambda}+\ldots
%&+{\cal O}(\omega^{-n})\,.
\end{align}
We now expand $f_\pm(z,\lambda,\omegap)$ and $g_\pm(z,\lambda,\omegap)$
in Laurent series in $\lambda$, $\lambda-z$ and $\omegap^{-1/2}$, and
Taylor expand $\rho(z+f_\pm(z,\lambda,\omega'))$. This produces terms of the
type $\rho(\lambda)\rho^{(a)}(z)\lambda^b(\lambda-z)^d{\omegap}^{-e/2}$
with $a,b,e\geq 0$ integers and $d$ a positive or negative integer. We
integrate this by parts $a$ times over $z$ so that the
integrand involves only $\rho(z)$,
and then write the full result $S^{(2)}(q,\omega)$ as one half of
itself plus one half of itself after swapping the dummy variables
$\lambda$ and $z$. We observe that there remain only positive powers
$d\geq 0$, and one obtains the first two terms of the expansion
${\omegap}^{-7/2},{\omegap}^{-9/2}$ stated in the text.   

%%%%%%%%%%%%%%%%%%%%%%%%

\end{appendix}

%%%%%%%%%%%%%%%%%%%%%%%%%%%%

\end{document}